\documentclass[aps,twocolumn,prb,preprintnumbers,amsmath,amssymb,superscriptaddress,floats]{revtex4}
\usepackage{txfonts}
\usepackage{graphicx}
\usepackage{dcolumn}
\usepackage{bm}
\usepackage{hyperref}
\newcommand{\upperRomannumeral}[1]{\uppercase\expandafter{\romannumeral#1}}

\begin{document}

\title{High-Resolution Faraday Rotation and Electron-Phonon Coupling in Surface States of the Bulk-Insulating Topological Insulator Cu$_{0.02}$Bi$_2$Se$_3$}

\author{Liang Wu}
\email{lwu29@jhu.edu}
\affiliation{The Institute for Quantum Matter, Department of Physics and Astronomy, The Johns Hopkins University, Baltimore, MD 21218 USA.}
\author{Wang-Kong Tse}
\affiliation{Theoretical Division, Los Alamos National Laboratory, Los Alamos, New Mexico 87545, USA.}
\affiliation{Department of Physics and Astronomy, MINT Center, University of Alabama, Tuscaloosa, Alabama 35487, USA}
\author{M. Brahlek}
\altaffiliation{Present address: Department of Materials Science and Engineering, Pennsylvania State University, University Park, Pennsylvania 16801, USA. }
\affiliation{Department of Physics and Astronomy, Rutgers the State University of New Jersey. Piscataway, NJ 08854 USA.}
\author{C. M. Morris}
\affiliation{The Institute for Quantum Matter, Department of Physics and Astronomy, The Johns Hopkins University, Baltimore, MD 21218 USA.}

\author{R. Vald\'es Aguilar}

\affiliation{The Institute for Quantum Matter, Department of Physics and Astronomy, The Johns Hopkins University, Baltimore, MD 21218 USA.}
\affiliation{Department of Physics, The Ohio State University, Columbus, Ohio 43210, USA.}
\author{N. Koirala}
\affiliation{Department of Physics and Astronomy, Rutgers the State University of New Jersey. Piscataway, NJ 08854 USA.}
\author{S. Oh}
\affiliation{Department of Physics and Astronomy, Rutgers the State University of New Jersey. Piscataway, NJ 08854 USA.}
 \author{N. P. Armitage}
 \email{npa@jhu.edu}
 \affiliation{The Institute for Quantum Matter, Department of Physics and Astronomy, The Johns Hopkins University, Baltimore, MD 21218 USA.}

\begin{abstract}
We have utilized time-domain magneto-terahertz spectroscopy to investigate the low frequency optical response of topological insulator Cu$_{0.02}$Bi$_2$Se$_3$ and Bi$_2$Se$_3$ films.  With both field and frequency dependence, such experiments give sufficient information to measure the mobility and carrier density of multiple conduction channels simultaneously.  We observe sharp cyclotron resonances (CRs) in both materials.  The small amount of Cu incorporated into the Cu$_{0.02}$Bi$_2$Se$_3$ induces a true bulk insulator with only a \textit{single} type of conduction with total sheet carrier density $\sim4.9\times10^{12}/$cm$^{2}$ and mobility as high as 4000 cm$^{2}/$V$\cdot$s.   This is consistent with conduction from two virtually identical topological surface states (TSSs) on top and bottom of the film with a chemical potential $\sim$145 meV above the Dirac point and in the bulk gap.     The CR broadens at high fields, an effect that we attribute to an electron-phonon interaction.  This assignment is supported by an extended Drude model analysis of the zero field Drude conductance.  In contrast, in normal Bi$_2$Se$_3$ films two  conduction channels were observed and we developed a self-consistent analysis method to distinguish the dominant TSSs and coexisting trivial bulk/2DEG states.  Our high-resolution Faraday rotation spectroscopy on Cu$_{0.02}$Bi$_2$Se$_3$ paves the way for the observation of quantized Faraday rotation under experimentally achievable conditions to push chemical potential in the lowest Landau Level. 
\end{abstract}

\date{\today}
\maketitle

Topological insulators (TIs) are a newly discovered class of materials characterized by an inverted band structure \cite{Hasan-Kane-10,Qi-Zhang-11} caused by strong spin-orbit coupling.  In the ideal case, they have an insulating bulk and only conduct via topologically protected massless Dirac topological surface states (TSSs). Spin-momentum locking in their electronic structure makes TIs promising platforms for spintronics applications \cite{PesinNatMat12}.   Progress in this field has been hampered by the fact that all discovered TIs to date are slightly doped and have a conducting bulk.  For instance, the proposed topological magneto-electric effect \cite{Qi08b} and quantized Faraday rotation \cite{Tse10a}  remains unobserved. Tuning the chemical potential towards the Dirac point and enhancing mobility was shown to be very successful in probing many-body interactions with plasmons and phonons in graphene \cite{BasovRMP14} and similar advancements are expected in TIs, but have not yet been realized. 

The band structure of Bi$_2$Se$_3$ is one of the simplest of the 3D TIs with only a single Dirac cone at the center of the Brillouin zone.    Unfortunately, native grown Bi$_2$Se$_3$ is known to have a conducting bulk due to defects from the growth. Suppression of the bulk carrier density has been achieved by chemical doping methods \cite{RenPRB2010, XiongPhysicaE2012}.  Nevertheless, these samples still have significant densities of bulk carriers or impurity states that are pinned near E$_F$.   Recently it was found that $\sim2\%$ Cu doping in thin films suppresses the bulk carriers and allows a true insulating state to be realized \cite{BrahlekPRL14}. Here we investigate these copper doped bulk insulating thin films and their decoupled TSSs \cite{BrahlekPRL14} via magneto-terahertz spectroscopy.   These films were capped by a thin insulating amorphous Se layer.     Details of the growth can be found in the supplementary information (SI) section \upperRomannumeral{1}  \cite{WuSI} and in Ref. \cite{BrahlekPRL14}.

Cyclotron resonance (CR) experiments using THz spectroscopy are a powerful tool to study Dirac fermions and probe many-body interactions \cite{Crassee10a, JiangPRL07a}.  CR is also one of the most accurate measures of effective mass \cite{KonoReview06}. In previous work, a large Kerr rotation in  bulk-conducting Bi$_2$Se$_3$ films was reported, but no obvious resonance was observed \cite{ValdesAguilarPRL12}.  Cyclotron resonance has been reported in In$_2$Se$_3$ capped films \cite{JenkinsPRB13}, but current understanding is that significant Indium diffusion from In$_2$Se$_3$ to Bi$_2$Se$_3$ \cite{LeeThinFilms14} destroys the simple non-TI/TI boundary at the interface, as a topological phase transition occurs at low Indium concentrations ($\sim 6\%$) \cite{WuNatPhys13}. 

\begin{figure*}[htp]
\includegraphics[trim = 10 5 5 5,width=5.5cm]{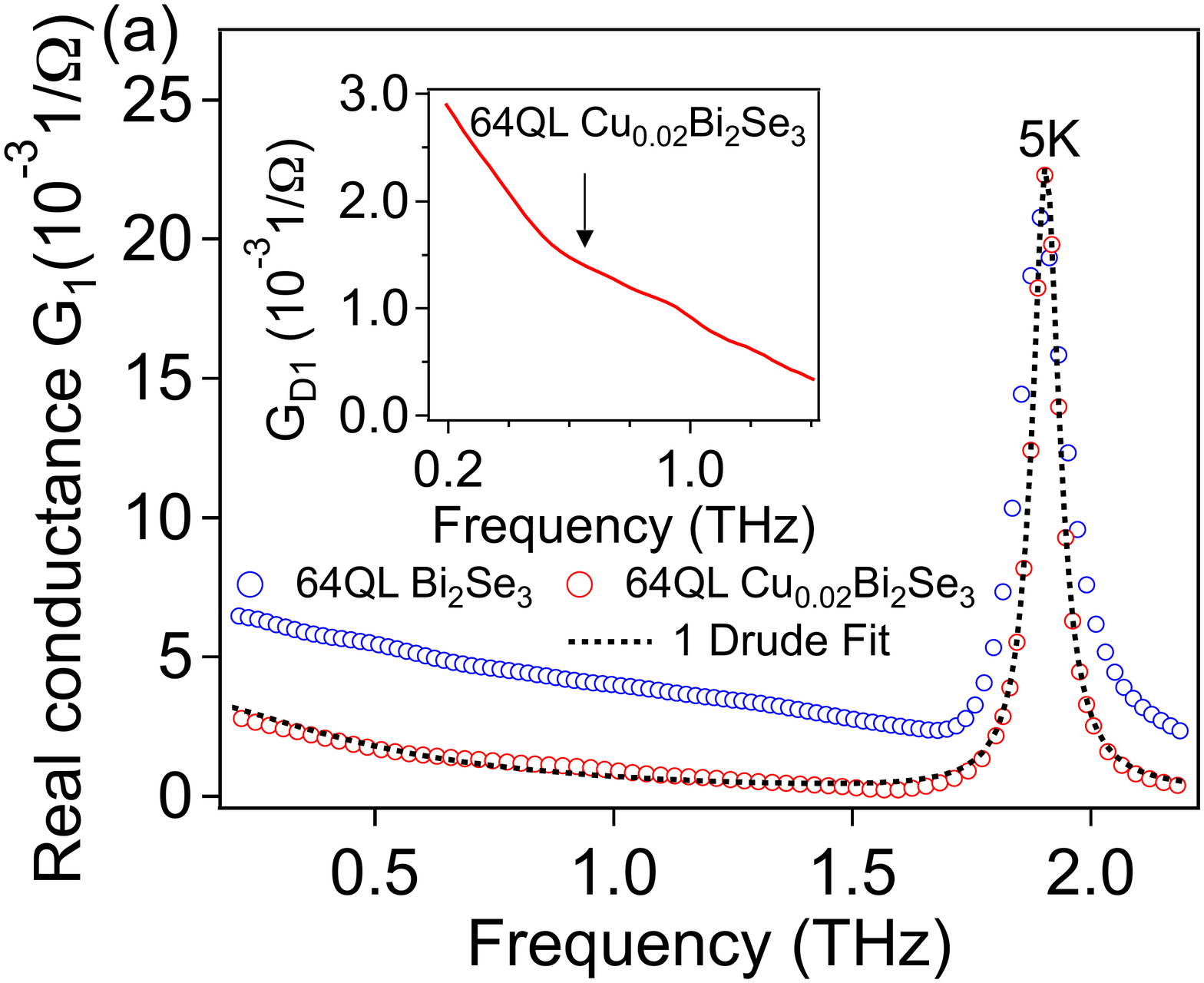}
\includegraphics[trim = 10 5 5 5,width=5.5cm]{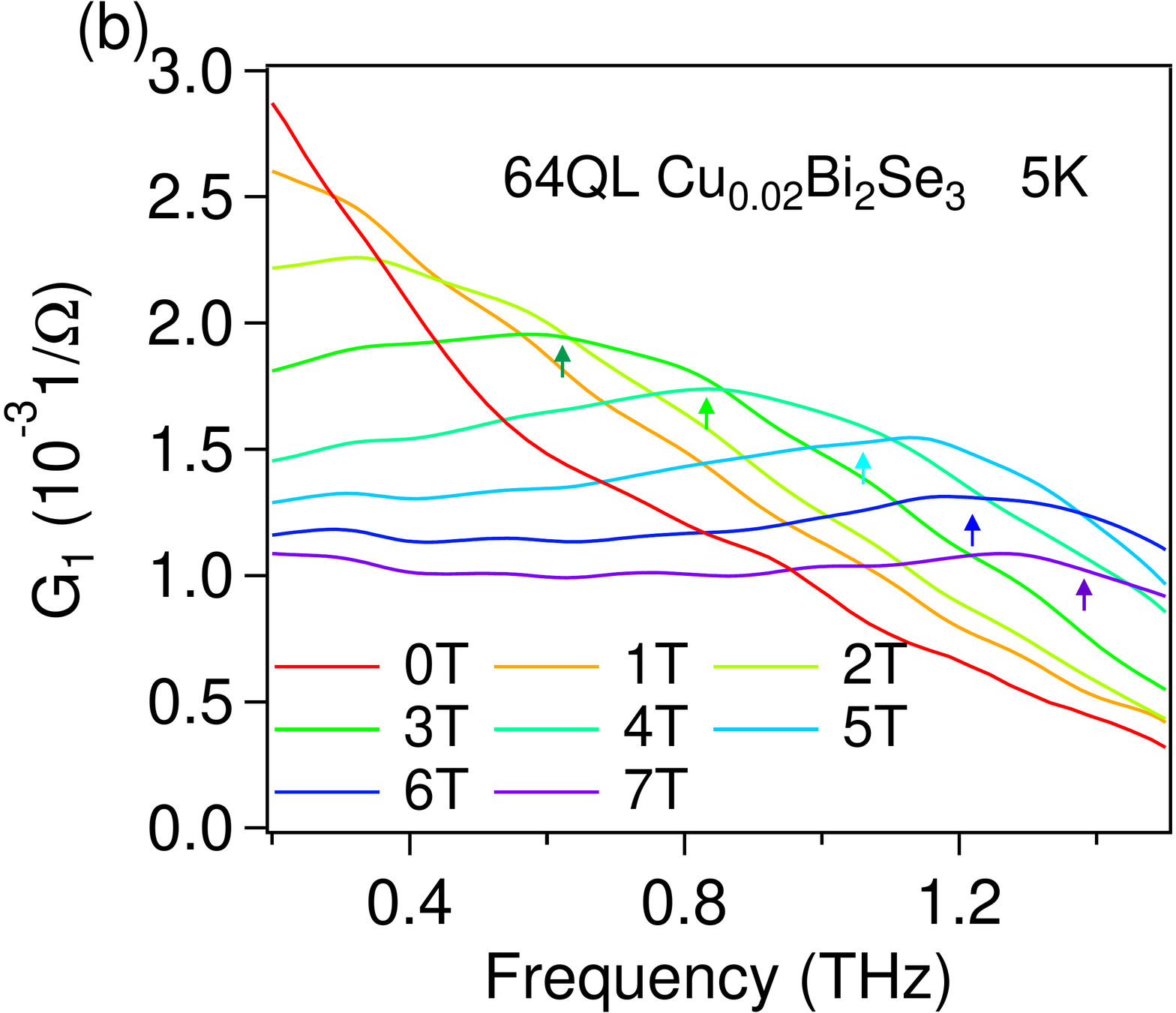}
\includegraphics[trim = 10 5 5 5,width=5.5cm]{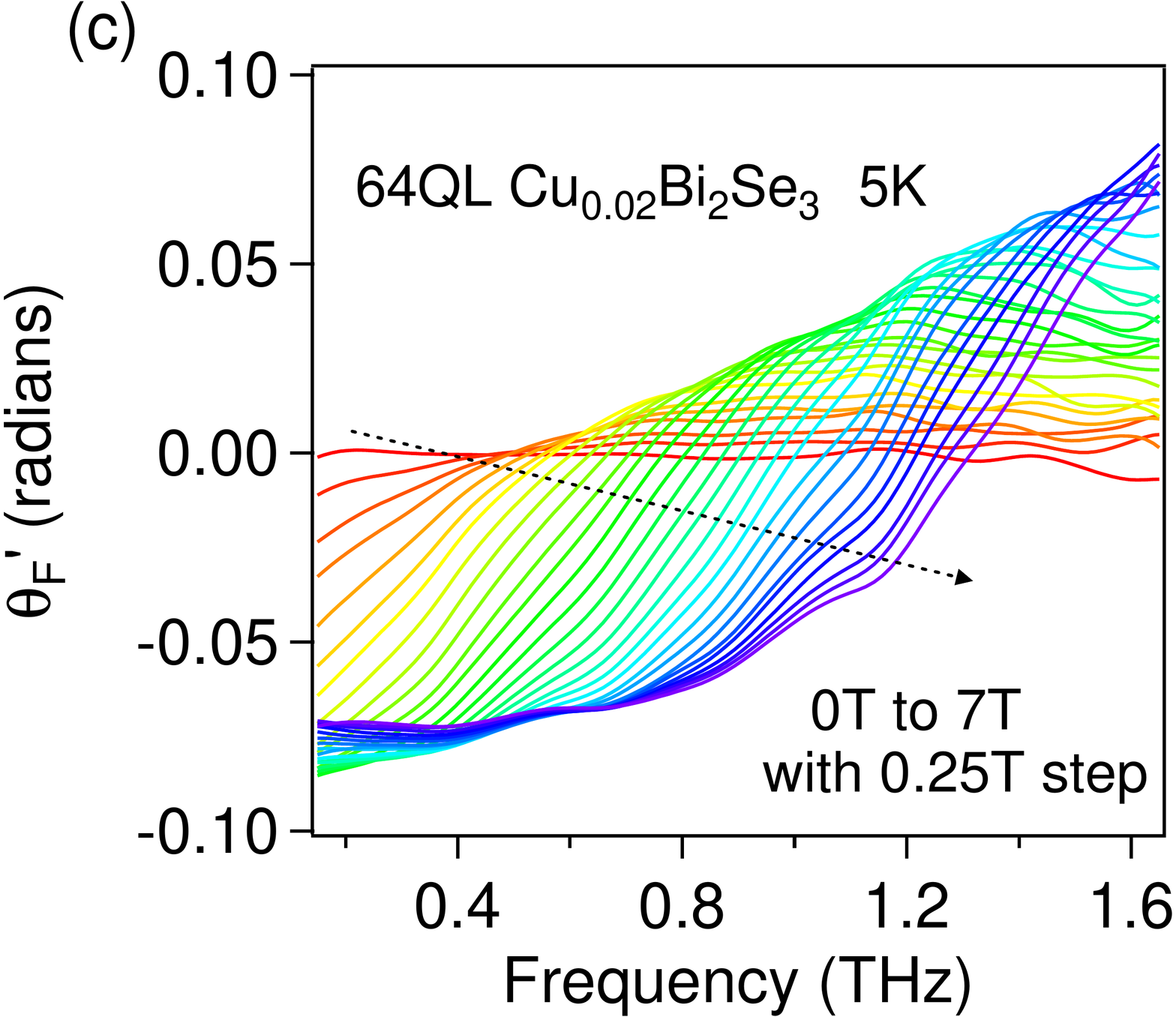}

\includegraphics[trim = 10 5 5 5,width=5.5cm]{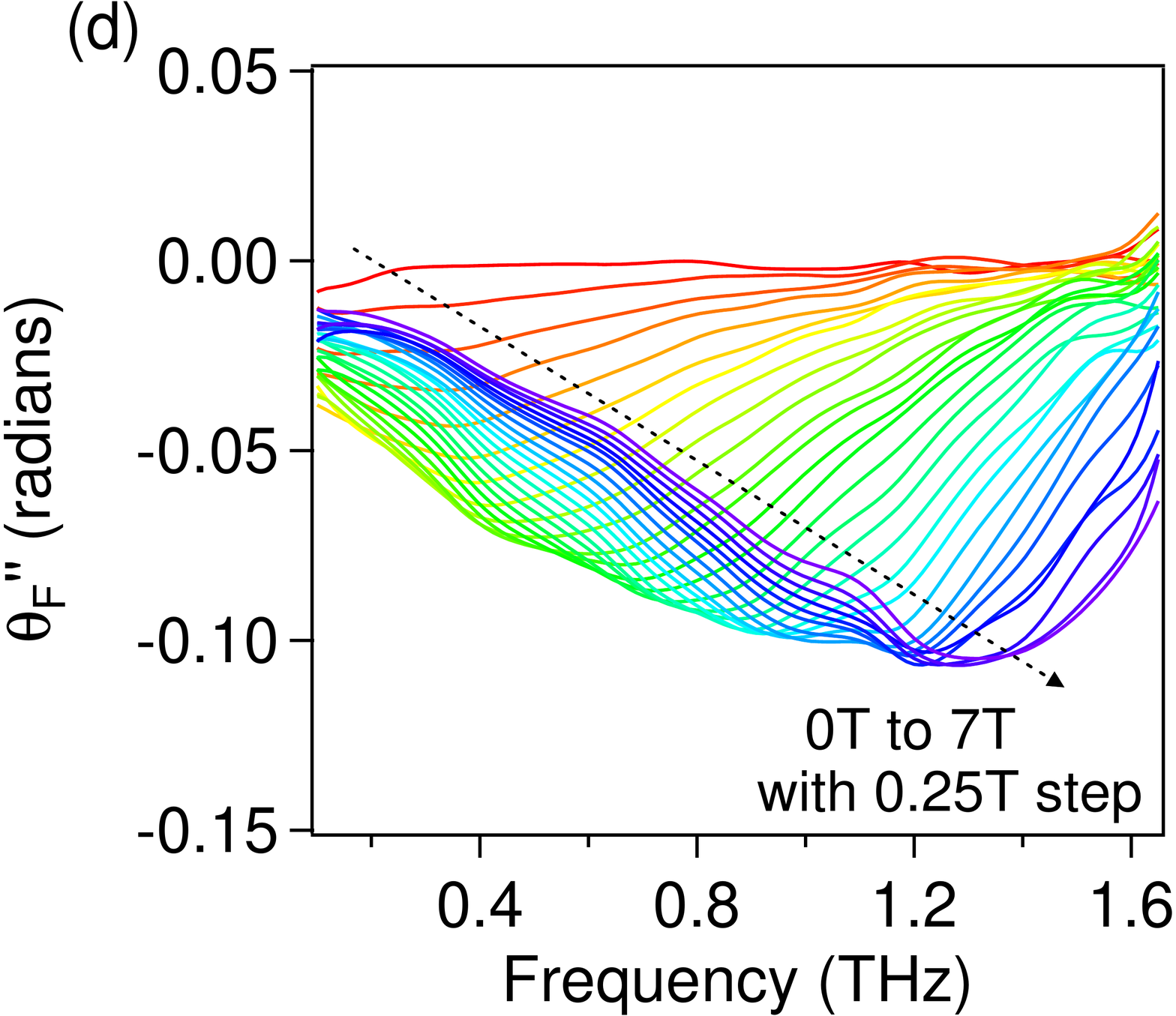}
\includegraphics[trim = 10 5 5 5,width=5.5cm]{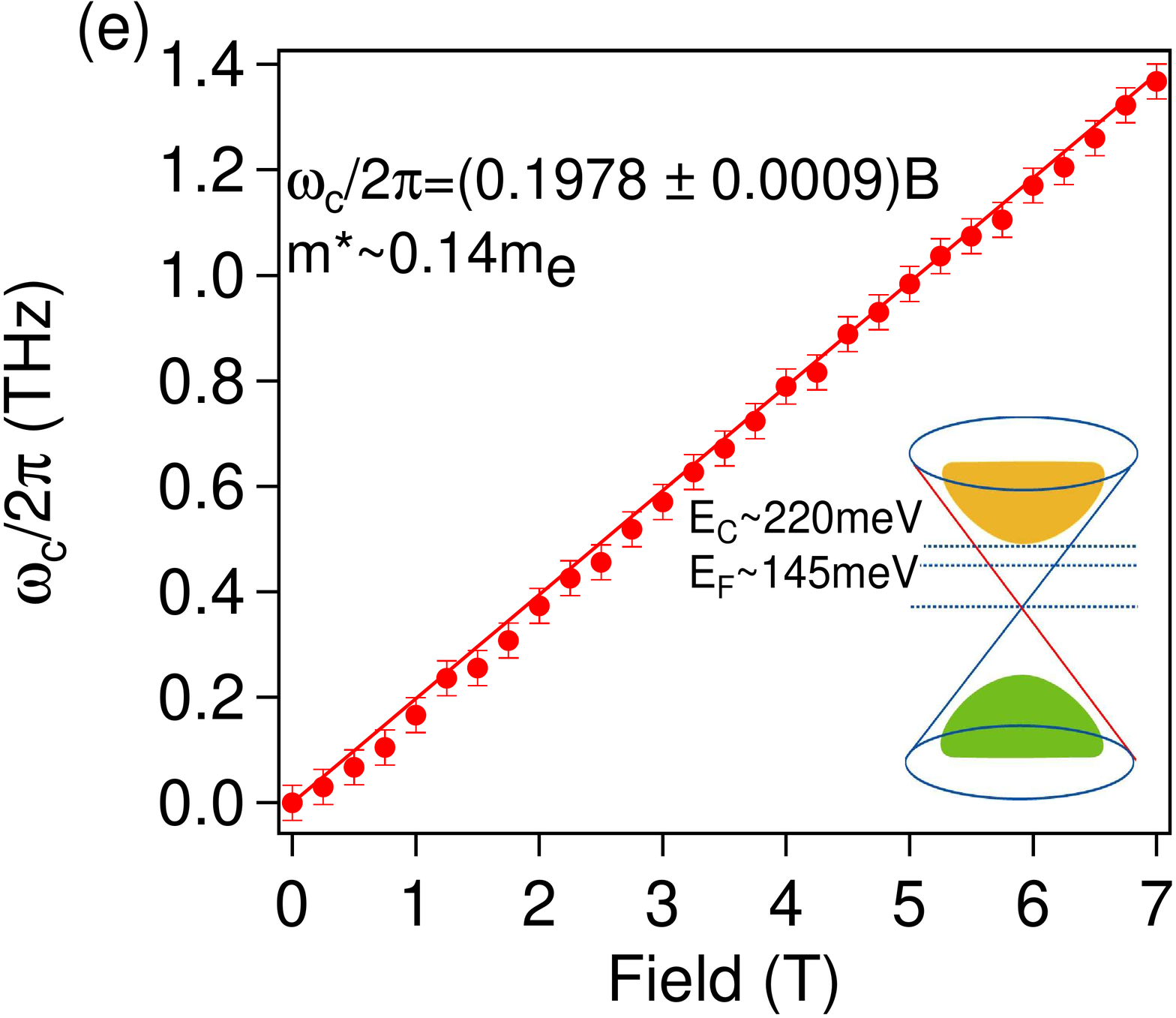}
\includegraphics[trim = 10 5 5 5,width=5.5cm]{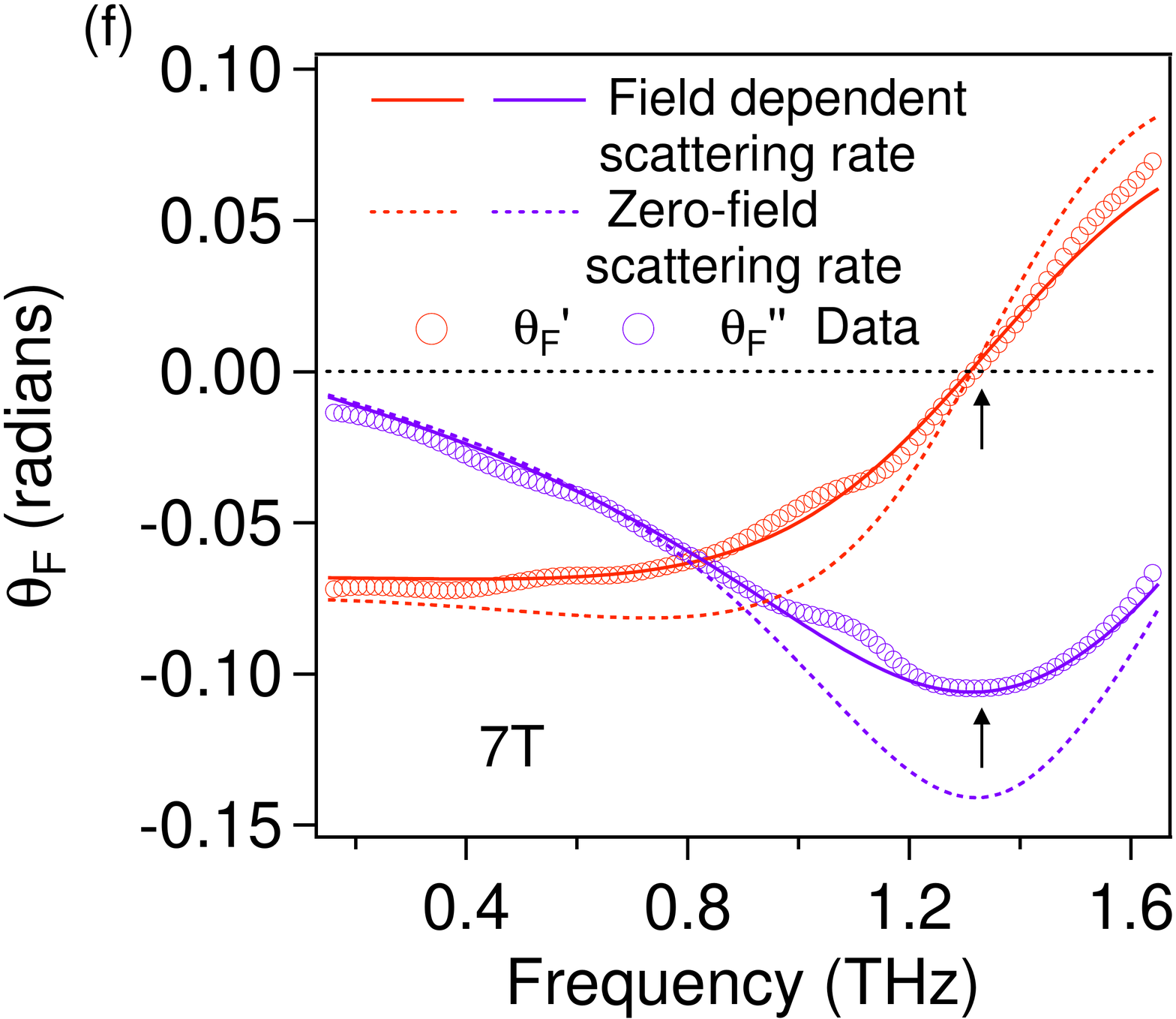}

\caption{(Color online) Data summary for 64 QL Cu$_{0.02}$Bi$_2$Se$_3$ (sample1). (a) Real sheet conductance of 64 QL Cu$_{0.02}$Bi$_2$Se$_3$ and pure Bi$_2$Se$_3$ films at 5K. Inset is the Drude conductance  $G_D$ of 64 QL Cu$_{0.02}$Bi$_2$Se$_3$ after subtracting the phonon and $\epsilon_{\infty}$ contributions.   The arrow indicates the deviation from pure Lorentzian form that arises from electron-phonon coupling.   (b) Field dependent conductance at 5K.  Arrows are guides to the eye for cyclotron frequencies. (c) Real and (d) Imaginary parts of the complex Faraday rotation data at different fields at 5K. (e) Cyclotron frequency versus field. Solid line is a linear fit. The inset is a cartoon indicating E$_F \sim$  145 meV above the Dirac point (75 meV below conduction band minimum). (f) Complex Faraday angle with fits at 7 T.  The solid curve is a fit with a field-dependent scattering rate.   The dashed curve is a fit using the zero-field scattering rate.  Arrows are guides to the eye for cyclotron frequency.}   \label{Fig1}
\end{figure*}   

In the present work, we used time-domain magneto-terahertz spectroscopy  with the polarization modulation technique \cite{MorrisOE12}  (0.5mrad resolution, see SI section \upperRomannumeral{1} for experimental details \cite{WuSI} )  to observe sharp cyclotron resonances in Faraday rotations from both Cu$_{0.02}$Bi$_2$Se$_3$ and pure Bi$_2$Se$_3$ thin films.  We demonstrate that Cu$_{0.02}$Bi$_2$Se$_3$ can be described by a $single$ Drude component with total carrier density $n_{2D} \sim4.9 \times10^{12}/$cm$^{2}$. This Drude contribution is consistent with pure surface state transport with an E$_F \sim 145$ meV above the Dirac point (75 meV below the conduction band edge), which makes Cu$_{0.02}$Bi$_2$Se$_3$ a true topological insulator.  CR broadening at high field is attributed to an electron phonon interaction. In contrast, we find two channel conduction in pure Bi$_2$Se$_3$ films.   We determine that the large Faraday rotation is induced by a dominant high-mobility TSSs channel with an E$_F \sim$ 350 meV.   However, a weaker low-mobility second Drude term is also required to fit the data. This subdominant term most likely derives from trivial states (bulk and/or 2DEG).

In Fig. \ref{Fig1}(a), we compare the zero field THz conductance of a 64 QL Cu$_{0.02}$Bi$_2$Se$_3$ film to a pure 64 QL  Bi$_2$Se$_3$ film. The Cu incorporated sample's spectra are characterized by a reduced total spectral weight and slightly lower scattering rate than the pure Bi$_2$Se$_3$.  The spectra can be well fit by an oscillator model with only a Drude term describing free electron-like motion, a Drude-Lorentz term modeling the phonon and a lattice polarizability $\epsilon_{\infty}$ term that originates from absorptions above the measured spectral range.

\small
\begin{equation}
 G(\omega)=   \epsilon_0 d \left(-\frac{\omega^{2}_{pD}}{i\omega-\Gamma_{D}}-\frac{i\omega\omega^{2}_{pDL}}{\omega^{2}_{DL}-\omega^{2}-i\omega\Gamma_{DL}}-i\left(\epsilon_{\infty}-1\right)\omega \right)
\end{equation}
\label{Eqa1}
\normalsize

\noindent Here $\Gamma$'s are scattering rates,  $\omega_p$'s are plasma frequencies, and  $d$ is the film thickness. The spectral weight ($\omega_{pD}^{2} d$) is proportional to the integrated area of each feature in the real part of the conductance.  It gives the ratio of carrier density to an effective transport mass. 

\small
\begin{equation}
\frac{2}{\pi\epsilon_{0}}\int G_{D1} d \omega = \omega_{pD}^{2} d =\frac{n_{2D}e^{2}}{m^{*}\epsilon_{0}}
\label{Eqa2}
\end{equation}

\normalsize

 \begin{figure*}[htp]
\includegraphics[trim = 10 5 5 5,width=5.5cm]{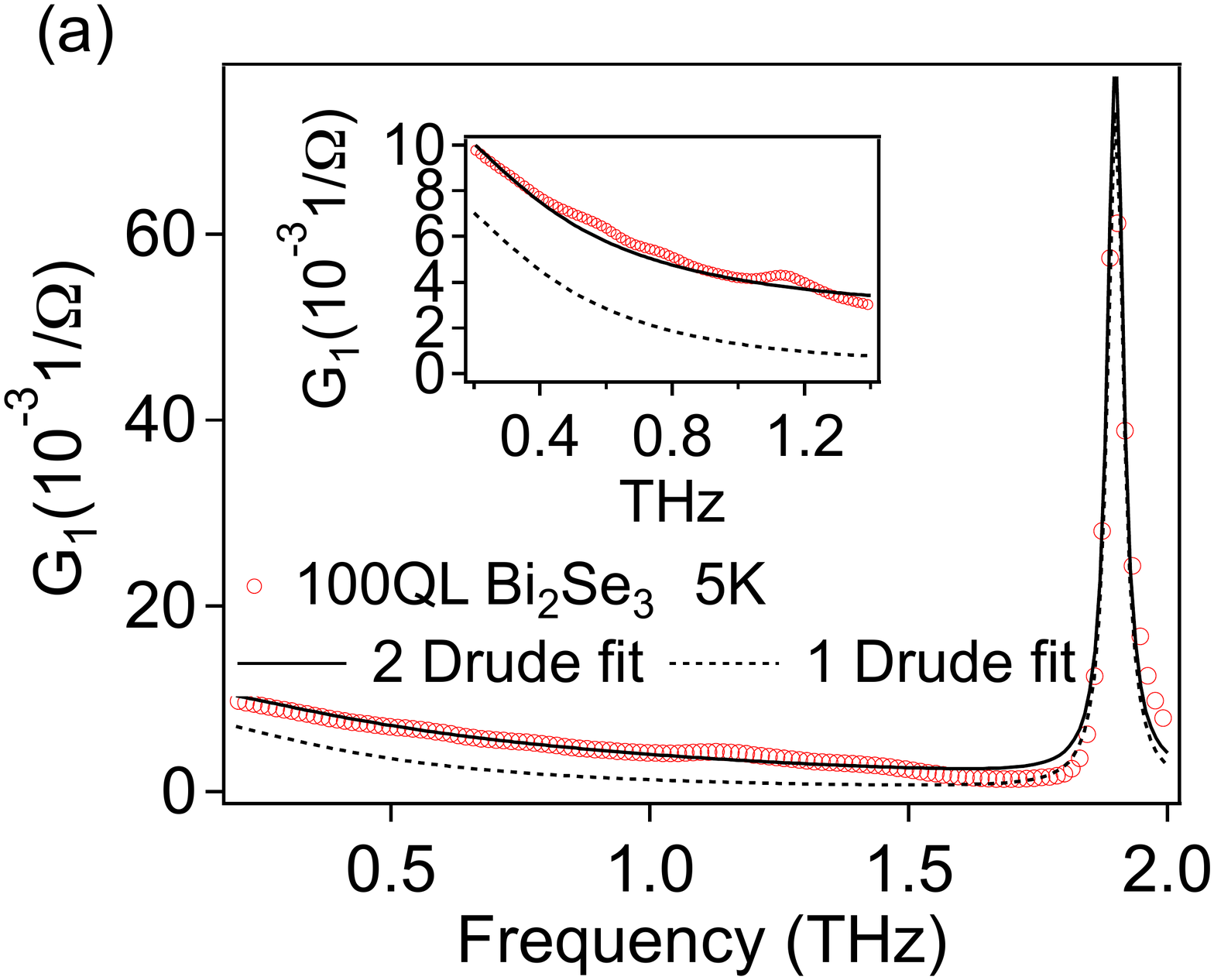}
\includegraphics[trim = 10 5 5 5,width=5.5cm]{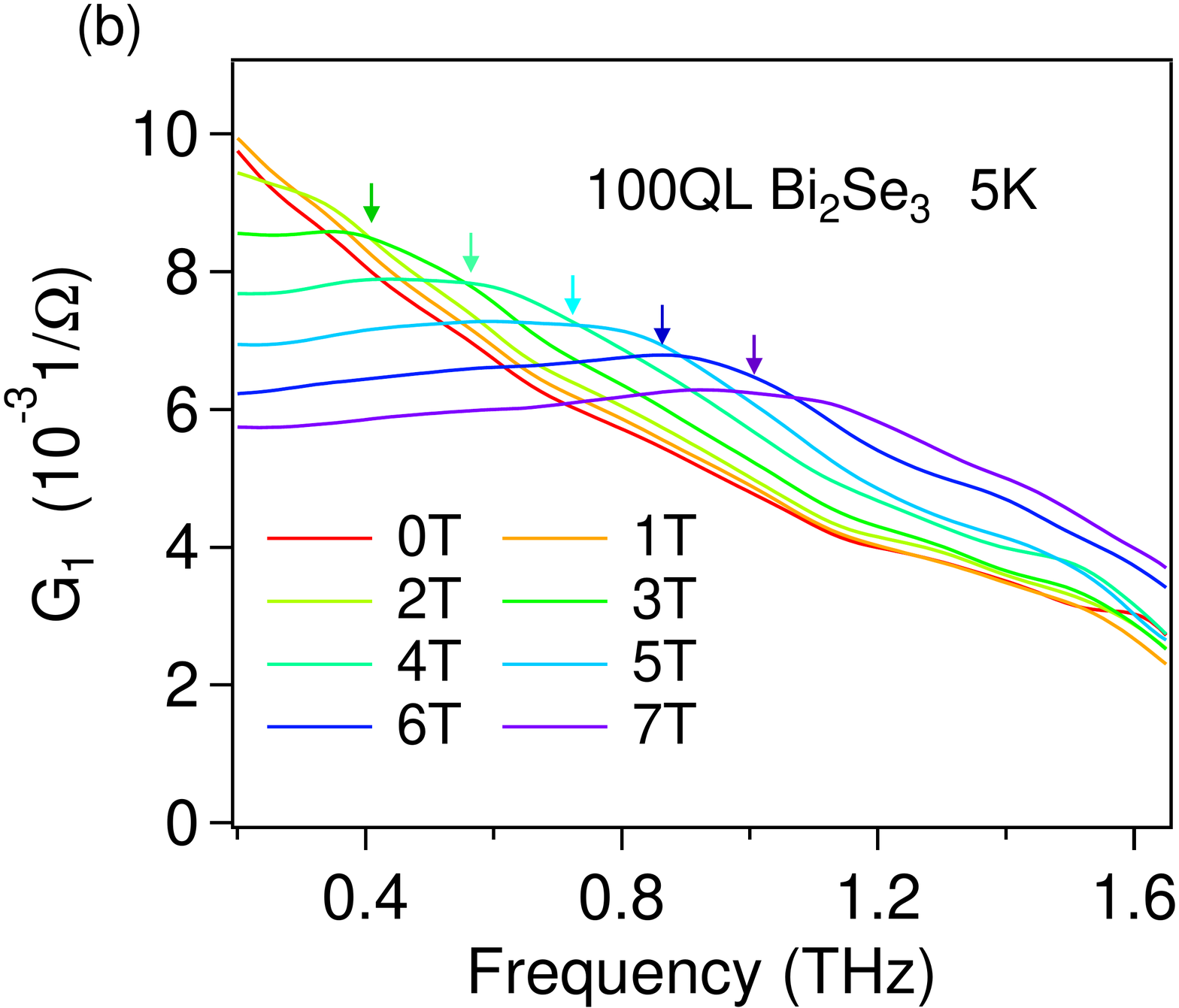}
\includegraphics[trim = 10 5 5 5,width=5.5cm]{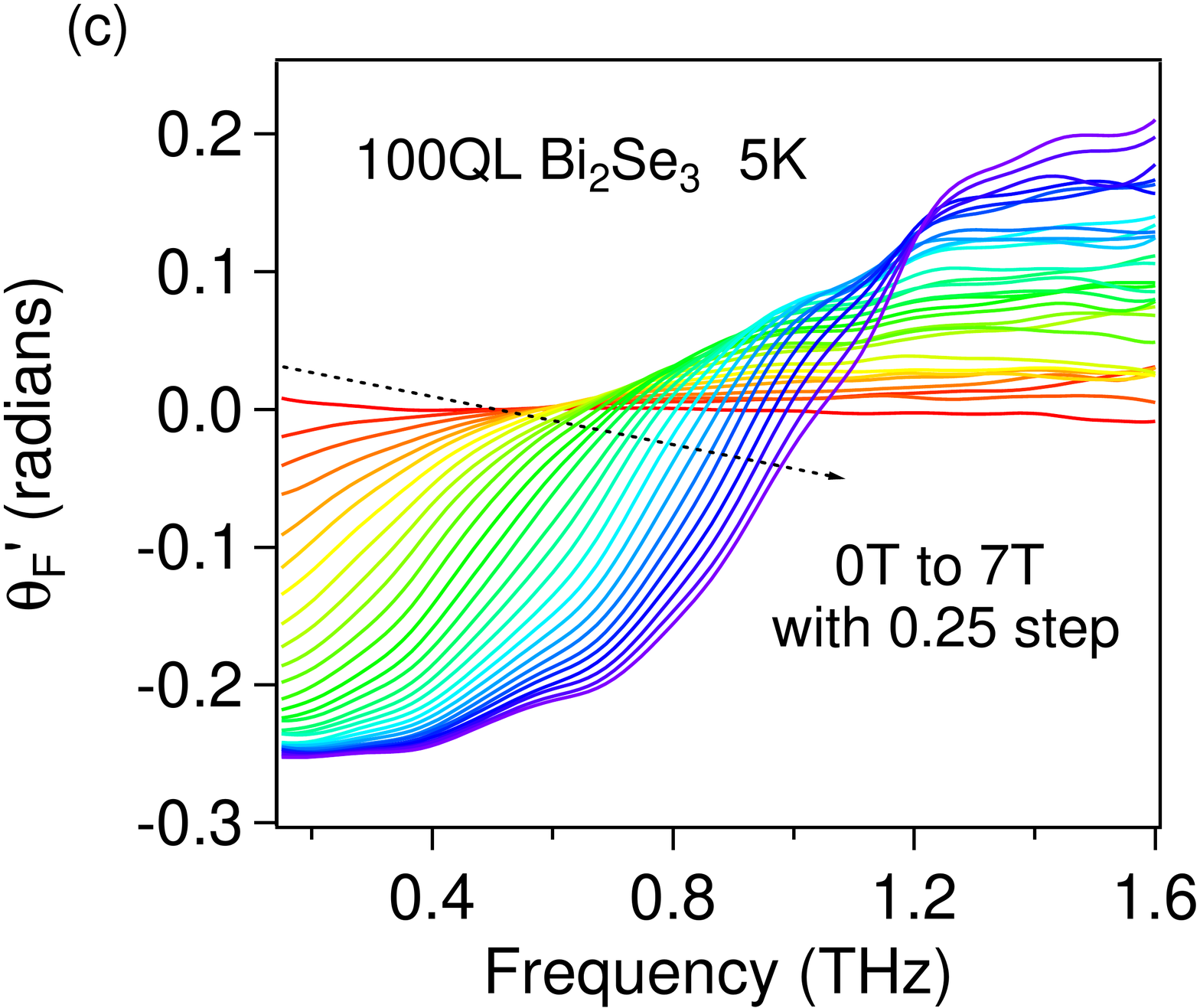}

\includegraphics[trim = 10 5 5 5,width=5.5cm]{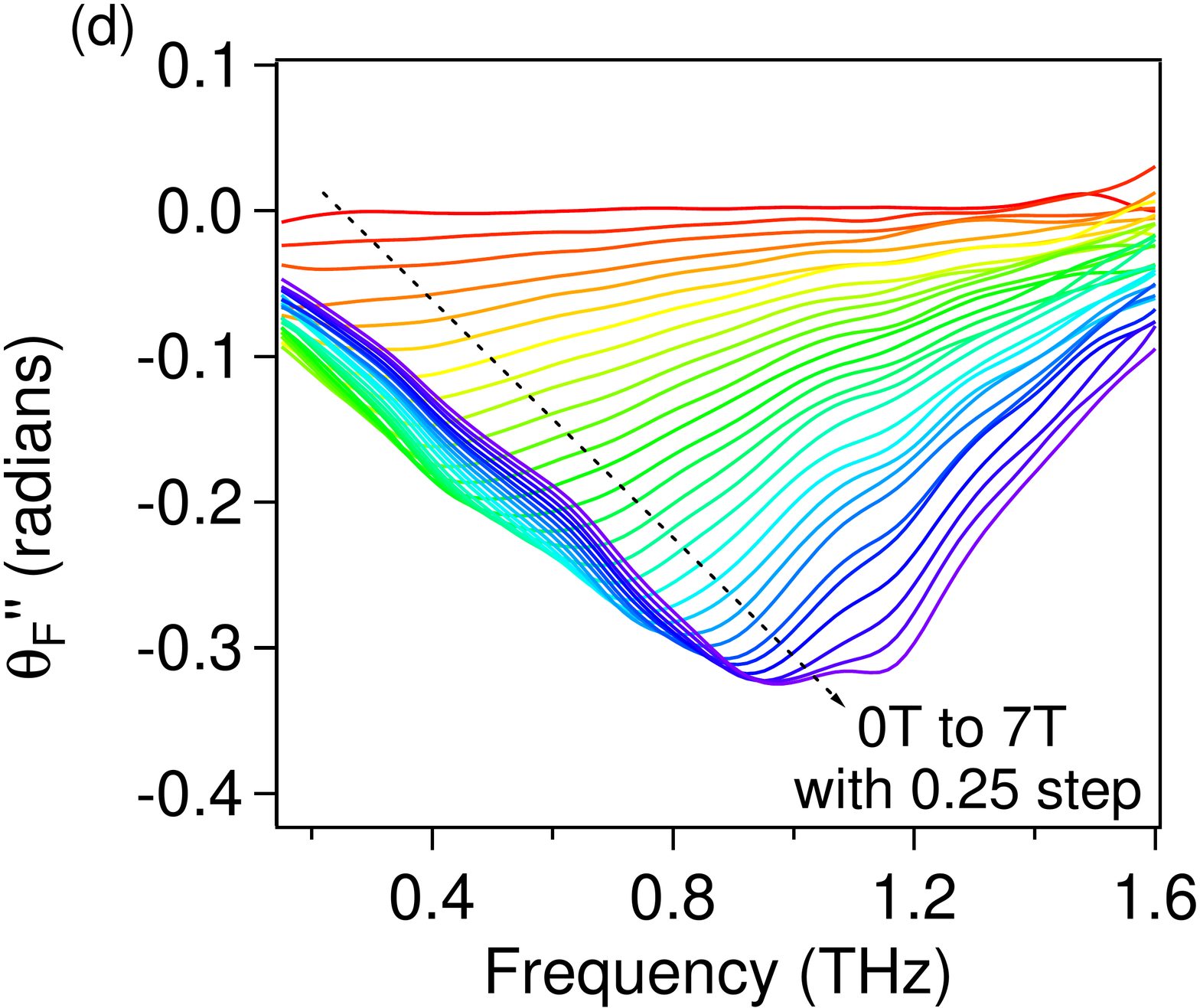}
\includegraphics[trim = 10 5 5 5,width=5.5cm]{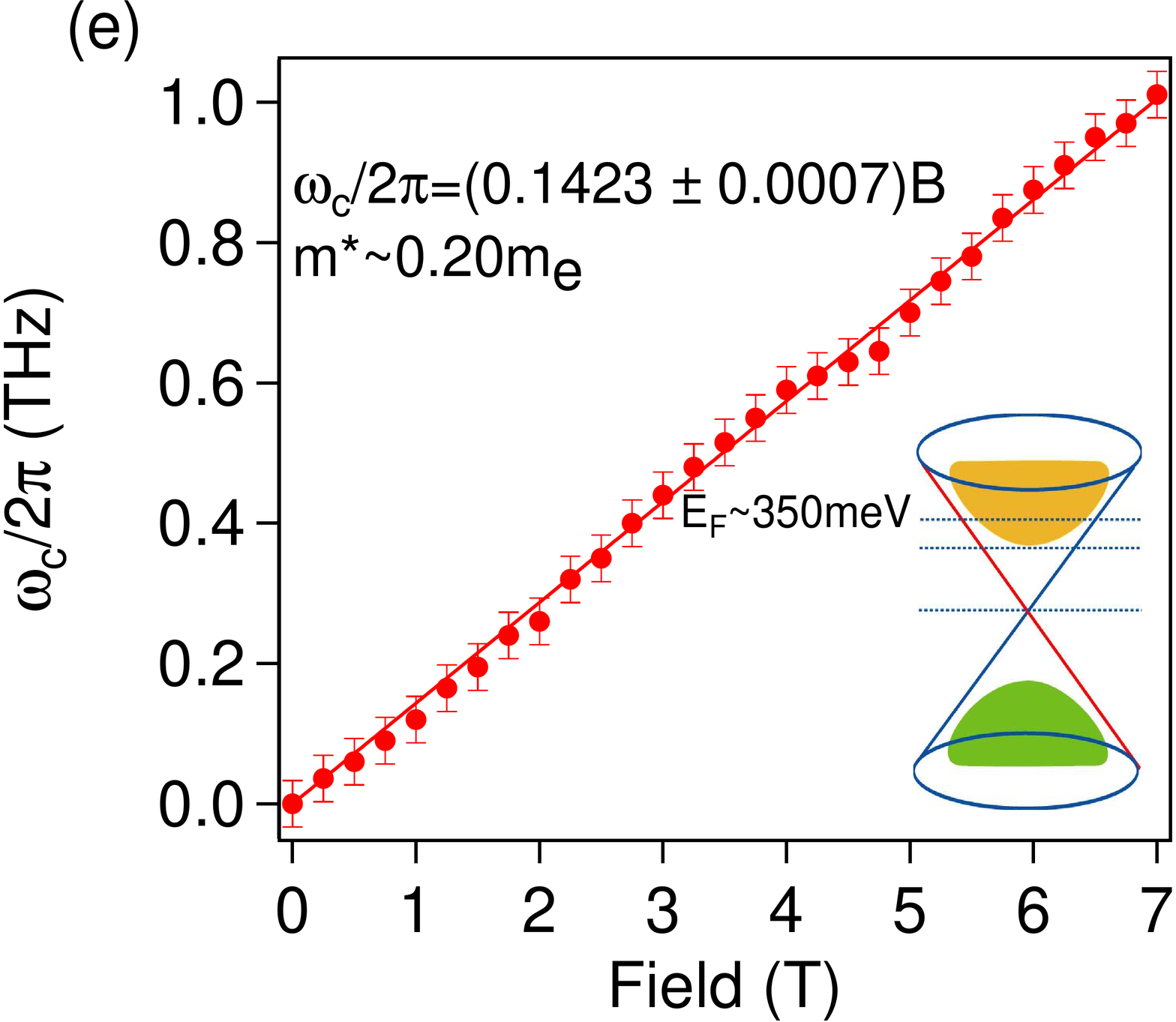}
\includegraphics[trim = 10 5 5 5,width=5.5cm]{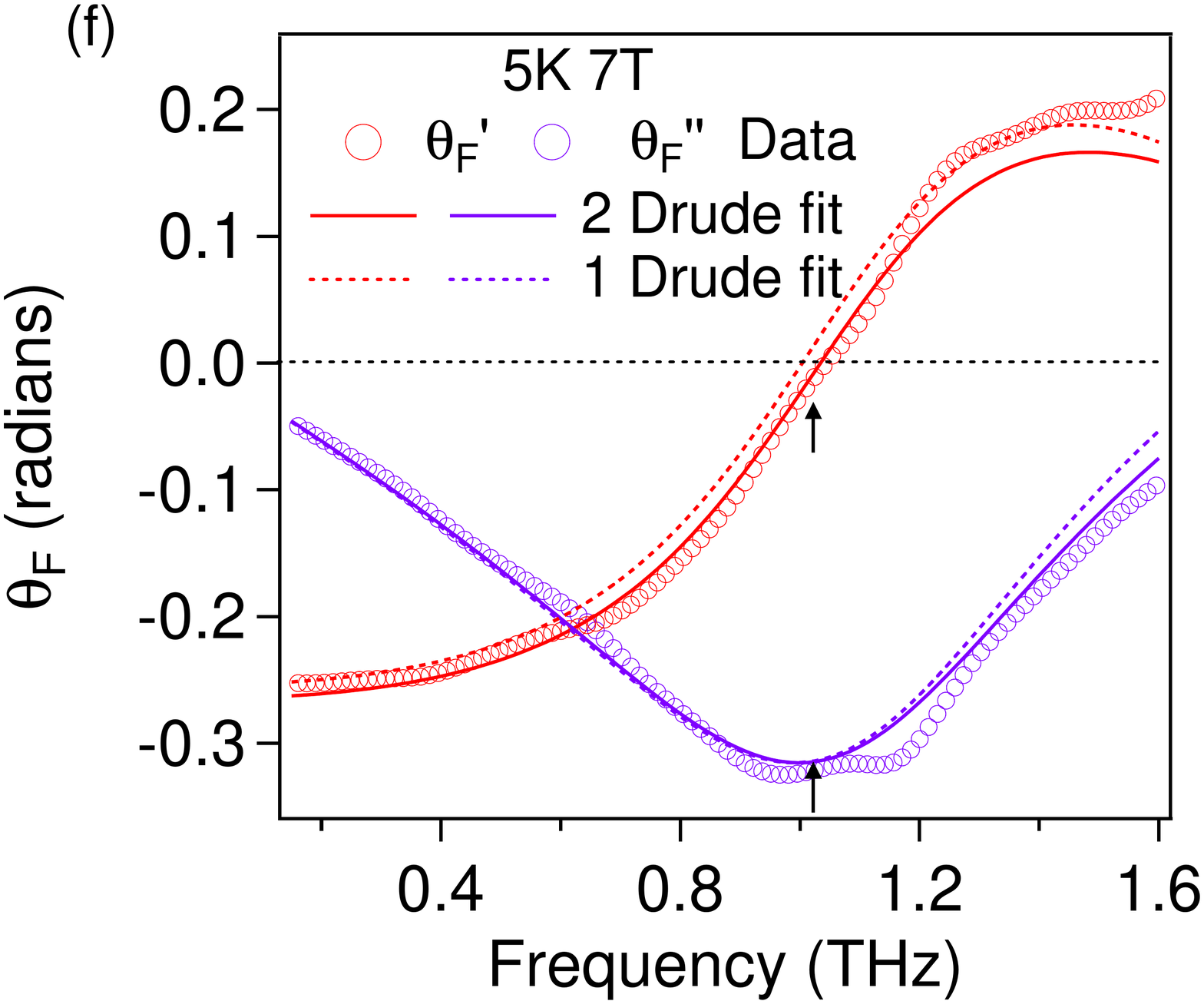}

\caption{(Color online) (a) Real conductance of 100 QL Bi$_2$Se$_3$ comparing single and two component Drude fits as detailed in text.  The inset enlarges the low frequency regime. (b) Field dependent conductance at 5K. (c) The real and (d) imaginary parts of the complex Faraday rotation data at different fields at 5K. (e) The cyclotron resonance frequency versus field. The inset is a cartoon indicating E$_F \sim$ 350 meV above the Dirac point. (f) Complex Faraday rotation data with one and two component Drude fits at 7 T.  The dashed and solid curves are one and two Drude fits respectively.} \label{Fig2}
\end{figure*} 

\noindent Here $m^*$ is defined as $\hbar k_F / v_F$ for ``massless Dirac fermions".  Considering the TSS dispersion up to quadratic corrections $E=Ak_F+Bk_F^2$, the spectral weight can be expressed in terms of $k_{F}$, where A and B are parameters obtained from ARPES (see SI section \upperRomannumeral{2} \cite{WuSI} ).

\small
\begin{equation}
 \omega_{pD}^{2} d =  \frac{k_F ( A + 2B k_F)e^2 }{ 2 \pi \hbar^2 \epsilon_0}
 \label{Eqa3}
\end{equation}

\normalsize

\noindent  This expression assumes the single channel conduction  originates in two nominally identical TSSs.  From the spectral weight analysis, having determined $k_F$, we can then calculate both  $n_{2D}$ (proportional to $k_F^2$) and m$^{*}$.  From these Drude-Lorentz fits, we find a \textit{total} sheet carrier density $n_{2D}$ $\sim$ 5.0 $\times$10$^{12}$/cm$^{2}$, $m^{*}\sim$ 0.14 $m_{e}$, and E$_F \sim$ 145 ($\pm5$) meV according to Eq.\ref{Eqa3}.

We can estimate the carrier density and mass by the zero field spectra alone because both can be expressed as a function of $k_{F}$.  Below we determine the mass in a model free fashion through CR experiments.  The signature of CR is a peak in the real part of the conductance (Fig. \ref{Fig1}(b)), an inflection point in real part of Faraday rotation $\theta_F^{'}$ (Fig. \ref{Fig1}(c)) and a dip in the imaginary part of Faraday rotation  $\theta_F^{''}$ (Fig. \ref{Fig1}(d)). Its full width at half maximum (FWHM) is the scattering rate. Field dependent complex Faraday rotation data are shown in Fig. \ref{Fig1}(c)(d). One can see that an edge feature around an inflection point in $\theta_F^{'}$ and a dip in $\theta_F^{''}$ shifts to higher frequency with increasing field, which is consistent with CR. We fit the data by a Drude-Lorentz model accounting for the field dependence of the Drude term and constraining the parameters of the phonon and $\epsilon_{\infty}$ by the values extracted from the zero-field conductance fits. The conductance in a magnetic field can be described by the expression:

\small 
\begin{equation}
G_{\pm}  = -i\epsilon_{0}\omega d  \left( \frac{\omega^{2}_{pD}}{-\omega^{2}-i\Gamma_{D}\omega\mp\omega_{c}\omega} +  \frac{\omega^{2}_{pDL}}{\omega^{2}_{DL}-\omega^{2}-i\omega\Gamma_{DL}} +\left(\epsilon_{\infty}-1\right) \right)
\label{Eqa4} 
\end{equation}
\normalsize 

\noindent Here the $\pm$ sign signifies the response to right/left-hand circularly polarized light, respectively.  $\omega_{c}$ is the CR frequency to be defined below.    The Faraday rotation can be  expressed as $\mathrm{tan}(\theta_{F})=-i (t_{+}-t_{-})/(t_{+}+t_{-}) $ \cite{note1}.   Note that the Faraday equation is a complex quantity because, in addition to rotations, phase shifts that are different for right/left-hand polarized light can be accumulated. The imaginary part is related to the ellipticity \cite{MorrisOE12}.

The fits to this model for the Faraday rotation are shown for a representative field of 7 T in Fig. \ref{Fig1}(f) (see SI section \upperRomannumeral{2} for fits to all fields \cite{WuSI} ). In this plot, the dashed curves are from a fit with the spectral weight and scattering rate $\sim$ 0.4 THz set by the zero-field conductance fit (e.g. only $\omega_{c}$ allowed to vary).  One can see that although the gross features of the spectra are reproduced, using the  zero field scattering rate  entirety fails to reproduce certain aspects of the Faraday rotation, including the value of the minimum in  $\theta_F^{''}$.  A much better fit (solid line) can be obtained by letting the scattering rate vary with field, while keeping other parameters (except for $\omega_{c}$) fixed. The origin of this field dependent scattering rate will be addressed below. The fits allow us to extract the cyclotron frequency as a function of field; it is exhibited by the raw spectra as the minimum in  $\theta_F^{''}$. As shown in Fig. \ref{Fig1}(e), a linear fitting using the expected relation between mass and the resonance frequency $\omega_c=eB/m^*$ gives an effective mass of 0.14 $m_e$. By using Eq. \ref{Eqa2} and spectral weight obtained from fitting Faraday rotation, we can extract a \textit{total} sheet carrier density  $n_{2D}=4.9 \pm0.1 \times10^{12}/$cm$^{2}$.     With the known band structure of Bi$_2$Se$_3$, this charge density, and the observation of a single kind of charge carrier is only consistent with two essentially identical TSSs and an $E_{F} \sim$145  meV above the Dirac point.  We can conclude that Cu$_{0.02}$Bi$_2$Se$_3$ has an insulating bulk and for the sample highlighted here a high mobility $\mu=e/\Gamma_{D}m^{*}\sim$ 4000cm$^{2}/$V$\cdot$ s.  The features were robust to sample aging as the samples (with the amorphous Se cap) maintain bulk-insulating and high mobility after sitting in air for eight months (see SI section \upperRomannumeral{2} \cite{WuSI} ).

Films of Cu$_{0.02}$Bi$_2$Se$_3$ can be contrasted with films of pure Bi$_2$Se$_3$, which is known to have the surface chemical potential pinned in the bulk conduction band \cite{Xia09a, BansalPRL12}.    We develop a self-consistent data analysis method and use our high-resolution magneto-terahertz spectroscopy to precisely determine the contribution of the subdominant bulk.  In Fig. \ref{Fig2}(a), we show zero field conductance spectra from a typical 100 QL Bi$_2$Se$_3$ film.   It has higher spectral weight than Cu$_{0.02}$Bi$_2$Se$_3$, consistent with a higher charge density.  For the same reason, the plateau-like Faraday rotation at 7 T at low frequencies in Fig. \ref{Fig2}(f) is as large as $\sim$0.25 radians  while the value for the Cu$_{0.02}$Bi$_2$Se$_3$ sample is $\sim$0.07 radians.

In Figs. \ref{Fig2}(b)-(d), one can see that the CR is exhibited at lower frequencies, which indicates that Bi2Se3 has a higher CR mass. In Fig. \ref{Fig2}(e), the linear fit of $\omega_{c}$ vs. B gives an effective mass $\sim$0.20 $m_{e}$.    We believe that this derives from TSSs, as this mass is inconsistent with the accepted values for the bulk bands\cite{Kohler1973} or band bending induced surface 2DEG bands \cite{BIanchi10}. Note that the value of CR mass in Bi$_2$Se$_3$ we determine here is different than that given in our previous work \cite{ValdesAguilarPRL12}.  The reason for this discrepancy is discussed at length in the SI section \upperRomannumeral{6} \cite{WuSI} . 

One can see from Fig. \ref{Fig2}(f) that fits of the Faraday rotation using only a single Drude term are reasonably good.  As before we use the spectral weight and CR mass by Eq. \ref{Eqa2} to extract a total sheet carrier density  $n_{2D}\sim1.9 \pm0.1 \times10^{13}/$cm$^{2}$.   This compares favorably to a density of $n_{2D}\sim2.0\times10^{13}/$cm$^{2}$, a mass of $\sim$0.20$m_{e}$, and an E$_{F}\sim$350 meV that we can determine purely from an analysis of the spectral weight using the TSS dispersion by Eq. \ref{Eqa3}.   We determine a mobility of $\mu\sim3200$ cm$^{2}/$V$\cdot$ s of the TSSs.

However, while the fits to the Faraday rotation with a single Drude term are excellent, significant discrepancies arise if we use the resulting parameters to fit the conductance data.  In Fig. \ref{Fig2}(a), one can see that the single Drude component fit significantly underestimates the conductance by a roughly constant amount over the entire spectral range.   As E$_{F}$ is in the bottom of the conduction band, it is reasonable to ascribe this difference to a subdominant low mobility conduction channel originating in bulk or 2DEG.  Adding a second term with a large scattering rate ($\Gamma/2\pi >$ 4 THz such that the contribution to the real conductance is a constant offset) improves the conductance fit dramatically.   If literature values for the bulk or 2DEG mass are used, this channel has carrier density $n_{2D}\sim8.0\times10^{12}/$cm$^{2}$ and low mobility $\mu<300$cm$^{2}/$V$\cdot$s. The second flat Drude term only adds a featureless small background to the Faraday rotation. The ratio $G_{1TSSs}/G_{1total}\geqslant90\%$ at low frequencies confirms that TSSs dominate transport as inferred in previous THz measurements \cite{ValdesAguilarPRL12, WuNatPhys13}.

\begin{figure}[htp]
\includegraphics[trim = 10 5 5 5,width=5.5cm]{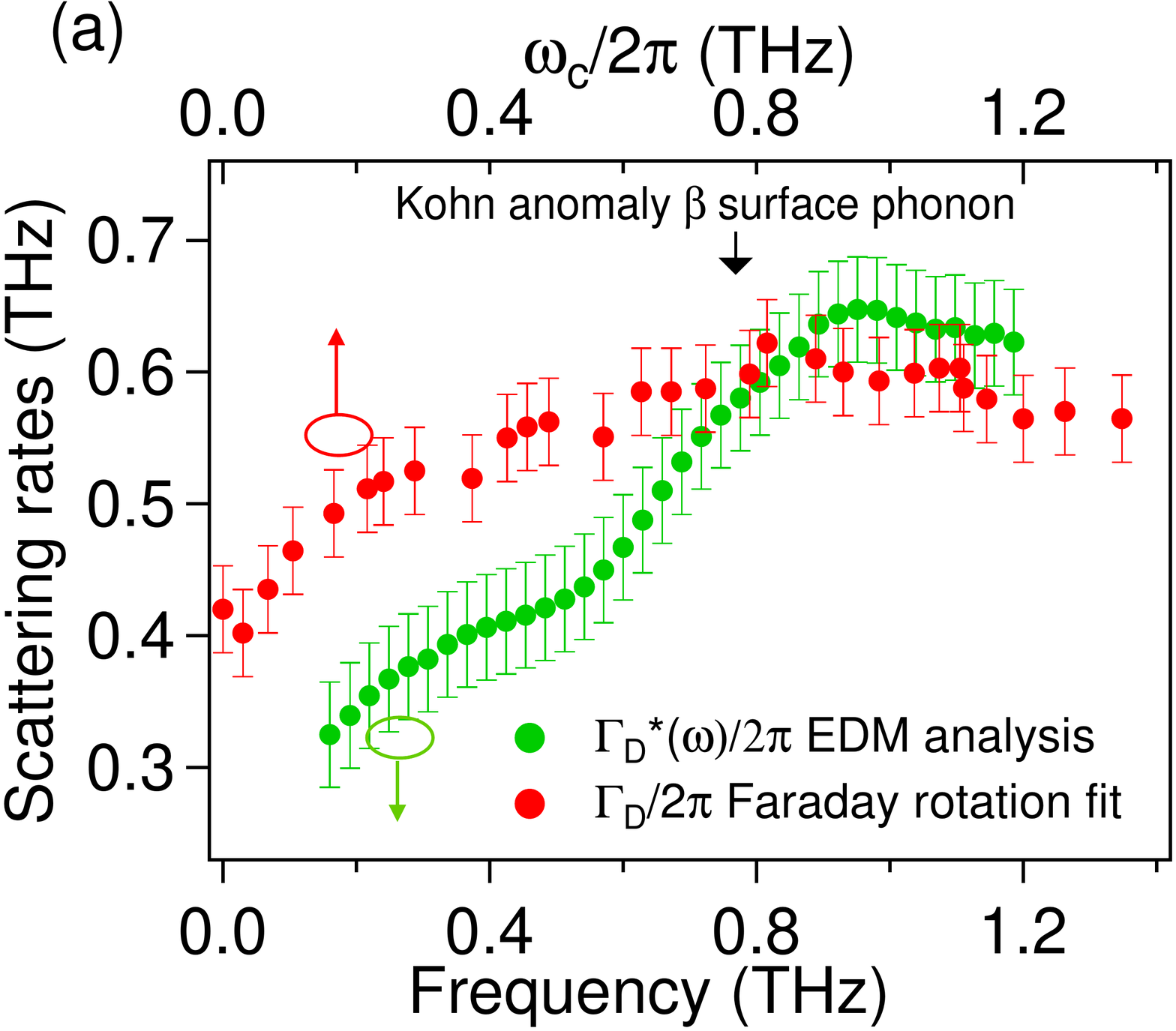}
\includegraphics[trim = 10 5 5 5,width=5.5cm]{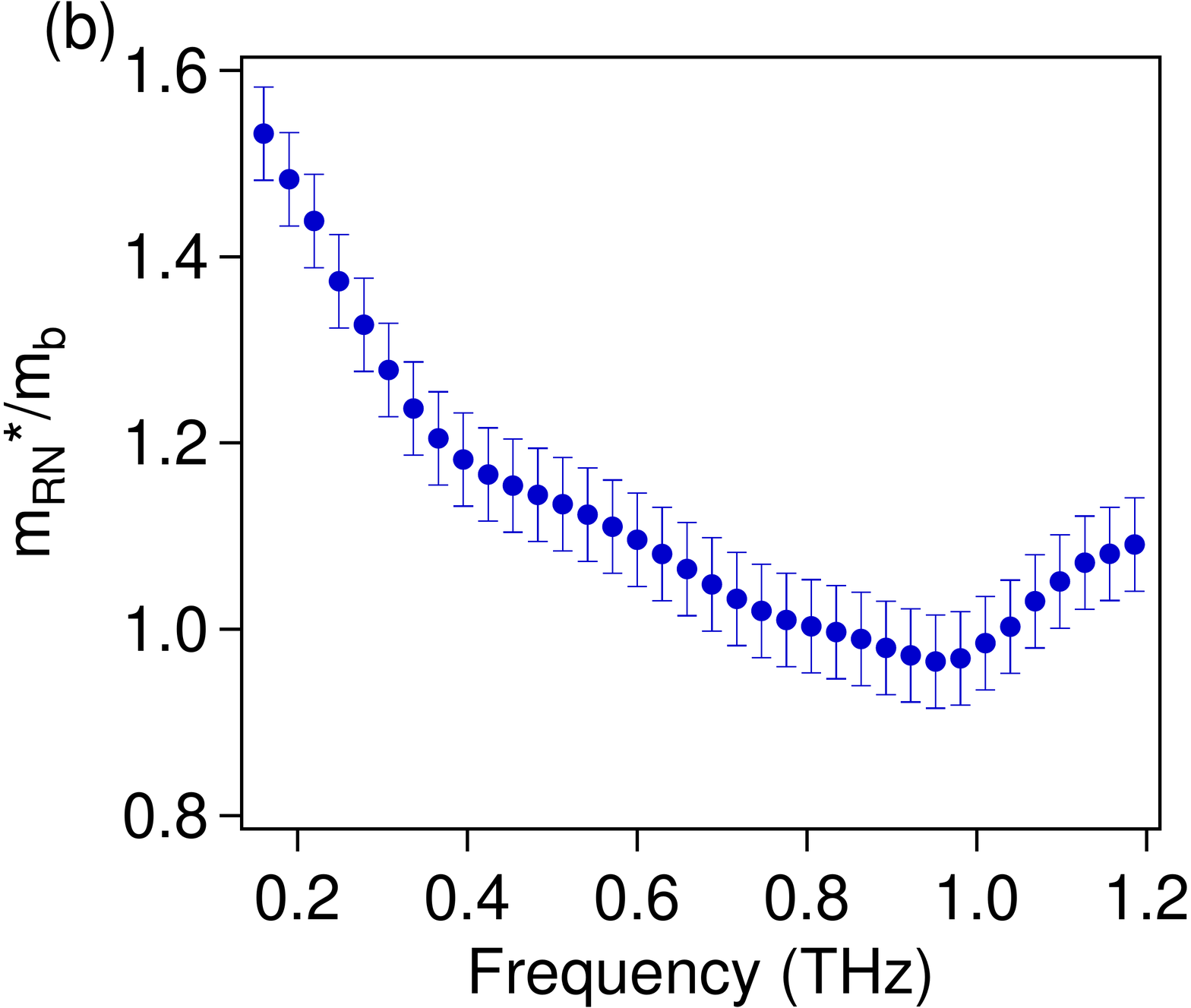}

\caption{(Color online) (a) Scattering rate as a function of cyclotron frequency (red). Fully renormalized scattering rate by mass through extended Drude analysis  as a function of frequency (green). Black arrow indicates $\beta$ surface phonon frequency at Kohn anomaly.  (b) Renormalized mass as a function of frequency.  The error bars express the uncertainty in $\omega_{pD}$.}.
 \label{Fig3}
\end{figure} 

Having identified the transport channels in these films and their relevant parameters, we can look in more detail at the magnetic field dependence of the scattering rates.  As shown in Fig. \ref{Fig3}(a) the scattering rate $\Gamma_D/2\pi$ in Cu$_{0.02}$Bi$_2$Se$_3$ increases with cyclotron frequency/field, displaying a maximum near a cyclotron frequency of 0.9 THz $\pm$ 0.1 THz.  As discussed in the SI section \upperRomannumeral{3} \cite{WuSI} , we can exclude magnetic field induced spin- or orbital-based electronic mechanisms for this broadening \cite{note3}.     We plot the scattering rate vs. the cyclotron frequency instead of vs. the magnetic field in order to allow a direct comparison with the measurement and phonon frequencies.  One can understand magnetic field as a energy sampling process as following. With increasing field, the Drude peak center frequency $\omega_c$ shifts to higher energies. As the Drude peak overlaps with the phonon frequency, one expects the scattering rate to reach its maximum value and saturate as one generally expects that the CR will become broader at frequencies above the relevant phonon frequency because new scattering channels emerge. Similar CR broadening has been observed in GaInAs quantum wells and graphene and ascribed to electron-phonon coupling\cite{Orlita_EPL_2010, Orlita_PRL_2012}. 

An extended Drude model (EDM) analysis \cite{AllenPRB77} of the \textit{zero field} Drude conductance supports the inference that the broadening is due to coupling to a low energy mode such as a phonon.  Such analysis allows one to quantify subtle deviations from pure Drude behavior (e.g s. pure Lorentzian form) as shown in Fig. \ref{Fig1}(a) inset in the form of frequency dependent mass renormalizations $m_{RN}^{*}(\omega)$ and scattering rates $\Gamma_D(\omega)/2\pi$ shown in Fig.\ref{Fig3}.  In such analysis one inverts the Drude intraband contribution to the conductance ($G_D$ ) via Eqs. \ref{Eqa5}.

\small
\begin{equation}
\Gamma_D(\omega)=\frac{1}{ \tau(\omega)}=\frac{\omega_{pD}^{2}d}{4\pi}\mathrm{Re}(\frac{1}{G_D}),  \; \;
\frac{m_{RN}^{*}(\omega)}{m_b}=-\frac{\omega_{pD}^{2}d}{4\pi \omega}\mathrm{Im}(\frac{1}{G_D})
\label{Eqa5}
\end{equation}
\normalsize

\noindent Here $m_b$ is the band mass without interaction renormalization.  An EDM analysis shows a frequency dependent scattering rate $\Gamma_D(\omega)$ and an electron-phonon coupling constant ($\lambda     =  \frac{m_{RN}^{*}}{m_b} - 1  $) defined at DC limit $\sim$ 0.55 $\pm$ 0.05. Since the mass renormalization  here is reasonably small, we approximate the bare plasma frequency with the value extracted from zero-field conductance or CR fits.

From the EDM parameters we can define the ``fully renormalized" scattering rate $\Gamma_D^{*}(\omega)=\frac{\Gamma_D(\omega)} {m_{RN}^{*}(\omega)/m_b}$ which is manifested as the actual width of the low frequency Drude feature (e.g.  the $\Gamma_D$ of Eq. \ref{Eqa4})\cite{note4}.  Plotting this quantity in Fig. \ref{Fig3}(a), we find the general trend and energy scales of the scattering rate from the EDM are about the same as those from the CR fit.  One can see that the result from EDM is a sharper version of the curve from the CR fit, which may be expected because the CR fit assumed a frequency independent scattering rate, and as such means it is an average over a (weakly) frequency dependent quantity.

For the first time we have demonstrated a correspondence between the spectral features in a CR experiment and EDM analyses, which indicates the magnetic field itself is not the source of the CR broadening.  We propose electron-phonon coupling as cause of these effects \cite{SobotaPRL14,ZhuPRL11,KimPRL12,GiraudPRB12} as the energy scale of the threshold in the scattering rate matches that of a number of electron-phonon scales in this system.   The value is close to the scale of the previously observed Kohn anomaly of the surface $\beta$ phonon, 0.75 THz at 2k$_F$ \cite{ZhuPRL11} \cite{note2}.  It also matches closely the scale of the maximum acoustic phonon energy that can couple to cross Fermi surface scattering  c $\times $2k$_F$/2$\pi$ $\sim$ 0.6 THz (where c  is the acoustic phonon velocity) \cite{KimPRL12}. The coupling strength $\lambda$ extracted by EDM is close to the value $\lambda$ $\sim$0.43 by analysis of the Kohn anomaly \cite{ZhuPRL12}, and also agrees with calculation $\lambda$ $\sim$ 0.42  for acoustic phonons in thin film geometry\cite{GiraudPRB12}. The coupling constant we observed is within the different values ARPES give\cite{HatchPRB11, PanPRL12, KondoPRL13}.  The energy scale of the 3 meV mode found via ARPES in Ref.\cite{KondoPRL13} agrees well with the characteristic energy here.  The lower coupling strength we observed may be due to the lower chemical potential of our sample than that used in Ref.\cite{KondoPRL13} .

Looking forward our results demonstrate the possibility of observing a quantized Faraday rotation in such films of a TI.  For the quantized Faraday rotation of a film on a substrate one has $\mathrm{tan}(\theta_F)=\frac{2\alpha}{1+n}(\nu+1/2)\sim3.5mrad(\nu+1/2)$ for each surface state in the quantum regime, where $\alpha$ is the fine structure constant, $\nu$ is the filling factor and $n$ is the substrate index of refraction.  The Landau level spectrum will be $E( \pm \nu)=\pm v_F \sqrt{2e \hbar B \mid \nu \mid}$. In order to reach the lowest LL  in this true topological insulator Cu$_{0.02}$Bi$_2$Se$_3$ will require 150T, which may be achievable for future THz experiments at pulse field facilities.  Electric field gating the present samples in a THz compatible arrangement would bring down this field threshold.  We hope our high-resolution Faraday rotation spectroscopy will enable the observation of the long-sought quantized Faraday rotation and topological magneto-electric effect\cite{Qi08b, Tse10a}.

We would like to thank A. Kuzmenko, G. Refael, and S. Valenzuela for discussions and L. Pan for assistance with the figures.  THz measurements at JHU and film growth and development at Rutgers were supported by NSF DMR-1308142, with additional support by the Gordon and Betty Moore Foundation through Grant No. GBMF2628 to NPA at JHU and EPiQS Initiative Grant GBMF4418 to SO at Rutgers, and by ONR-N000141210456 at Rutgers. Work at  Los Alamos National Laboratory(LANL) was carried out under the auspices of the U.S. Department of Energy at LANL under Contract No. DE-AC52-06NA25396.

\bibliography{TopoIns}

\begin{thebibliography}{52}
\expandafter\ifx\csname natexlab\endcsname\relax\def\natexlab#1{#1}\fi
\expandafter\ifx\csname bibnamefont\endcsname\relax
  \def\bibnamefont#1{#1}\fi
\expandafter\ifx\csname bibfnamefont\endcsname\relax
  \def\bibfnamefont#1{#1}\fi
\expandafter\ifx\csname citenamefont\endcsname\relax
  \def\citenamefont#1{#1}\fi
\expandafter\ifx\csname url\endcsname\relax
  \def\url#1{\texttt{#1}}\fi
\expandafter\ifx\csname urlprefix\endcsname\relax\def\urlprefix{URL }\fi
\providecommand{\bibinfo}[2]{#2}
\providecommand{\eprint}[2][]{\url{#2}}

\bibitem[{\citenamefont{Hasan and Kane}(2010)}]{Hasan-Kane-10}
\bibinfo{author}{\bibfnamefont{M.~Z.} \bibnamefont{Hasan}} \bibnamefont{and}
  \bibinfo{author}{\bibfnamefont{C.~L.} \bibnamefont{Kane}},
  \bibinfo{journal}{Rev. Mod. Phys.} \textbf{\bibinfo{volume}{82}},
  \bibinfo{pages}{3045} (\bibinfo{year}{2010}).

\bibitem[{\citenamefont{Qi and Zhang}(2011)}]{Qi-Zhang-11}
\bibinfo{author}{\bibfnamefont{X.-L.} \bibnamefont{Qi}} \bibnamefont{and}
  \bibinfo{author}{\bibfnamefont{S.-C.} \bibnamefont{Zhang}},
  \bibinfo{journal}{Rev. Mod. Phys.} \textbf{\bibinfo{volume}{83}},
  \bibinfo{pages}{1057} (\bibinfo{year}{2011}).

\bibitem[{\citenamefont{Pesin and MacDonald}(2012)}]{PesinNatMat12}
\bibinfo{author}{\bibfnamefont{D.}~\bibnamefont{Pesin}} \bibnamefont{and}
  \bibinfo{author}{\bibfnamefont{A.~H.} \bibnamefont{MacDonald}},
  \bibinfo{journal}{Nat. Mat.} \textbf{\bibinfo{volume}{11}},
  \bibinfo{pages}{409} (\bibinfo{year}{2012}).

\bibitem[{\citenamefont{Qi et~al.}(2008)\citenamefont{Qi, Hughes, and
  Zhang}}]{Qi08b}
\bibinfo{author}{\bibfnamefont{X.-L.} \bibnamefont{Qi}},
  \bibinfo{author}{\bibfnamefont{T.~L.} \bibnamefont{Hughes}},
  \bibnamefont{and} \bibinfo{author}{\bibfnamefont{S.-C.} \bibnamefont{Zhang}},
  \bibinfo{journal}{Phys. Rev. B} \textbf{\bibinfo{volume}{78}},
  \bibinfo{pages}{195424} (\bibinfo{year}{2008}).

\bibitem[{\citenamefont{Tse and MacDonald}(2010)}]{Tse10a}
\bibinfo{author}{\bibfnamefont{W.-K.} \bibnamefont{Tse}} \bibnamefont{and}
  \bibinfo{author}{\bibfnamefont{A.~H.} \bibnamefont{MacDonald}},
  \bibinfo{journal}{Phys. Rev. Lett.} \textbf{\bibinfo{volume}{105}},
  \bibinfo{pages}{057401} (\bibinfo{year}{2010}).

\bibitem[{\citenamefont{Basov et~al.}(2014)\citenamefont{Basov, Fogler,
  Lanzara, Wang, and Zhang}}]{BasovRMP14}
\bibinfo{author}{\bibfnamefont{D.}~\bibnamefont{Basov}},
  \bibinfo{author}{\bibfnamefont{M.}~\bibnamefont{Fogler}},
  \bibinfo{author}{\bibfnamefont{A.}~\bibnamefont{Lanzara}},
  \bibinfo{author}{\bibfnamefont{F.}~\bibnamefont{Wang}}, \bibnamefont{and}
  \bibinfo{author}{\bibfnamefont{Y.}~\bibnamefont{Zhang}},
  \bibinfo{journal}{Reviews of Modern Physics} \textbf{\bibinfo{volume}{86}},
  \bibinfo{pages}{959} (\bibinfo{year}{2014}).

\bibitem[{\citenamefont{Ren et~al.}(2010)\citenamefont{Ren, Taskin, Sasaki,
  Segawa, and Ando}}]{RenPRB2010}
\bibinfo{author}{\bibfnamefont{Z.}~\bibnamefont{Ren}},
  \bibinfo{author}{\bibfnamefont{A.~A.} \bibnamefont{Taskin}},
  \bibinfo{author}{\bibfnamefont{S.}~\bibnamefont{Sasaki}},
  \bibinfo{author}{\bibfnamefont{K.}~\bibnamefont{Segawa}}, \bibnamefont{and}
  \bibinfo{author}{\bibfnamefont{Y.}~\bibnamefont{Ando}},
  \bibinfo{journal}{Phys. Rev. B} \textbf{\bibinfo{volume}{82}},
  \bibinfo{pages}{241306} (\bibinfo{year}{2010}).

\bibitem[{\citenamefont{Xiong et~al.}(2012)\citenamefont{Xiong, Petersen, Qu,
  Hor, Cava, and Ong}}]{XiongPhysicaE2012}
\bibinfo{author}{\bibfnamefont{J.}~\bibnamefont{Xiong}},
  \bibinfo{author}{\bibfnamefont{A.}~\bibnamefont{Petersen}},
  \bibinfo{author}{\bibfnamefont{D.}~\bibnamefont{Qu}},
  \bibinfo{author}{\bibfnamefont{Y.}~\bibnamefont{Hor}},
  \bibinfo{author}{\bibfnamefont{R.}~\bibnamefont{Cava}}, \bibnamefont{and}
  \bibinfo{author}{\bibfnamefont{N.}~\bibnamefont{Ong}},
  \bibinfo{journal}{Physica E} \textbf{\bibinfo{volume}{44}},
  \bibinfo{pages}{917 } (\bibinfo{year}{2012}).

\bibitem[{\citenamefont{Brahlek et~al.}(2014)\citenamefont{Brahlek, Koirala,
  Salehi, Bansal, and Oh}}]{BrahlekPRL14}
\bibinfo{author}{\bibfnamefont{M.}~\bibnamefont{Brahlek}},
  \bibinfo{author}{\bibfnamefont{N.}~\bibnamefont{Koirala}},
  \bibinfo{author}{\bibfnamefont{M.}~\bibnamefont{Salehi}},
  \bibinfo{author}{\bibfnamefont{N.}~\bibnamefont{Bansal}}, \bibnamefont{and}
  \bibinfo{author}{\bibfnamefont{S.}~\bibnamefont{Oh}}, \bibinfo{journal}{Phys.
  Rev. Lett.} \textbf{\bibinfo{volume}{113}}, \bibinfo{pages}{026801}
  (\bibinfo{year}{2014}).

\bibitem[{WuS()}]{WuSI}
\bibinfo{note}{See Supplemental Material [url], which includes Refs.
  \cite{ValdesAguilarJAP13, WuNatPhys13SI, CaoNatPhys13, ZhangY10, YangPRB10,
  QKE_ref, Dimi, SCBA, Ando1, Zudov_RMP}}.

\bibitem[{\citenamefont{Crassee et~al.}(2010)\citenamefont{Crassee, Levallois,
  Walter, Ostler, Bostwick, Rotenberg, Seyller, Van Der~Marel, and
  Kuzmenko}}]{Crassee10a}
\bibinfo{author}{\bibfnamefont{I.}~\bibnamefont{Crassee}},
  \bibinfo{author}{\bibfnamefont{J.}~\bibnamefont{Levallois}},
  \bibinfo{author}{\bibfnamefont{A.}~\bibnamefont{Walter}},
  \bibinfo{author}{\bibfnamefont{M.}~\bibnamefont{Ostler}},
  \bibinfo{author}{\bibfnamefont{A.}~\bibnamefont{Bostwick}},
  \bibinfo{author}{\bibfnamefont{E.}~\bibnamefont{Rotenberg}},
  \bibinfo{author}{\bibfnamefont{T.}~\bibnamefont{Seyller}},
  \bibinfo{author}{\bibfnamefont{D.}~\bibnamefont{Van Der~Marel}},
  \bibnamefont{and} \bibinfo{author}{\bibfnamefont{A.}~\bibnamefont{Kuzmenko}},
  \bibinfo{journal}{Nature Physics} \textbf{\bibinfo{volume}{7}},
  \bibinfo{pages}{48} (\bibinfo{year}{2010}).

\bibitem[{\citenamefont{Jiang et~al.}(2007)\citenamefont{Jiang, Henriksen,
  Tung, Wang, Schwartz, Han, Kim, and Stormer}}]{JiangPRL07a}
\bibinfo{author}{\bibfnamefont{Z.}~\bibnamefont{Jiang}},
  \bibinfo{author}{\bibfnamefont{E.}~\bibnamefont{Henriksen}},
  \bibinfo{author}{\bibfnamefont{L.}~\bibnamefont{Tung}},
  \bibinfo{author}{\bibfnamefont{Y.-J.} \bibnamefont{Wang}},
  \bibinfo{author}{\bibfnamefont{M.}~\bibnamefont{Schwartz}},
  \bibinfo{author}{\bibfnamefont{M.}~\bibnamefont{Han}},
  \bibinfo{author}{\bibfnamefont{P.}~\bibnamefont{Kim}}, \bibnamefont{and}
  \bibinfo{author}{\bibfnamefont{H.}~\bibnamefont{Stormer}},
  \bibinfo{journal}{Phys. Rev. Lett.} \textbf{\bibinfo{volume}{98}},
  \bibinfo{pages}{197403} (\bibinfo{year}{2007}).

\bibitem[{\citenamefont{Kono and Miura}(2006)}]{KonoReview06}
\bibinfo{author}{\bibfnamefont{J.}~\bibnamefont{Kono}} \bibnamefont{and}
  \bibinfo{author}{\bibfnamefont{N.}~\bibnamefont{Miura}},
  \bibinfo{journal}{High Magnetic Fields: Science and Technology, Volume III,
  World Scientific, Singapore}  (\bibinfo{year}{2006}).

\bibitem[{\citenamefont{Vald\'es~Aguilar
  et~al.}(2012)\citenamefont{Vald\'es~Aguilar, Stier, Liu, Bilbro, George,
  Bansal, Wu, Cerne, Markelz, Oh et~al.}}]{ValdesAguilarPRL12}
\bibinfo{author}{\bibfnamefont{R.}~\bibnamefont{Vald\'es~Aguilar}},
  \bibinfo{author}{\bibfnamefont{A.~V.} \bibnamefont{Stier}},
  \bibinfo{author}{\bibfnamefont{W.}~\bibnamefont{Liu}},
  \bibinfo{author}{\bibfnamefont{L.~S.} \bibnamefont{Bilbro}},
  \bibinfo{author}{\bibfnamefont{D.~K.} \bibnamefont{George}},
  \bibinfo{author}{\bibfnamefont{N.}~\bibnamefont{Bansal}},
  \bibinfo{author}{\bibfnamefont{L.}~\bibnamefont{Wu}},
  \bibinfo{author}{\bibfnamefont{J.}~\bibnamefont{Cerne}},
  \bibinfo{author}{\bibfnamefont{A.~G.} \bibnamefont{Markelz}},
  \bibinfo{author}{\bibfnamefont{S.}~\bibnamefont{Oh}}, \bibnamefont{et~al.},
  \bibinfo{journal}{Phys. Rev. Lett.} \textbf{\bibinfo{volume}{108}},
  \bibinfo{pages}{087403} (\bibinfo{year}{2012}).

\bibitem[{\citenamefont{Jenkins et~al.}(2013)\citenamefont{Jenkins, Schmadel,
  Sushkov, Drew, Bichler, Koblmueller, Brahlek, Bansal, and Oh}}]{JenkinsPRB13}
\bibinfo{author}{\bibfnamefont{G.~S.} \bibnamefont{Jenkins}},
  \bibinfo{author}{\bibfnamefont{D.~C.} \bibnamefont{Schmadel}},
  \bibinfo{author}{\bibfnamefont{A.~B.} \bibnamefont{Sushkov}},
  \bibinfo{author}{\bibfnamefont{H.~D.} \bibnamefont{Drew}},
  \bibinfo{author}{\bibfnamefont{M.}~\bibnamefont{Bichler}},
  \bibinfo{author}{\bibfnamefont{G.}~\bibnamefont{Koblmueller}},
  \bibinfo{author}{\bibfnamefont{M.}~\bibnamefont{Brahlek}},
  \bibinfo{author}{\bibfnamefont{N.}~\bibnamefont{Bansal}}, \bibnamefont{and}
  \bibinfo{author}{\bibfnamefont{S.}~\bibnamefont{Oh}}, \bibinfo{journal}{Phys.
  Rev. B} \textbf{\bibinfo{volume}{87}}, \bibinfo{pages}{155126}
  (\bibinfo{year}{2013}).

\bibitem[{\citenamefont{Lee et~al.}(2014)\citenamefont{Lee, Xu, Shubeita,
  Brahlek, Koirala, Oh, and Gustafsson}}]{LeeThinFilms14}
\bibinfo{author}{\bibfnamefont{H.}~\bibnamefont{Lee}},
  \bibinfo{author}{\bibfnamefont{C.}~\bibnamefont{Xu}},
  \bibinfo{author}{\bibfnamefont{S.}~\bibnamefont{Shubeita}},
  \bibinfo{author}{\bibfnamefont{M.}~\bibnamefont{Brahlek}},
  \bibinfo{author}{\bibfnamefont{N.}~\bibnamefont{Koirala}},
  \bibinfo{author}{\bibfnamefont{S.}~\bibnamefont{Oh}}, \bibnamefont{and}
  \bibinfo{author}{\bibfnamefont{T.}~\bibnamefont{Gustafsson}},
  \bibinfo{journal}{Thin Solid Films} \textbf{\bibinfo{volume}{556}},
  \bibinfo{pages}{322} (\bibinfo{year}{2014}).

\bibitem[{\citenamefont{Wu et~al.}(2013{\natexlab{a}})\citenamefont{Wu,
  Brahlek, Aguilar, Stier, Morris, Lubashevsky, Bilbro, Bansal, Oh, and
  Armitage}}]{WuNatPhys13}
\bibinfo{author}{\bibfnamefont{L.}~\bibnamefont{Wu}},
  \bibinfo{author}{\bibfnamefont{M.}~\bibnamefont{Brahlek}},
  \bibinfo{author}{\bibfnamefont{R.~V.} \bibnamefont{Aguilar}},
  \bibinfo{author}{\bibfnamefont{A.~V.} \bibnamefont{Stier}},
  \bibinfo{author}{\bibfnamefont{C.~M.} \bibnamefont{Morris}},
  \bibinfo{author}{\bibfnamefont{Y.}~\bibnamefont{Lubashevsky}},
  \bibinfo{author}{\bibfnamefont{L.~S.} \bibnamefont{Bilbro}},
  \bibinfo{author}{\bibfnamefont{N.}~\bibnamefont{Bansal}},
  \bibinfo{author}{\bibfnamefont{S.}~\bibnamefont{Oh}}, \bibnamefont{and}
  \bibinfo{author}{\bibfnamefont{N.~P.} \bibnamefont{Armitage}},
  \bibinfo{journal}{Nature Physics} \textbf{\bibinfo{volume}{9}},
  \bibinfo{pages}{410} (\bibinfo{year}{2013}{\natexlab{a}}).

\bibitem[{\citenamefont{Morris et~al.}(2012)\citenamefont{Morris,
  Vald\'es~Aguilar, Stier, and Armitage}}]{MorrisOE12}
\bibinfo{author}{\bibfnamefont{C.}~\bibnamefont{Morris}},
  \bibinfo{author}{\bibfnamefont{R.}~\bibnamefont{Vald\'es~Aguilar}},
  \bibinfo{author}{\bibfnamefont{A.}~\bibnamefont{Stier}}, \bibnamefont{and}
  \bibinfo{author}{\bibfnamefont{N.}~\bibnamefont{Armitage}},
  \bibinfo{journal}{Optics Express} \textbf{\bibinfo{volume}{20}},
  \bibinfo{pages}{12303} (\bibinfo{year}{2012}).

\bibitem[{not({\natexlab{a}})}]{note1}
\bibinfo{note}{The transmission can be analyzed according to the thin film
  transmission equation for each polarization handedness separately. $t_{\pm}$
  is the transmission for right/left-hand circularly polarized light. $t_{\pm}=
  t_{vs}e^{i\phi_{s}}t_{sfv\pm}$ is the transmission for right/left hand
  circularly polarized light, where $t_{vs}$ is the transmission from vacuum to
  substrate $t_{vs}=2/(1+n_{s})$, $t_{sfv\pm}$ is the transmission from
  substrate thought film then to vacuum
  $t_{sfv\pm}=2n_{s}/(1+n_{s}+Z_{0}G_{\pm})$. $e^{i\phi_{s}}$ is the phase
  accumulation inside the substrate\cite{HancockPRL11}.}

\bibitem[{\citenamefont{Xia et~al.}(2009)\citenamefont{Xia, Qian, Hsieh, Wray,
  Pal, Lin, Bansil, Grauer, Hor, Cava et~al.}}]{Xia09a}
\bibinfo{author}{\bibfnamefont{Y.}~\bibnamefont{Xia}},
  \bibinfo{author}{\bibfnamefont{D.}~\bibnamefont{Qian}},
  \bibinfo{author}{\bibfnamefont{D.}~\bibnamefont{Hsieh}},
  \bibinfo{author}{\bibfnamefont{L.}~\bibnamefont{Wray}},
  \bibinfo{author}{\bibfnamefont{A.}~\bibnamefont{Pal}},
  \bibinfo{author}{\bibfnamefont{H.}~\bibnamefont{Lin}},
  \bibinfo{author}{\bibfnamefont{A.}~\bibnamefont{Bansil}},
  \bibinfo{author}{\bibfnamefont{D.}~\bibnamefont{Grauer}},
  \bibinfo{author}{\bibfnamefont{Y.}~\bibnamefont{Hor}},
  \bibinfo{author}{\bibfnamefont{R.}~\bibnamefont{Cava}}, \bibnamefont{et~al.},
  \bibinfo{journal}{Nature Physics} \textbf{\bibinfo{volume}{5}},
  \bibinfo{pages}{398} (\bibinfo{year}{2009}).

\bibitem[{\citenamefont{Bansal et~al.}(2012)\citenamefont{Bansal, Kim,
  Brahleck, Edrey, and Oh}}]{BansalPRL12}
\bibinfo{author}{\bibfnamefont{N.}~\bibnamefont{Bansal}},
  \bibinfo{author}{\bibfnamefont{Y.}~\bibnamefont{Kim}},
  \bibinfo{author}{\bibfnamefont{M.}~\bibnamefont{Brahleck}},
  \bibinfo{author}{\bibfnamefont{E.}~\bibnamefont{Edrey}}, \bibnamefont{and}
  \bibinfo{author}{\bibfnamefont{S.}~\bibnamefont{Oh}}, \bibinfo{journal}{Phys.
  Rev. Lett.} \textbf{\bibinfo{volume}{109}}, \bibinfo{pages}{116804}
  (\bibinfo{year}{2012}).

\bibitem[{\citenamefont{Hohler}(1973)}]{Kohler1973}
\bibinfo{author}{\bibfnamefont{H.}~\bibnamefont{Hohler}},
  \bibinfo{journal}{Physica Status Solidi (b)} \textbf{\bibinfo{volume}{58}},
  \bibinfo{pages}{91} (\bibinfo{year}{1973}).

\bibitem[{\citenamefont{Bianchi et~al.}(2010)\citenamefont{Bianchi, Guan, Bao,
  Mi, Iversen, King, and Hofmann}}]{BIanchi10}
\bibinfo{author}{\bibfnamefont{M.}~\bibnamefont{Bianchi}},
  \bibinfo{author}{\bibfnamefont{D.}~\bibnamefont{Guan}},
  \bibinfo{author}{\bibfnamefont{S.}~\bibnamefont{Bao}},
  \bibinfo{author}{\bibfnamefont{J.}~\bibnamefont{Mi}},
  \bibinfo{author}{\bibfnamefont{B.~B.} \bibnamefont{Iversen}},
  \bibinfo{author}{\bibfnamefont{P.~D.~C.} \bibnamefont{King}},
  \bibnamefont{and} \bibinfo{author}{\bibfnamefont{P.}~\bibnamefont{Hofmann}},
  \bibinfo{journal}{Nature Communications} \textbf{\bibinfo{volume}{1}},
  \bibinfo{pages}{128} (\bibinfo{year}{2010}).

\bibitem[{not({\natexlab{b}})}]{note3}
\bibinfo{note}{In principle, the Zeeman effect and orbital effect should
  increase the scattering rate as well. Neverthless, Zeeman effect should be
  much smaller than a 50$\%$ increase at low field. Orbital effects will lead
  to magneto-oscillations of transport lifetime (inverse scattering rate) above
  3 T when $\omega_c \tau \geq$ 1. (See SI section \upperRomannumeral{3} for
  detailed calculation)}.

\bibitem[{\citenamefont{Orlita et~al.}(2010)\citenamefont{Orlita, Faugeras,
  Martinez, Studenikin, and Poole}}]{Orlita_EPL_2010}
\bibinfo{author}{\bibfnamefont{M.}~\bibnamefont{Orlita}},
  \bibinfo{author}{\bibfnamefont{C.}~\bibnamefont{Faugeras}},
  \bibinfo{author}{\bibfnamefont{G.}~\bibnamefont{Martinez}},
  \bibinfo{author}{\bibfnamefont{S.~A.} \bibnamefont{Studenikin}},
  \bibnamefont{and} \bibinfo{author}{\bibfnamefont{P.~J.} \bibnamefont{Poole}},
  \bibinfo{journal}{EPL} \textbf{\bibinfo{volume}{92}}, \bibinfo{pages}{37002}
  (\bibinfo{year}{2010}).

\bibitem[{\citenamefont{Orlita et~al.}(2012)\citenamefont{Orlita, Tan,
  Potemski, Sprinkle, Berger, de~Heer, Louie, and Martinez}}]{Orlita_PRL_2012}
\bibinfo{author}{\bibfnamefont{M.}~\bibnamefont{Orlita}},
  \bibinfo{author}{\bibfnamefont{L.}~\bibnamefont{Tan}},
  \bibinfo{author}{\bibfnamefont{M.}~\bibnamefont{Potemski}},
  \bibinfo{author}{\bibfnamefont{M.}~\bibnamefont{Sprinkle}},
  \bibinfo{author}{\bibfnamefont{C.}~\bibnamefont{Berger}},
  \bibinfo{author}{\bibfnamefont{W.}~\bibnamefont{de~Heer}},
  \bibinfo{author}{\bibfnamefont{S.}~\bibnamefont{Louie}}, \bibnamefont{and}
  \bibinfo{author}{\bibfnamefont{G.}~\bibnamefont{Martinez}},
  \bibinfo{journal}{Phys. Rev. Lett.} \textbf{\bibinfo{volume}{108}},
  \bibinfo{pages}{247401} (\bibinfo{year}{2012}).

\bibitem[{\citenamefont{Allen and Mikkelsen}(1977)}]{AllenPRB77}
\bibinfo{author}{\bibfnamefont{J.~W.} \bibnamefont{Allen}} \bibnamefont{and}
  \bibinfo{author}{\bibfnamefont{J.~C.} \bibnamefont{Mikkelsen}},
  \bibinfo{journal}{Phys. Rev. B} \textbf{\bibinfo{volume}{15}},
  \bibinfo{pages}{2952} (\bibinfo{year}{1977}).

\bibitem[{not({\natexlab{c}})}]{note4}
\bibinfo{note}{$\Gamma_D^{*}(\omega)$ includes the renormalization effects of
  the lifetime as well as the mass.
  $\Gamma_D^{*}(\omega)=\frac{\Gamma_D(\omega)} {m_{RN}^{*}(\omega)/m_b}=
  \omega \frac{\mathrm{Re}(G_D)}{\mathrm{Im}(G_D)}$, which tells us the
  frequency-dependent scattering rate. Note that $\Gamma_D^{*}(\omega)$ does
  not depend on plasma frequency.}

\bibitem[{\citenamefont{Sobota et~al.}(2014)\citenamefont{Sobota, Yang,
  Leuenberger, Kemper, Analytis, Fisher, Kirchmann, Devereaux, and
  Shen}}]{SobotaPRL14}
\bibinfo{author}{\bibfnamefont{J.~A.} \bibnamefont{Sobota}},
  \bibinfo{author}{\bibfnamefont{S.-L.} \bibnamefont{Yang}},
  \bibinfo{author}{\bibfnamefont{D.}~\bibnamefont{Leuenberger}},
  \bibinfo{author}{\bibfnamefont{A.~F.} \bibnamefont{Kemper}},
  \bibinfo{author}{\bibfnamefont{J.~G.} \bibnamefont{Analytis}},
  \bibinfo{author}{\bibfnamefont{I.~R.} \bibnamefont{Fisher}},
  \bibinfo{author}{\bibfnamefont{P.~S.} \bibnamefont{Kirchmann}},
  \bibinfo{author}{\bibfnamefont{T.~P.} \bibnamefont{Devereaux}},
  \bibnamefont{and} \bibinfo{author}{\bibfnamefont{Z.-X.} \bibnamefont{Shen}},
  \bibinfo{journal}{Phys. Rev. Lett.} \textbf{\bibinfo{volume}{113}},
  \bibinfo{pages}{157401} (\bibinfo{year}{2014}).

\bibitem[{\citenamefont{Zhu et~al.}(2011)\citenamefont{Zhu, Santos, Sankar,
  Chikara, Howard, Chou, Chamon, and El-Batanouny}}]{ZhuPRL11}
\bibinfo{author}{\bibfnamefont{X.}~\bibnamefont{Zhu}},
  \bibinfo{author}{\bibfnamefont{L.}~\bibnamefont{Santos}},
  \bibinfo{author}{\bibfnamefont{R.}~\bibnamefont{Sankar}},
  \bibinfo{author}{\bibfnamefont{S.}~\bibnamefont{Chikara}},
  \bibinfo{author}{\bibfnamefont{C.}~\bibnamefont{Howard}},
  \bibinfo{author}{\bibfnamefont{F.}~\bibnamefont{Chou}},
  \bibinfo{author}{\bibfnamefont{C.}~\bibnamefont{Chamon}}, \bibnamefont{and}
  \bibinfo{author}{\bibfnamefont{M.}~\bibnamefont{El-Batanouny}},
  \bibinfo{journal}{Phys. Rev. Lett.} \textbf{\bibinfo{volume}{107}},
  \bibinfo{pages}{186102} (\bibinfo{year}{2011}).

\bibitem[{\citenamefont{Kim et~al.}(2012)\citenamefont{Kim, Li, Syers, Butch,
  Paglione, Sarma, and Fuhrer}}]{KimPRL12}
\bibinfo{author}{\bibfnamefont{D.}~\bibnamefont{Kim}},
  \bibinfo{author}{\bibfnamefont{Q.}~\bibnamefont{Li}},
  \bibinfo{author}{\bibfnamefont{P.}~\bibnamefont{Syers}},
  \bibinfo{author}{\bibfnamefont{N.~P.} \bibnamefont{Butch}},
  \bibinfo{author}{\bibfnamefont{J.}~\bibnamefont{Paglione}},
  \bibinfo{author}{\bibfnamefont{S.~D.} \bibnamefont{Sarma}}, \bibnamefont{and}
  \bibinfo{author}{\bibfnamefont{M.~S.} \bibnamefont{Fuhrer}},
  \bibinfo{journal}{Phys. Rev. Lett.} \textbf{\bibinfo{volume}{109}},
  \bibinfo{pages}{166801} (\bibinfo{year}{2012}).

\bibitem[{\citenamefont{Giraud et~al.}(2012)\citenamefont{Giraud, Kundu, and
  Egger}}]{GiraudPRB12}
\bibinfo{author}{\bibfnamefont{S.}~\bibnamefont{Giraud}},
  \bibinfo{author}{\bibfnamefont{A.}~\bibnamefont{Kundu}}, \bibnamefont{and}
  \bibinfo{author}{\bibfnamefont{R.}~\bibnamefont{Egger}},
  \bibinfo{journal}{Phys. Rev. B} \textbf{\bibinfo{volume}{85}},
  \bibinfo{pages}{035441} (\bibinfo{year}{2012}).

\bibitem[{not({\natexlab{d}})}]{note2}
\bibinfo{note}{Ref. \cite{ZhuPRL11} reported the surface $\beta$ phonon has
  frequency $\sim$1.8 THz and surface spring constant is 75$\%$ of the bulk
  value. Therefore, the surface $\beta$ phonon could be a softening bulk
  A$_{1g}$ Raman active phonon (2.23 THz bulk value). The surface $\beta$
  phonon shows a Kohn anomaly at 2 k$_f$ with $\sim$ 0.75THz phonon frequency.}

\bibitem[{\citenamefont{Zhu et~al.}(2012)\citenamefont{Zhu, Santos, Howard,
  Sankar, Chou, Chamon, and El-Batanouny}}]{ZhuPRL12}
\bibinfo{author}{\bibfnamefont{X.}~\bibnamefont{Zhu}},
  \bibinfo{author}{\bibfnamefont{L.}~\bibnamefont{Santos}},
  \bibinfo{author}{\bibfnamefont{C.}~\bibnamefont{Howard}},
  \bibinfo{author}{\bibfnamefont{R.}~\bibnamefont{Sankar}},
  \bibinfo{author}{\bibfnamefont{F.}~\bibnamefont{Chou}},
  \bibinfo{author}{\bibfnamefont{C.}~\bibnamefont{Chamon}}, \bibnamefont{and}
  \bibinfo{author}{\bibfnamefont{M.}~\bibnamefont{El-Batanouny}},
  \bibinfo{journal}{Phys. Rev. Lett.} \textbf{\bibinfo{volume}{108}},
  \bibinfo{pages}{185501} (\bibinfo{year}{2012}).

\bibitem[{\citenamefont{Hatch et~al.}(2011)\citenamefont{Hatch, Bianchi, Guan,
  Bao, Mi, Iversen, Nilsson, Hornek\ae{}r, and Hofmann}}]{HatchPRB11}
\bibinfo{author}{\bibfnamefont{R.~C.} \bibnamefont{Hatch}},
  \bibinfo{author}{\bibfnamefont{M.}~\bibnamefont{Bianchi}},
  \bibinfo{author}{\bibfnamefont{D.}~\bibnamefont{Guan}},
  \bibinfo{author}{\bibfnamefont{S.}~\bibnamefont{Bao}},
  \bibinfo{author}{\bibfnamefont{J.}~\bibnamefont{Mi}},
  \bibinfo{author}{\bibfnamefont{B.~B.} \bibnamefont{Iversen}},
  \bibinfo{author}{\bibfnamefont{L.}~\bibnamefont{Nilsson}},
  \bibinfo{author}{\bibfnamefont{L.}~\bibnamefont{Hornek\ae{}r}},
  \bibnamefont{and} \bibinfo{author}{\bibfnamefont{P.}~\bibnamefont{Hofmann}},
  \bibinfo{journal}{Phys. Rev. B} \textbf{\bibinfo{volume}{83}},
  \bibinfo{pages}{241303} (\bibinfo{year}{2011}).

\bibitem[{\citenamefont{Pan et~al.}(2012)\citenamefont{Pan, Fedorov, Gardner,
  Lee, Chu, and Valla}}]{PanPRL12}
\bibinfo{author}{\bibfnamefont{Z.-H.} \bibnamefont{Pan}},
  \bibinfo{author}{\bibfnamefont{A.~V.} \bibnamefont{Fedorov}},
  \bibinfo{author}{\bibfnamefont{D.}~\bibnamefont{Gardner}},
  \bibinfo{author}{\bibfnamefont{Y.~S.} \bibnamefont{Lee}},
  \bibinfo{author}{\bibfnamefont{S.}~\bibnamefont{Chu}}, \bibnamefont{and}
  \bibinfo{author}{\bibfnamefont{T.}~\bibnamefont{Valla}},
  \bibinfo{journal}{Phys. Rev. Lett.} \textbf{\bibinfo{volume}{108}},
  \bibinfo{pages}{187001} (\bibinfo{year}{2012}).

\bibitem[{\citenamefont{Kondo et~al.}(2013)\citenamefont{Kondo, Nakashima, Ota,
  Ishida, Malaeb, Okazaki, Shin, Kriener, Sasaki, Segawa et~al.}}]{KondoPRL13}
\bibinfo{author}{\bibfnamefont{T.}~\bibnamefont{Kondo}},
  \bibinfo{author}{\bibfnamefont{Y.}~\bibnamefont{Nakashima}},
  \bibinfo{author}{\bibfnamefont{Y.}~\bibnamefont{Ota}},
  \bibinfo{author}{\bibfnamefont{Y.}~\bibnamefont{Ishida}},
  \bibinfo{author}{\bibfnamefont{W.}~\bibnamefont{Malaeb}},
  \bibinfo{author}{\bibfnamefont{K.}~\bibnamefont{Okazaki}},
  \bibinfo{author}{\bibfnamefont{S.}~\bibnamefont{Shin}},
  \bibinfo{author}{\bibfnamefont{M.}~\bibnamefont{Kriener}},
  \bibinfo{author}{\bibfnamefont{S.}~\bibnamefont{Sasaki}},
  \bibinfo{author}{\bibfnamefont{K.}~\bibnamefont{Segawa}},
  \bibnamefont{et~al.}, \bibinfo{journal}{Phys. Rev. Lett.}
  \textbf{\bibinfo{volume}{110}}, \bibinfo{pages}{217601}
  (\bibinfo{year}{2013}).

\bibitem[{\citenamefont{Vald\'es~Aguilar
  et~al.}(2013)\citenamefont{Vald\'es~Aguilar, Wu, Stier, Bilbro, Brahlek,
  Bansal, Oh, and Armitage}}]{ValdesAguilarJAP13}
\bibinfo{author}{\bibfnamefont{R.}~\bibnamefont{Vald\'es~Aguilar}},
  \bibinfo{author}{\bibfnamefont{L.}~\bibnamefont{Wu}},
  \bibinfo{author}{\bibfnamefont{A.~V.} \bibnamefont{Stier}},
  \bibinfo{author}{\bibfnamefont{L.~S.} \bibnamefont{Bilbro}},
  \bibinfo{author}{\bibfnamefont{M.}~\bibnamefont{Brahlek}},
  \bibinfo{author}{\bibfnamefont{N.}~\bibnamefont{Bansal}},
  \bibinfo{author}{\bibfnamefont{S.}~\bibnamefont{Oh}}, \bibnamefont{and}
  \bibinfo{author}{\bibfnamefont{N.~P.} \bibnamefont{Armitage}},
  \bibinfo{journal}{J. Appl. Phys.} \textbf{\bibinfo{volume}{113}},
  \bibinfo{pages}{153702} (\bibinfo{year}{2013}).

\bibitem[{\citenamefont{Wu et~al.}(2013{\natexlab{b}})\citenamefont{Wu,
  Brahlek, Aguilar, Stier, Morris, Lubashevsky, Bilbro, Bansal, Oh, and
  Armitage}}]{WuNatPhys13SI}
\bibinfo{author}{\bibfnamefont{L.}~\bibnamefont{Wu}},
  \bibinfo{author}{\bibfnamefont{M.}~\bibnamefont{Brahlek}},
  \bibinfo{author}{\bibfnamefont{R.~V.} \bibnamefont{Aguilar}},
  \bibinfo{author}{\bibfnamefont{A.~V.} \bibnamefont{Stier}},
  \bibinfo{author}{\bibfnamefont{C.~M.} \bibnamefont{Morris}},
  \bibinfo{author}{\bibfnamefont{Y.}~\bibnamefont{Lubashevsky}},
  \bibinfo{author}{\bibfnamefont{L.~S.} \bibnamefont{Bilbro}},
  \bibinfo{author}{\bibfnamefont{N.}~\bibnamefont{Bansal}},
  \bibinfo{author}{\bibfnamefont{S.}~\bibnamefont{Oh}}, \bibnamefont{and}
  \bibinfo{author}{\bibfnamefont{N.~P.} \bibnamefont{Armitage}},
  \bibinfo{journal}{Nature Physics} \textbf{\bibinfo{volume}{9}},
  \bibinfo{pages}{410} (\bibinfo{year}{2013}{\natexlab{b}}),
  \bibinfo{note}{supplemetary Information}.

\bibitem[{\citenamefont{Cao et~al.}(2013)\citenamefont{Cao, Waugh, Zhang, Luo,
  Wang, Reber, Mo, Xu, Yang, Schneeloch et~al.}}]{CaoNatPhys13}
\bibinfo{author}{\bibfnamefont{Y.}~\bibnamefont{Cao}},
  \bibinfo{author}{\bibfnamefont{J.}~\bibnamefont{Waugh}},
  \bibinfo{author}{\bibfnamefont{X.}~\bibnamefont{Zhang}},
  \bibinfo{author}{\bibfnamefont{J.}~\bibnamefont{Luo}},
  \bibinfo{author}{\bibfnamefont{Q.}~\bibnamefont{Wang}},
  \bibinfo{author}{\bibfnamefont{T.}~\bibnamefont{Reber}},
  \bibinfo{author}{\bibfnamefont{S.}~\bibnamefont{Mo}},
  \bibinfo{author}{\bibfnamefont{Z.}~\bibnamefont{Xu}},
  \bibinfo{author}{\bibfnamefont{A.}~\bibnamefont{Yang}},
  \bibinfo{author}{\bibfnamefont{J.}~\bibnamefont{Schneeloch}},
  \bibnamefont{et~al.}, \bibinfo{journal}{Nature Physics}
  \textbf{\bibinfo{volume}{9}}, \bibinfo{pages}{499} (\bibinfo{year}{2013}).

\bibitem[{\citenamefont{Zhang et~al.}(2010)\citenamefont{Zhang, He, Chang,
  Song, Wang, Chen, Jia, Fang, Dai, Shan et~al.}}]{ZhangY10}
\bibinfo{author}{\bibfnamefont{Y.}~\bibnamefont{Zhang}},
  \bibinfo{author}{\bibfnamefont{K.}~\bibnamefont{He}},
  \bibinfo{author}{\bibfnamefont{C.}~\bibnamefont{Chang}},
  \bibinfo{author}{\bibfnamefont{C.}~\bibnamefont{Song}},
  \bibinfo{author}{\bibfnamefont{L.}~\bibnamefont{Wang}},
  \bibinfo{author}{\bibfnamefont{X.}~\bibnamefont{Chen}},
  \bibinfo{author}{\bibfnamefont{J.}~\bibnamefont{Jia}},
  \bibinfo{author}{\bibfnamefont{Z.}~\bibnamefont{Fang}},
  \bibinfo{author}{\bibfnamefont{X.}~\bibnamefont{Dai}},
  \bibinfo{author}{\bibfnamefont{W.}~\bibnamefont{Shan}}, \bibnamefont{et~al.},
  \bibinfo{journal}{Nature Physics} \textbf{\bibinfo{volume}{6}},
  \bibinfo{pages}{584} (\bibinfo{year}{2010}).

\bibitem[{\citenamefont{Yang et~al.}(2010)\citenamefont{Yang, Peeters, and
  Xu}}]{YangPRB10}
\bibinfo{author}{\bibfnamefont{C.~H.} \bibnamefont{Yang}},
  \bibinfo{author}{\bibfnamefont{F.~M.} \bibnamefont{Peeters}},
  \bibnamefont{and} \bibinfo{author}{\bibfnamefont{W.}~\bibnamefont{Xu}},
  \bibinfo{journal}{Phys. Rev. B} \textbf{\bibinfo{volume}{82}},
  \bibinfo{pages}{075401} (\bibinfo{year}{2010}).

\bibitem[{\citenamefont{Raichev and Vasko}(2005)}]{QKE_ref}
\bibinfo{author}{\bibfnamefont{O.~E.} \bibnamefont{Raichev}} \bibnamefont{and}
  \bibinfo{author}{\bibfnamefont{F.~T.} \bibnamefont{Vasko}},
  \emph{\bibinfo{title}{Quantum Kinetic Theory and Applications}}
  (\bibinfo{publisher}{Springer}, \bibinfo{year}{2005}).

\bibitem[{Dim()}]{Dimi}
\bibinfo{note}{D. Culcer and S. Das Sarma, Phys. Rev. B \textbf{83}, 245441
  (2011); D. Culcer, E. H. Hwang, T. D. Stanescu, and S. Das Sarma, Phys. Rev.
  B \textbf{82}, 155457 (2010).}

\bibitem[{SCB()}]{SCBA}
\bibinfo{note}{T. Ando and Y. Uemura, J. Phys. Soc. Jpn. \textbf{36} 959,
  (1974); T. Ando, J. Phys. Soc. Jpn. \textbf{36} 1521 (1974); \textbf{37} 622
  (1974); \textbf{37} 1233 (1974).}

\bibitem[{And()}]{Ando1}
\bibinfo{note}{N. H. Shon and T. Ando, J. Phys. Soc. Jpn \textbf{67}, 2421
  (1998); Y. Zheng and T. Ando, Phys. Rev. B\textbf{65}, 245420 (2002).}

\bibitem[{\citenamefont{Dmitriev et~al.}(2012)\citenamefont{Dmitriev, Mirlin,
  Polyakov, and Zudov}}]{Zudov_RMP}
\bibinfo{author}{\bibfnamefont{I.}~\bibnamefont{Dmitriev}},
  \bibinfo{author}{\bibfnamefont{A.}~\bibnamefont{Mirlin}},
  \bibinfo{author}{\bibfnamefont{D.}~\bibnamefont{Polyakov}}, \bibnamefont{and}
  \bibinfo{author}{\bibfnamefont{M.}~\bibnamefont{Zudov}},
  \bibinfo{journal}{Reviews of Modern Physics} \textbf{\bibinfo{volume}{84}},
  \bibinfo{pages}{1709} (\bibinfo{year}{2012}).

\bibitem[{\citenamefont{Liu et~al.}(2010)\citenamefont{Liu, Qi, Zhang, Dai,
  Fang, and Zhang}}]{LiuPRB10}
\bibinfo{author}{\bibfnamefont{C.-X.} \bibnamefont{Liu}},
  \bibinfo{author}{\bibfnamefont{X.-L.} \bibnamefont{Qi}},
  \bibinfo{author}{\bibfnamefont{H.}~\bibnamefont{Zhang}},
  \bibinfo{author}{\bibfnamefont{X.}~\bibnamefont{Dai}},
  \bibinfo{author}{\bibfnamefont{Z.}~\bibnamefont{Fang}}, \bibnamefont{and}
  \bibinfo{author}{\bibfnamefont{S.-C.} \bibnamefont{Zhang}},
  \bibinfo{journal}{Phys. Rev. B} \textbf{\bibinfo{volume}{82}},
  \bibinfo{pages}{045122} (\bibinfo{year}{2010}).

\bibitem[{Ref()}]{RefaelPriviteComm}
\bibinfo{note}{G. Refael, Privite Communication}.

\bibitem[{\citenamefont{Xu et~al.}(2014)\citenamefont{Xu, Miotkowski, Liu,
  Tian, Nam, Alidoust, Hu, Shih, Hasan, and Chen}}]{XuNatPhys14}
\bibinfo{author}{\bibfnamefont{Y.}~\bibnamefont{Xu}},
  \bibinfo{author}{\bibfnamefont{I.}~\bibnamefont{Miotkowski}},
  \bibinfo{author}{\bibfnamefont{C.}~\bibnamefont{Liu}},
  \bibinfo{author}{\bibfnamefont{J.}~\bibnamefont{Tian}},
  \bibinfo{author}{\bibfnamefont{H.}~\bibnamefont{Nam}},
  \bibinfo{author}{\bibfnamefont{N.}~\bibnamefont{Alidoust}},
  \bibinfo{author}{\bibfnamefont{J.}~\bibnamefont{Hu}},
  \bibinfo{author}{\bibfnamefont{C.-K.} \bibnamefont{Shih}},
  \bibinfo{author}{\bibfnamefont{M.~Z.} \bibnamefont{Hasan}}, \bibnamefont{and}
  \bibinfo{author}{\bibfnamefont{Y.~P.} \bibnamefont{Chen}},
  \bibinfo{journal}{Nature Physics} \textbf{\bibinfo{volume}{10}},
  \bibinfo{pages}{956} (\bibinfo{year}{2014}).

\bibitem[{\citenamefont{Brahlek et~al.}(2012)\citenamefont{Brahlek, Bansal,
  Koirala, Xu, Neupane, Liu, Hasan, and Oh}}]{BrahlekPRL12}
\bibinfo{author}{\bibfnamefont{M.}~\bibnamefont{Brahlek}},
  \bibinfo{author}{\bibfnamefont{N.}~\bibnamefont{Bansal}},
  \bibinfo{author}{\bibfnamefont{N.}~\bibnamefont{Koirala}},
  \bibinfo{author}{\bibfnamefont{S.-Y.} \bibnamefont{Xu}},
  \bibinfo{author}{\bibfnamefont{M.}~\bibnamefont{Neupane}},
  \bibinfo{author}{\bibfnamefont{C.}~\bibnamefont{Liu}},
  \bibinfo{author}{\bibfnamefont{M.~Z.} \bibnamefont{Hasan}}, \bibnamefont{and}
  \bibinfo{author}{\bibfnamefont{S.}~\bibnamefont{Oh}}, \bibinfo{journal}{Phys.
  Rev. Lett.} \textbf{\bibinfo{volume}{109}}, \bibinfo{pages}{186403}
  (\bibinfo{year}{2012}).

\bibitem[{\citenamefont{Analytis et~al.}(2010)\citenamefont{Analytis, Chu,
  Chen, Corredor, McDonald, Shen, and Fisher}}]{AnalytisPRB2010}
\bibinfo{author}{\bibfnamefont{J.~G.} \bibnamefont{Analytis}},
  \bibinfo{author}{\bibfnamefont{J.-H.} \bibnamefont{Chu}},
  \bibinfo{author}{\bibfnamefont{Y.}~\bibnamefont{Chen}},
  \bibinfo{author}{\bibfnamefont{F.}~\bibnamefont{Corredor}},
  \bibinfo{author}{\bibfnamefont{R.~D.} \bibnamefont{McDonald}},
  \bibinfo{author}{\bibfnamefont{Z.~X.} \bibnamefont{Shen}}, \bibnamefont{and}
  \bibinfo{author}{\bibfnamefont{I.~R.} \bibnamefont{Fisher}},
  \bibinfo{journal}{Phys. Rev. B} \textbf{\bibinfo{volume}{81}},
  \bibinfo{pages}{205407} (\bibinfo{year}{2010}).

\end{thebibliography}

\newpage

\setcounter{figure}{0}
\setcounter{equation}{0}
\setcounter{section}{0}
\begin{widetext}

\textbf{Supplementary Information for ``High-resolution Faraday rotation and electron phonon coupling in surface states of the bulk-insulating topological insulator Cu$_{0.02}$Bi$_2$Se$_3$'''}

\bigskip

\section{Materials and Experimental Methods}

Standard time-domain THz spectroscopy (TDTS) in a transmission geometry was performed with a custom
home-built THz spectrometer.   In this technique an approximately single-cycle picosecond pulse of light is transmitted through the sample and the substrate. The complex transmission is obtained from the ratio of a Fourier transformed time-domain sample pulse over  a Fourier transformed substrate pulse. The complex conductance can be directly inverted from the transmission equation in the thin film limit \cite{ValdesAguilarPRL12, WuNatPhys13}:

\begin{equation}
\tilde{T}(\omega)=\frac{1+n}{1+n+Z_0G(\omega)} e^{i\omega (n-1)\Delta L/c} 
\end{equation}

\noindent where  $\Delta L$ is the small difference in thickness between the sample and reference substrates, $n$ is the real  index of refraction of substrate and $Z_0$ is the vacuum impedance,  376.7 $\Omega$.  By measuring both the magnitude and phase of the transmission, this inversion to conductance is done directly and does not require Kramers-Kronig transformation. TDTS is an ideal tool to study the low frequency response of these materials with both the metallic Drude peak and a E$_{1u}$ infrared active phonon visible.  

Thin films of Cu$_{0.02}$Bi$_2$Se$_3$ were grown at Rutgers University by molecular beam epitaxy (MBE) on 0.5 mm thick crystalline Al$_2$O$_3$ substrates. Details on the growth can be found elsewhere \cite{BansalPRL12, BrahlekPRL14}.  Films grow a quintuple layer (1 unit cell) at a time (1 QL$\sim$1 nm).  After film growth,  a 100 nm amorphous Se cap is deposited to reduce aging effects \cite{ValdesAguilarJAP13}. Se capping was shown to have a negligible contribution to the optics at THz frequencies \cite{WuNatPhys13}, yet it serves a very important protection layer as Cu doped Bi$_2$Se$_3$ is much less stable in air than pure Bi$_2$Se$_3$.   3-4$\%$ optimal Cu concentration (Cu/Bi$\times$100$\%$) was incorporated during the film deposition and the concentration can be controlled at better than the 1\% level. Therefore, the `x' in  Cu$_{x}$Bi$_2$Se$_3$ formula is 0.015-0.02.  Samples were sealed in vacuum and sent immediately to JHU and low temperature TDTS measurements began within 24 hours of their growth.  The samples were mounted inside an optical helium flow cryostat and cooled down to 5 K within an hour.

\begin{figure}[htp]
\includegraphics[width=0.5\columnwidth,angle=0]{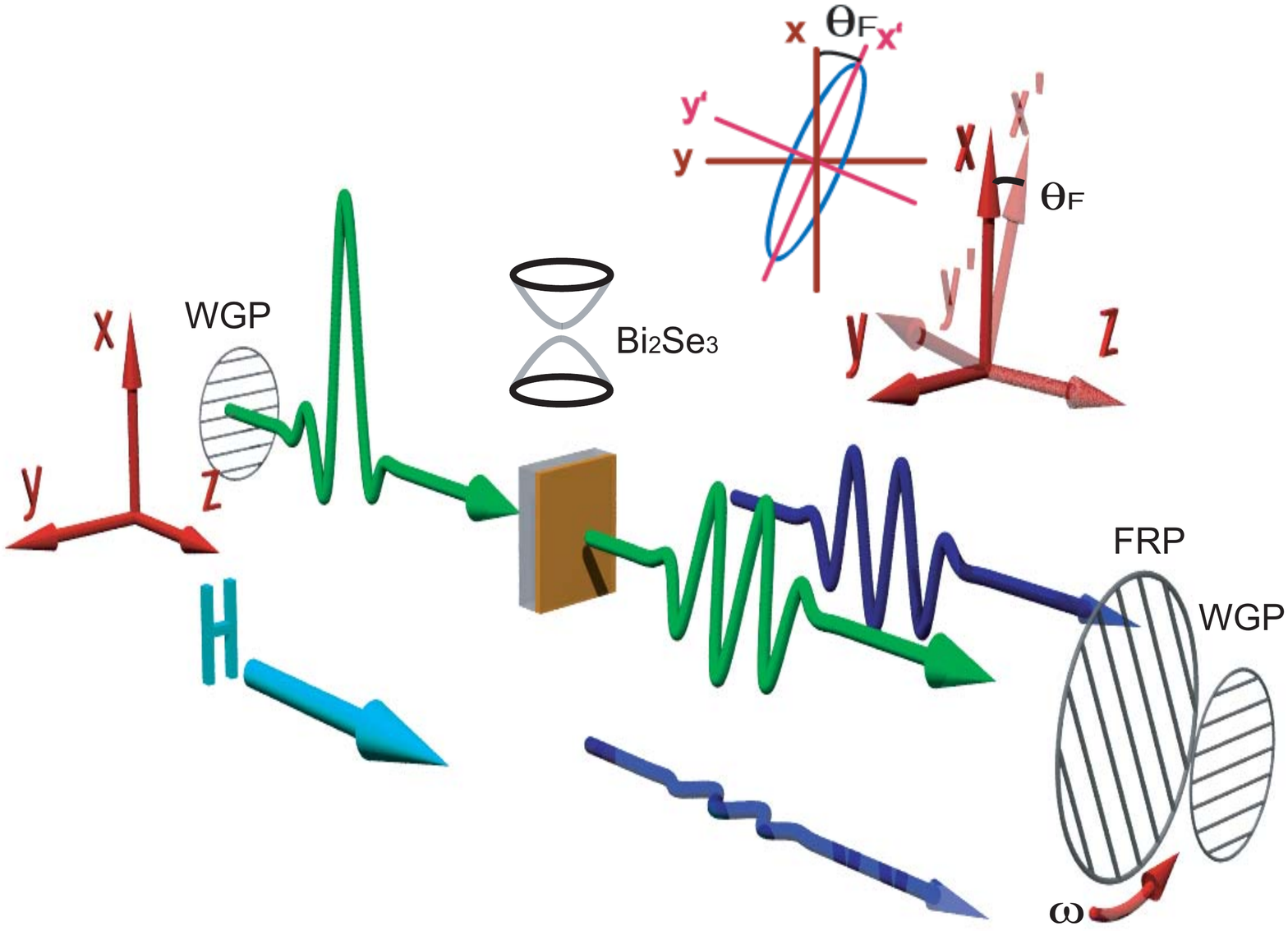}
\caption{(Color online) Demonstration of experimental setup for Faraday rotation experiments.}
 \label{SIFig1}
\end{figure} 

Complex Faraday rotation measurements were performed in a closed-cycle 7 T superconducting magnet at 5 K. We use the polarization modulation technique to measure the polarization states accurately \cite{MorrisOE12}.  As shown in Fig. \ref{SIFig1}, a static wire-grid polarizer (WGP1) is placed before the sample. After the sample, a fast rotating polarizer (FRP) unit and another static WGP2 are used. WGP1 and WGP2 transmit vertically polarized light. With this combination, in the polarization modulation technique, $E_{x}(t)$ and $E_{y}(t)$ (blue pulses) can be measured simultaneously in a single scan by reading off the in- and out-of-phase outputs from a lockin. Complex Faraday rotation $\theta_{F}=\theta_{F}^{'}+i\theta_{F}^{''}$ can be obtained by $\theta_{F}$=atan[$E_{y}$($\omega$)/$E_{x}$($\omega$)] after Fourier transforming into the frequency domain. Linearly polarized light becomes elliptically polarized after passing through the sample with a Faraday rotation $\theta_{F}^{'}$ (real part). The imaginary part of the Faraday rotation $\theta_{F}^{''}$ is related to the ellipticity in the small rotation angle regime \cite{MorrisOE12}.   In these measurements, a small background rotation from misalignment was subtracted by measuring a blank substrate.

\section{More data and analysis for C\lowercase{u}$_{0.02}$B\lowercase{i}$_2$S\lowercase{e}$_3$ and B\lowercase{i}$_2$S\lowercase{e}$_3$}

\begin{figure}[htp]
\includegraphics[trim = 10 5 5 5,width=5.5cm]{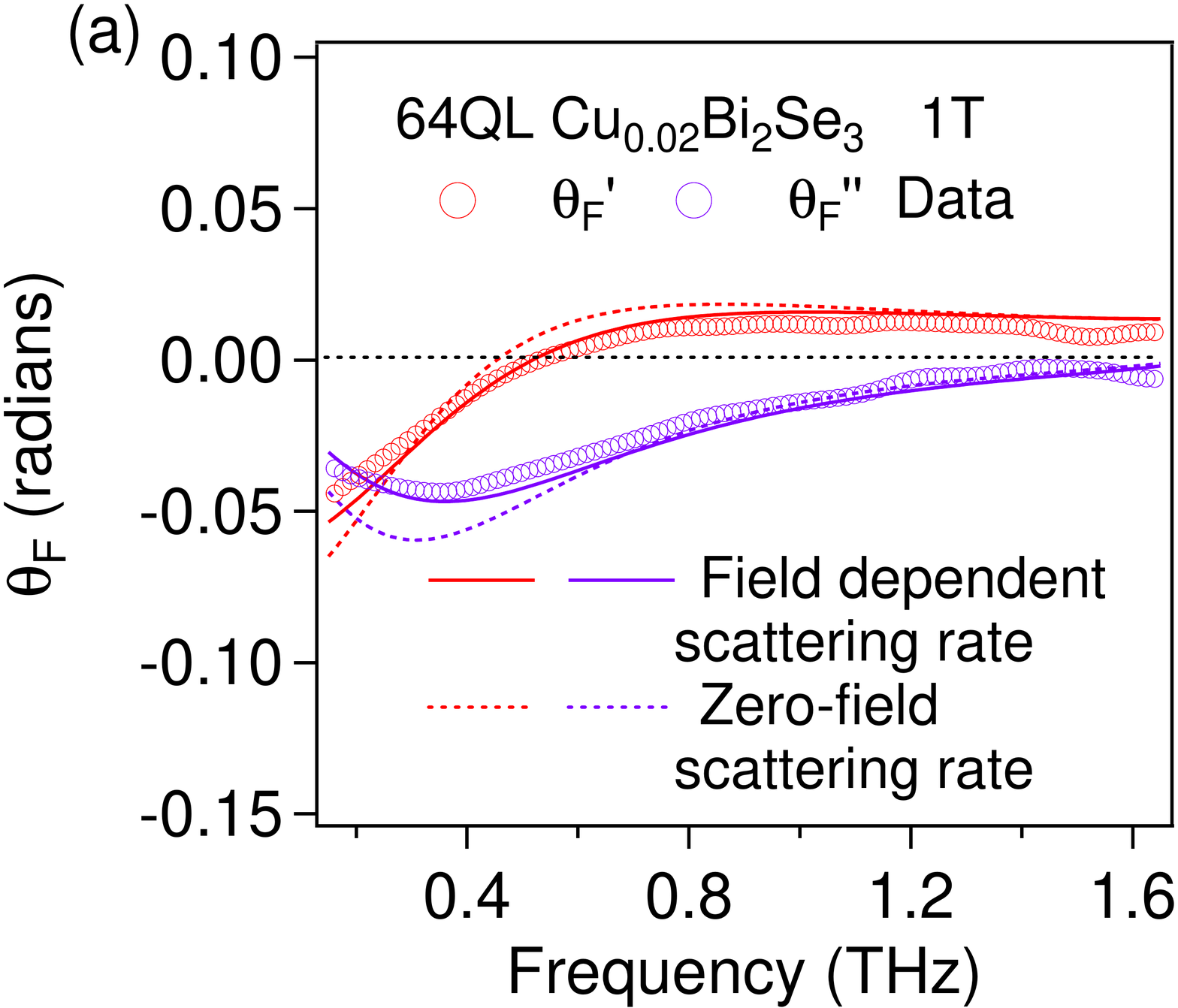}
\includegraphics[trim = 10 5 5 5,width=5.5cm]{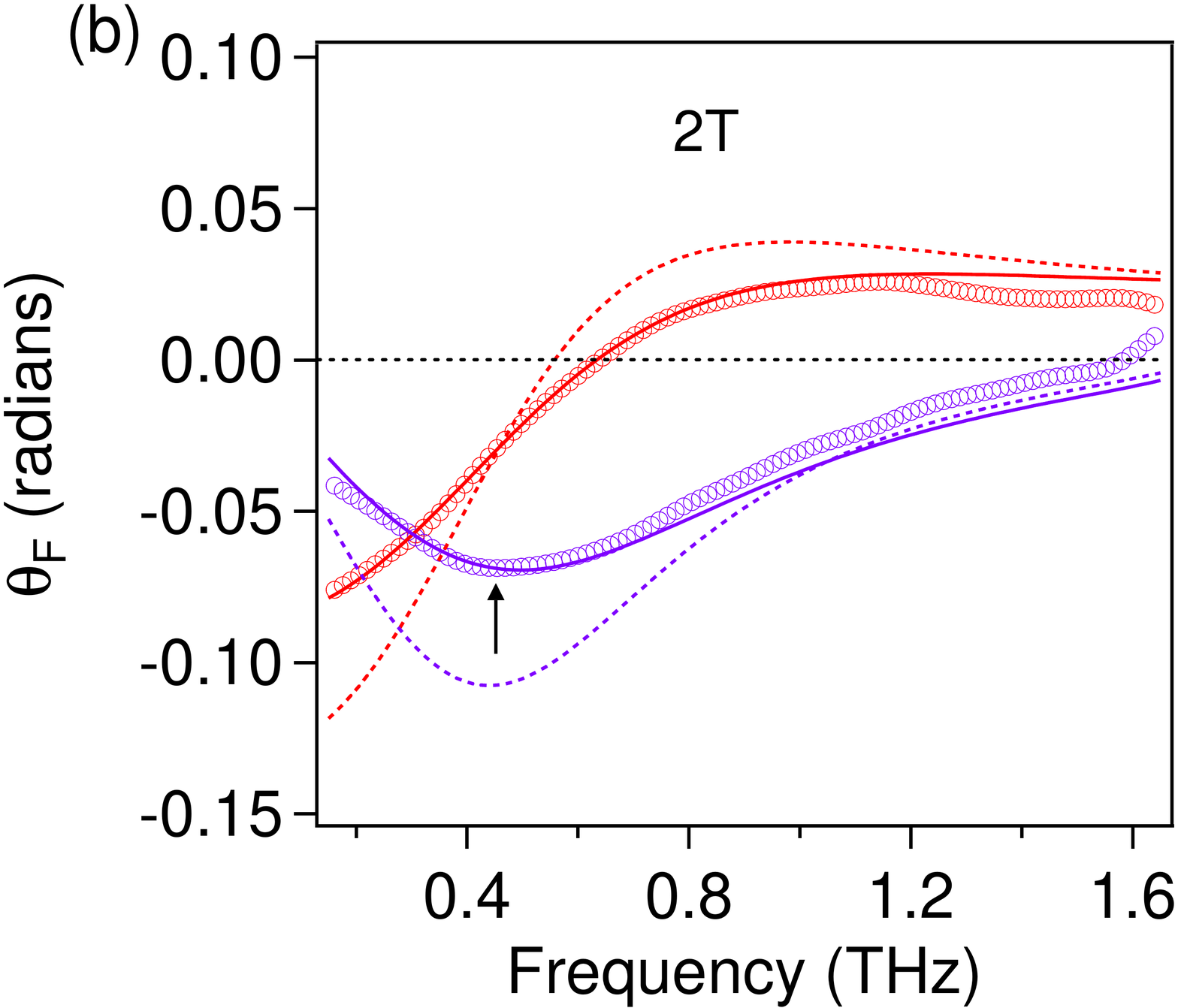}
\includegraphics[trim = 10 5 5 5,width=5.5cm]{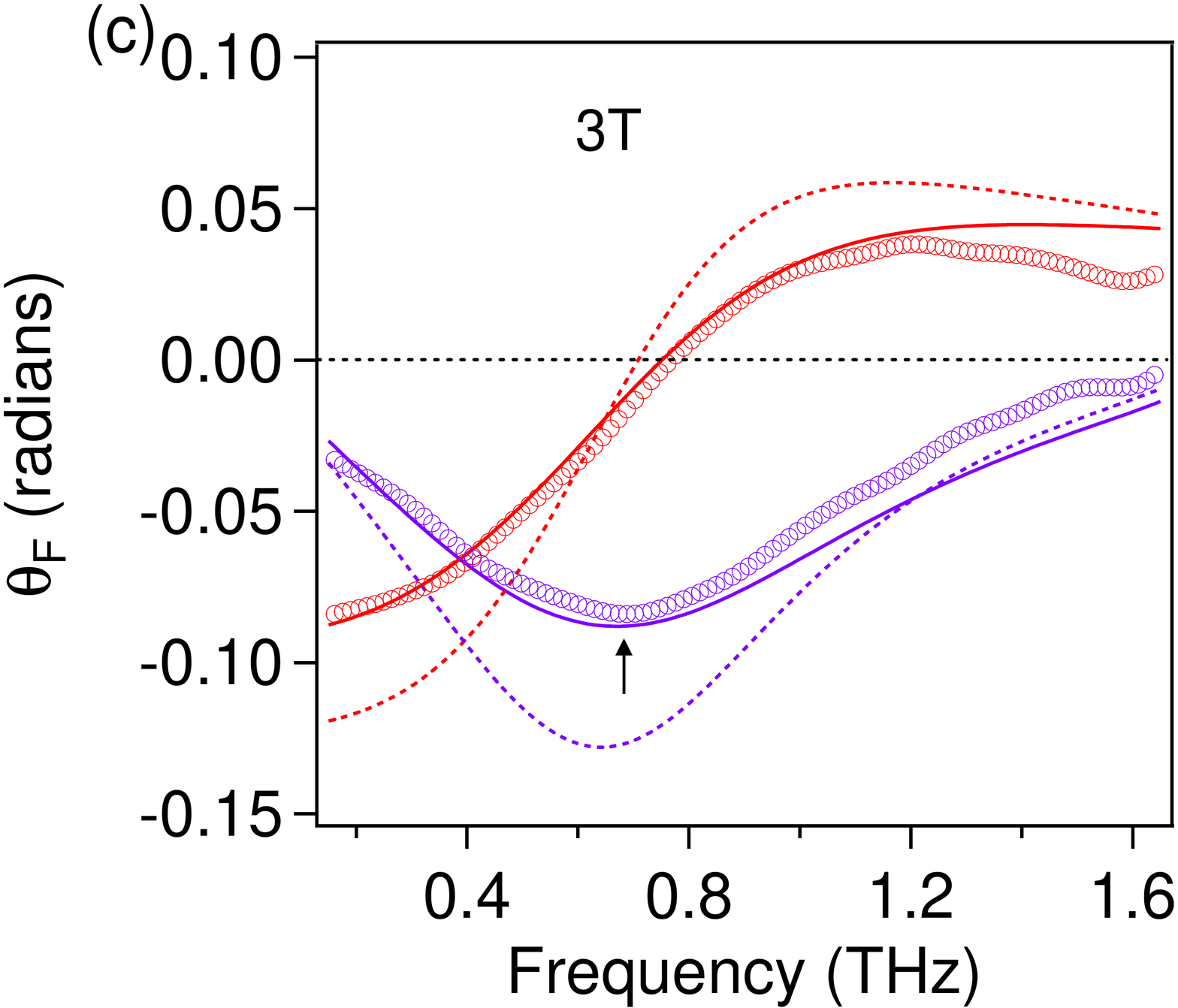}
\includegraphics[trim = 10 5 5 5,width=5.5cm]{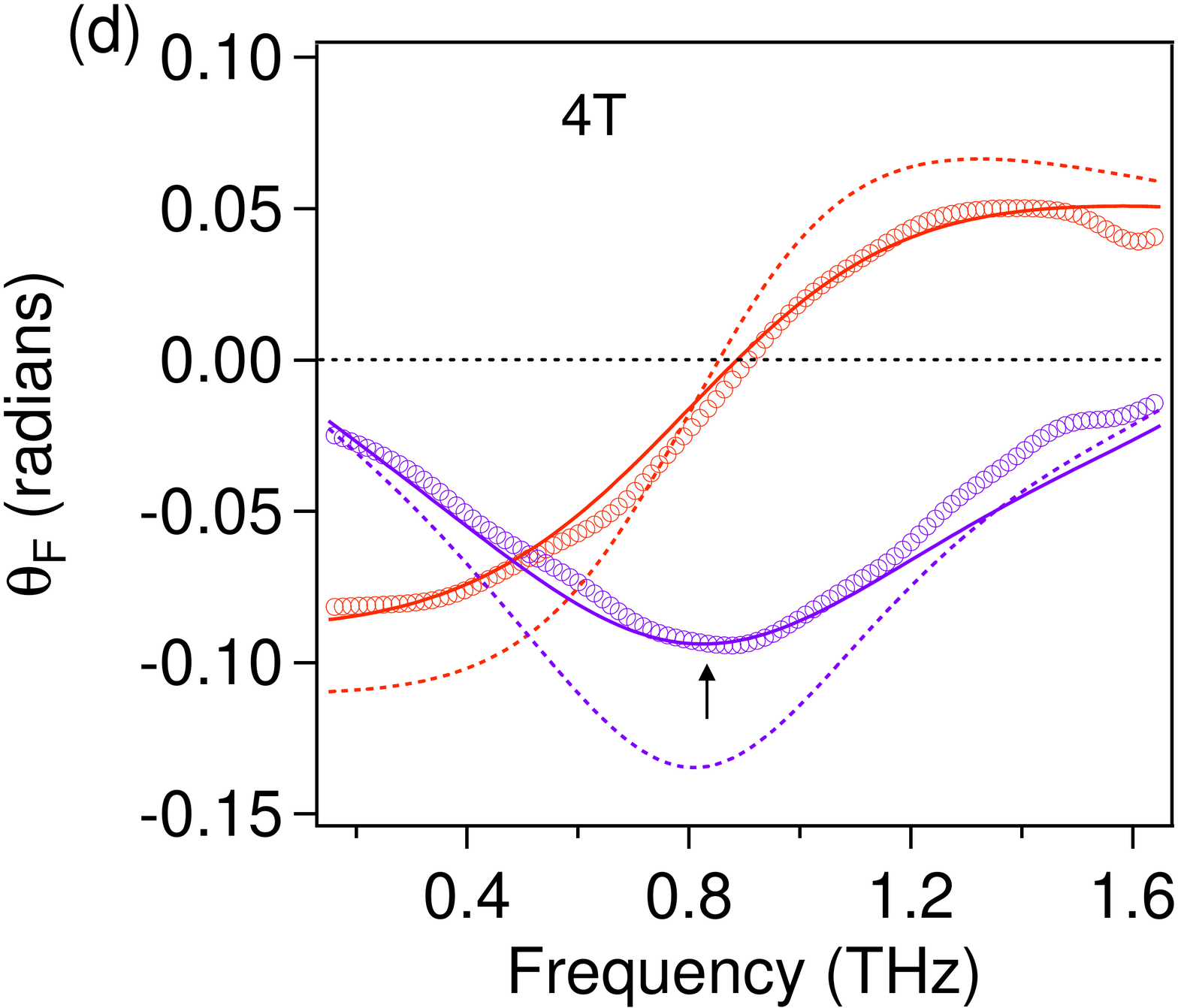}
\includegraphics[trim = 10 5 5 5,width=5.5cm]{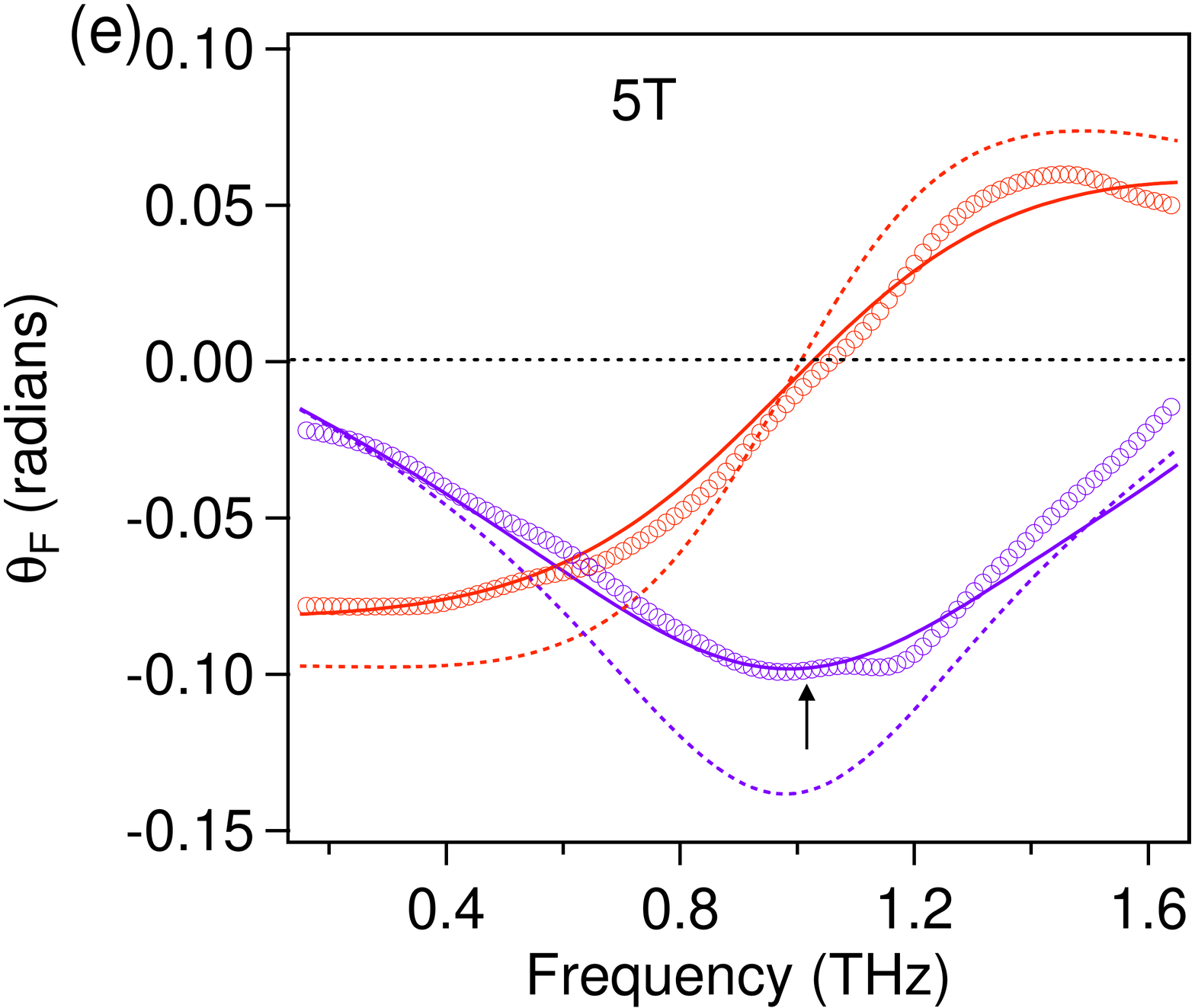}
\includegraphics[trim = 10 5 5 5,width=5.5cm]{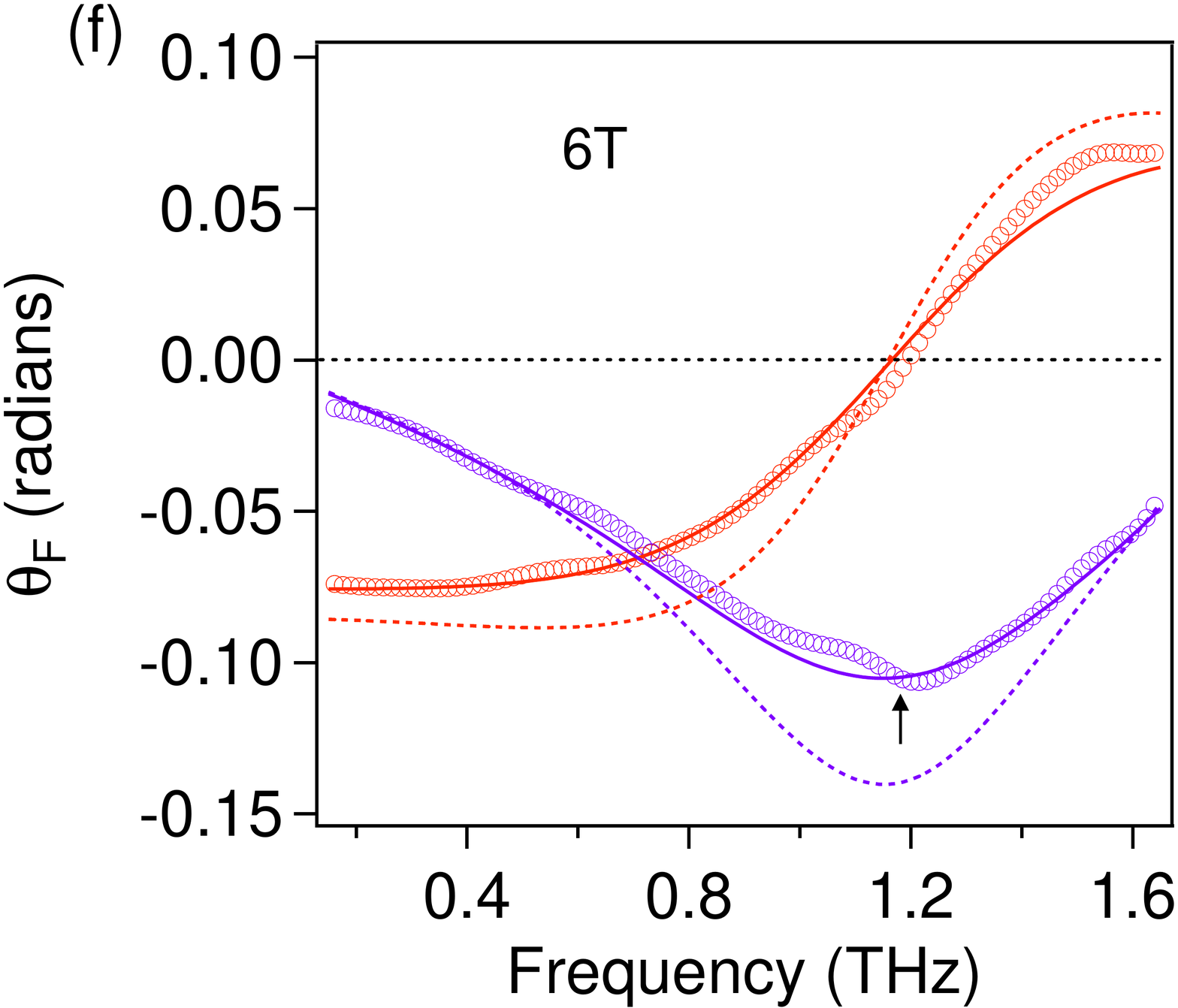}
\caption{(Color online) Fit quality on 64 QL Cu$_{0.02}$Bi$_2$Se$_3$ (sample 1) at different fields. Arrows are guides to the eye for the cyclotron frequencies. Accurate numbers were determined by fits. Solid lines are fits with field-dependent scattering rates. Dashed lines are fits using the zero-field scattering rates.}
\label{SIFig2}
\end{figure} 

The carrier density of each TSS can be calculated by the usual relation $n_{2D}=k_{F}^{2}/4\pi$. An effective transport mass  $m^{*}=\hbar k_{F}/v_{F}$ can still be defined for even for `massless' Dirac fermions where the Fermi velocity is determined by $v_{F}=\partial E_{F}/\hbar\partial k$.  In our analysis we will consider up to quadratic dispersion for surface states ($E = A k_F + B k_F^2$) and model the two surface states as identical with same carrier density.  Considering the TSS dispersion up to quadratic correction, the spectral weight can be expressed in terms of $k_{F} $ \cite{WuNatPhys13SI}.

\begin{equation}
\frac{2}{\pi\epsilon_{0}}\int G_{D1} d \omega = \omega_{pD}^{2} d  =  \frac{k_F ( A + 2B k_F)e^2 }{ 2 \pi \hbar^2 \epsilon_0}
\label{SIEqa2}
\end{equation}

\noindent Therefore, lower spectral weight means lower $k_F$. Lower $k_F$ means lower carrier density and smaller mass. By fitting the zero-field conductance, we find $(\omega_{pD}/2\pi)^{2} d$ in the 64 QL Cu$_{0.02}$Bi$_2$Se$_3$ sample 1  (the one discussed in main text) equal to 3.0$\pm$0.2 $\times$ 10$^{4}$ THz$^{2}$ $\cdot$ nm. By using Eq. \ref{SIEqa2}, we obtain $k_F$= 0.056$\pm$0.003 $\AA^{-1}$, $E_F$ = 145 $\pm$5 meV,  $m^{*}=$0.135$\pm$0.005 $m_{e}$ and a total sheet carrier density n$_{2D}$=5.0 $\pm0.3$ $\times$10$^{12}$/cm$^{2}$. From the Faraday rotation fit, we get the spectral weight  $(\omega_{pD}/2\pi)^{2} d =2.8 \pm 0.1  \times 10 ^{4}$ THz$^{2} \cdot$ nm more accurately.  By using the relation $\omega_{pD}^{2} d =\frac{n_{2D}e^{2}}{m^{*}\epsilon_{0}}$ and effective mass from CR experiments, we can get carrier density $n_{2D}=4.9 \pm0.1 \times10^{12}/$cm$^{2}$. These two analysis agree self-consistently. Here we have shown a new way to self-consistently study surface state transport.

Here we show all the data and fits for both 64 QL Cu$_{0.02}$Bi$_2$Se$_3$ samples (sample 1 and sample 2) .  One can see in Fig. \ref{SIFig2} for Cu$_{0.02}$Bi$_2$Se$_3$ (sample 1) that the difference in fits between field independent and field dependent scattering rates are highly distinguishable at finite fields, which is another indication of increasing scattering rate due to electron-phonon coupling.

We observed similar phenomena on another 64 QL Cu$_{0.02}$Bi$_2$Se$_3$ (sample 2) and reached a similar conclusion to sample 1. This sample 2 remained bulk-insulating and with high-mobility after 8 months of exposure to air, which demonstrates the realization of a robust bulk-insulating TI by the protection of Se capping. This sample also had  $\sim5.0\times10^{12}/$cm$^{2}$ but slightly lower mobility $\sim2800$ cm$^{2}/$V$\cdot$s. Data and fitting results are shown in Fig. \ref{SIFig3}. Here we subtracted the zero-field `rotation' as a small background from misalignment, which gives the same result as a substrate background is subtracted. We also performed extended Drude analysis and reached a similar conclusion to sample 1. Suppression of the scattering rate at low frequency was observed and a coupling constant in the DC limit $\lambda \sim$ 0.50 $\pm$ 0.05 is extracted. As discussed in the main text, electron-phonon coupling is not sensitive to magnetic field itself. This is what we showed by showing the extended Drude model analysis at zero magnetic field gives a overall similar frequency dependence.  The magnetic field rather, changes how electron phonon coupling manifests itself in the spectra.  One generally expects that such coupling causes broadening of some spectral feature at frequencies overlapping with a characteristic phonon energy.  By putting on a magnetic field one pushes the zero frequency Drude peak to higher frequency and causes it to broaden.  The electron-phonon coupling can be seen either by pushing the Drude peak across the characteristic phonon frequencies with magnetic field or by looking at subtle deviations from a Lorentzian lineshape at frequencies above the phonon frequencies in zero field.  Basically, we can view the magnetic field as a energy sampling tool.

\begin{figure}[htp]
\includegraphics[trim = 10 5 5 5,width=5.5cm]{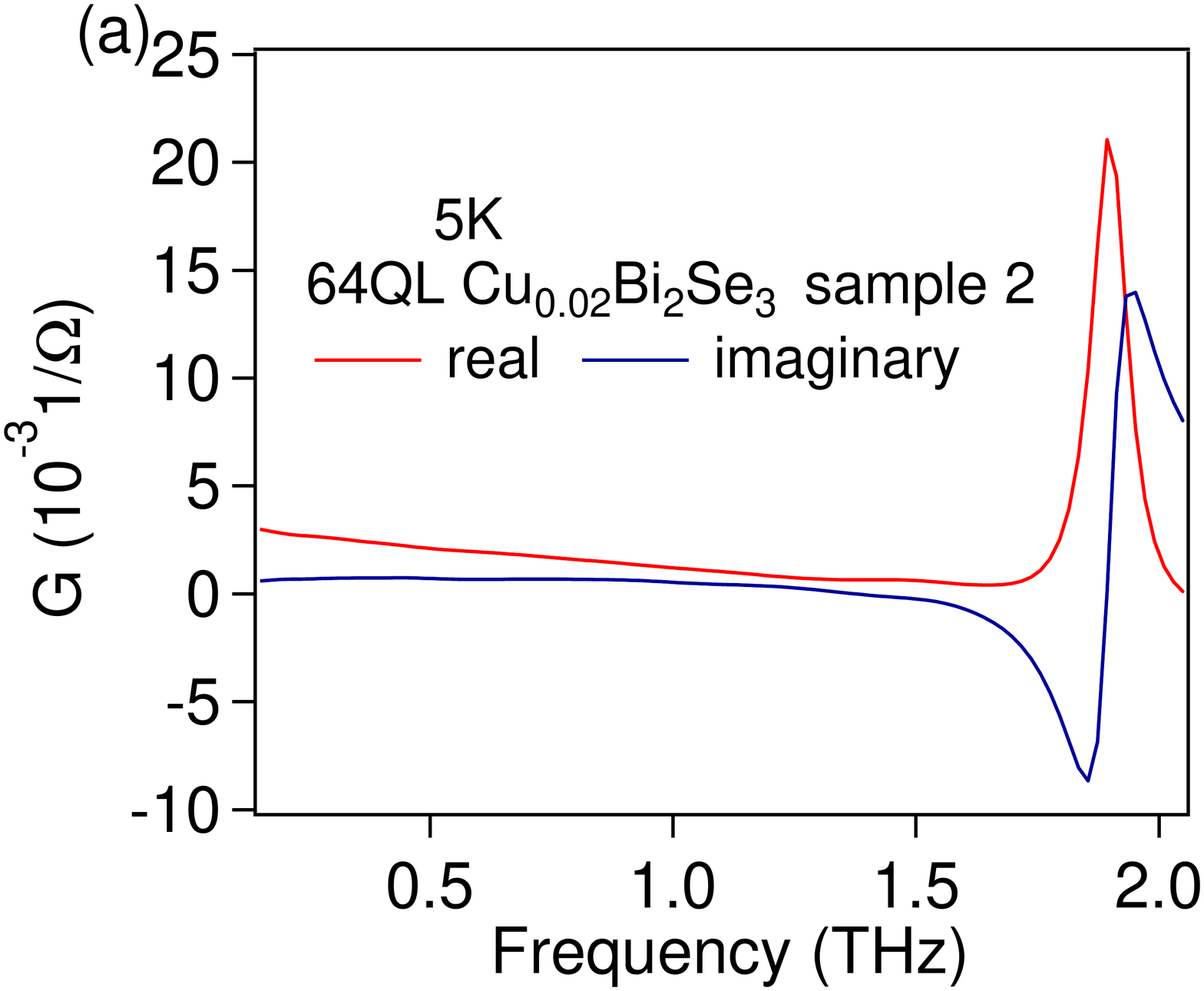}
\includegraphics[trim = 10 5 5 5,width=5.5cm]{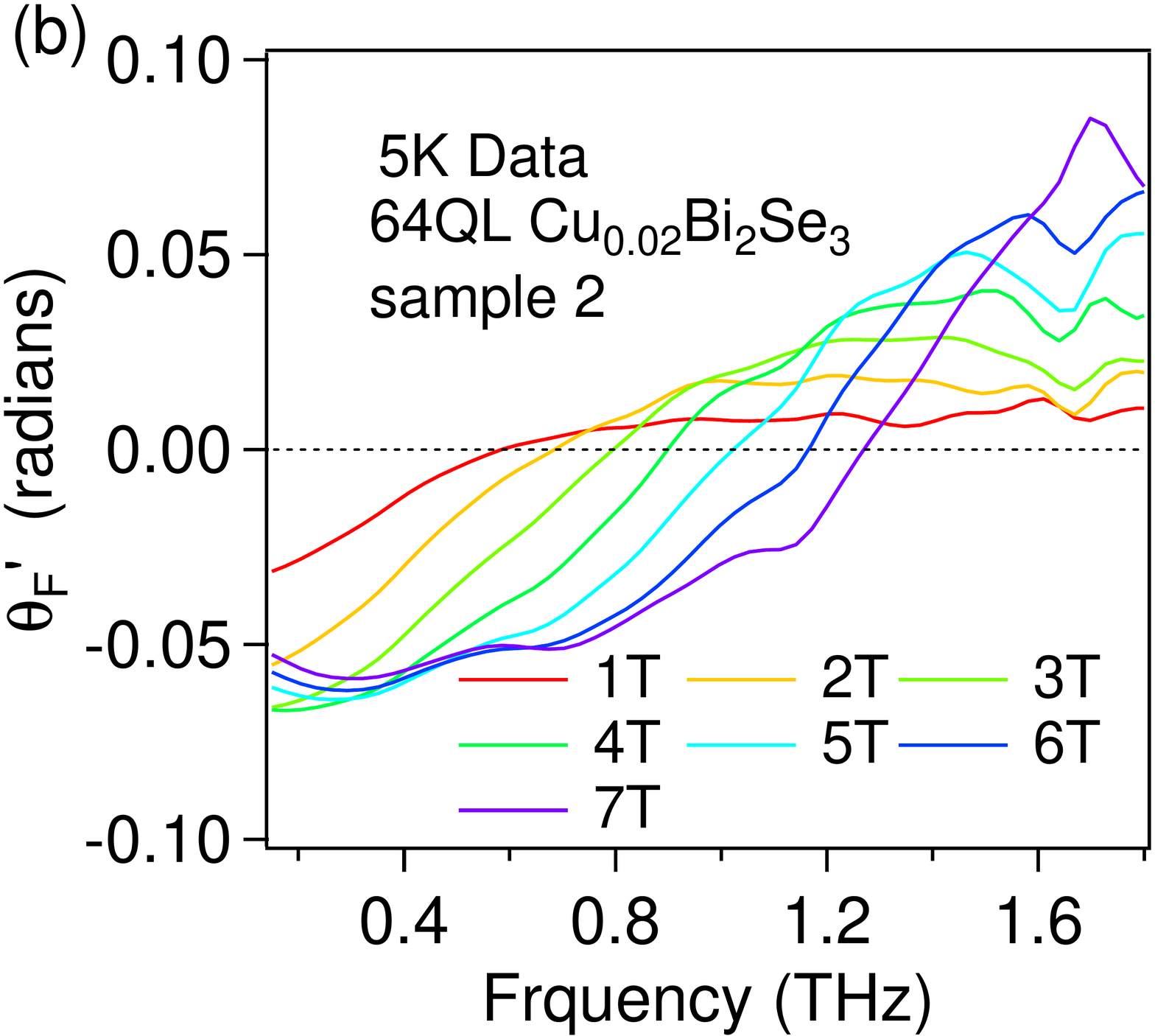}
\includegraphics[trim = 10 5 5 5,width=5.5cm]{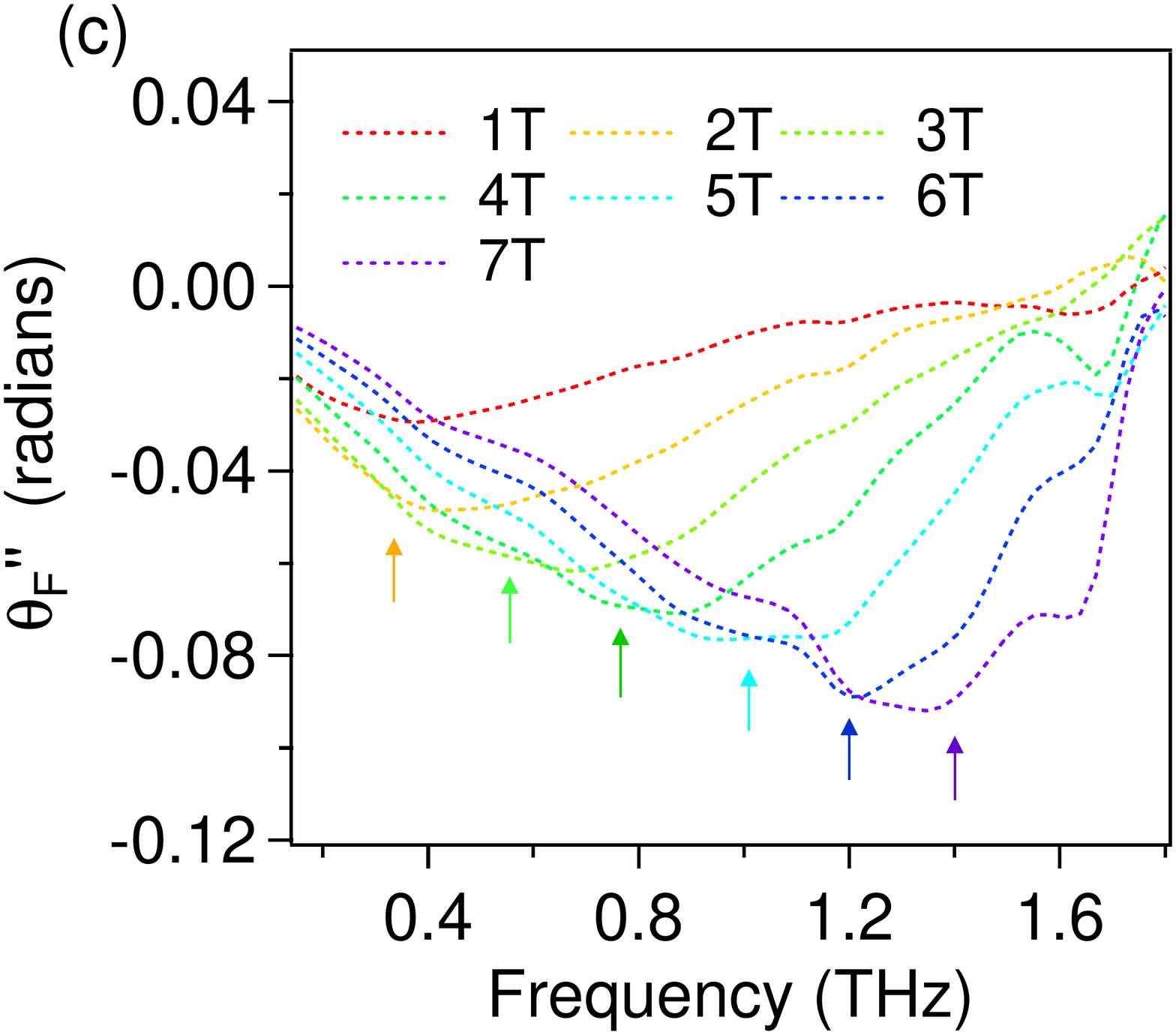}
\includegraphics[trim = 10 5 5 5,width=5.5cm]{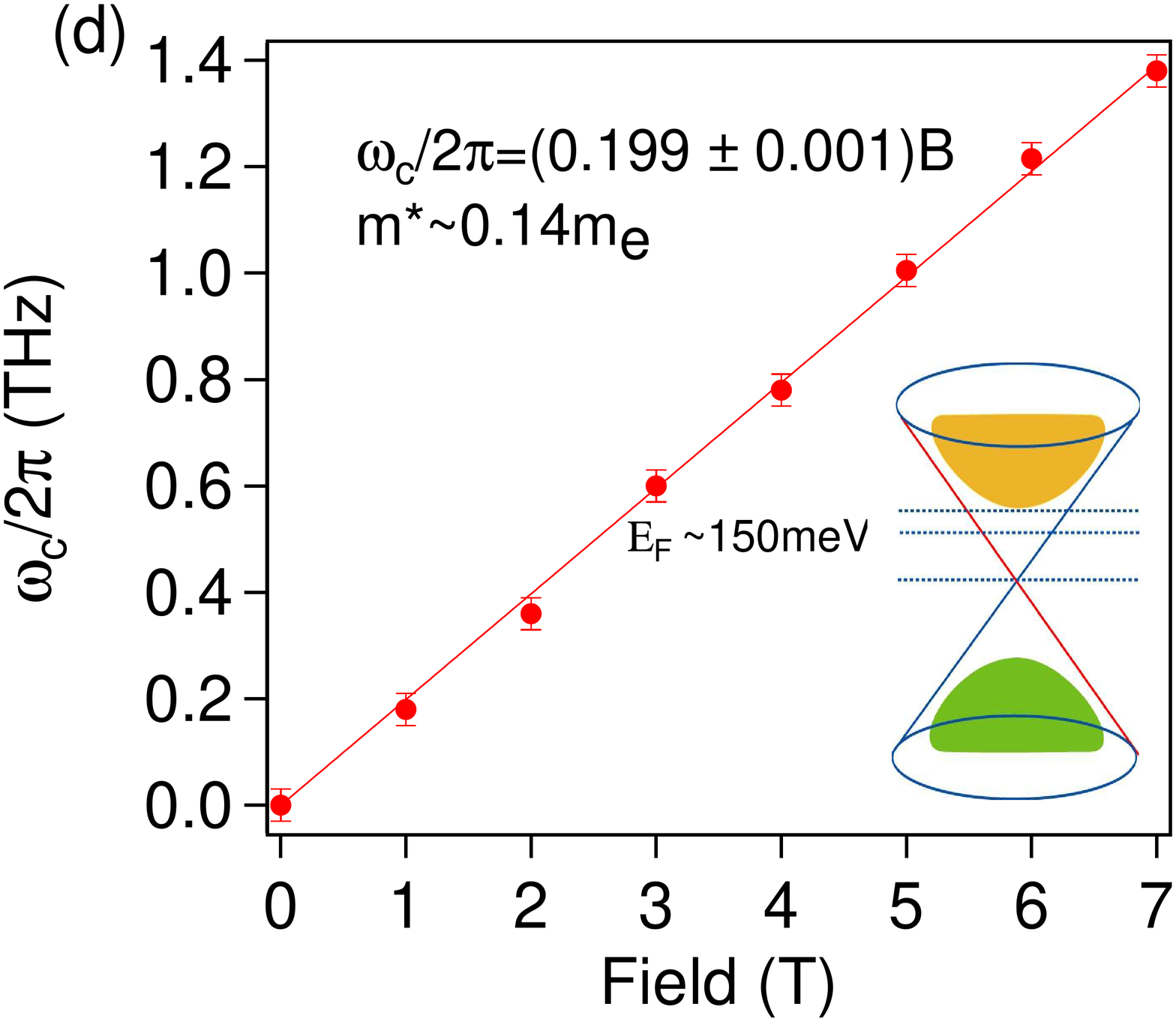}
\includegraphics[trim = 10 5 5 5,width=5.5cm]{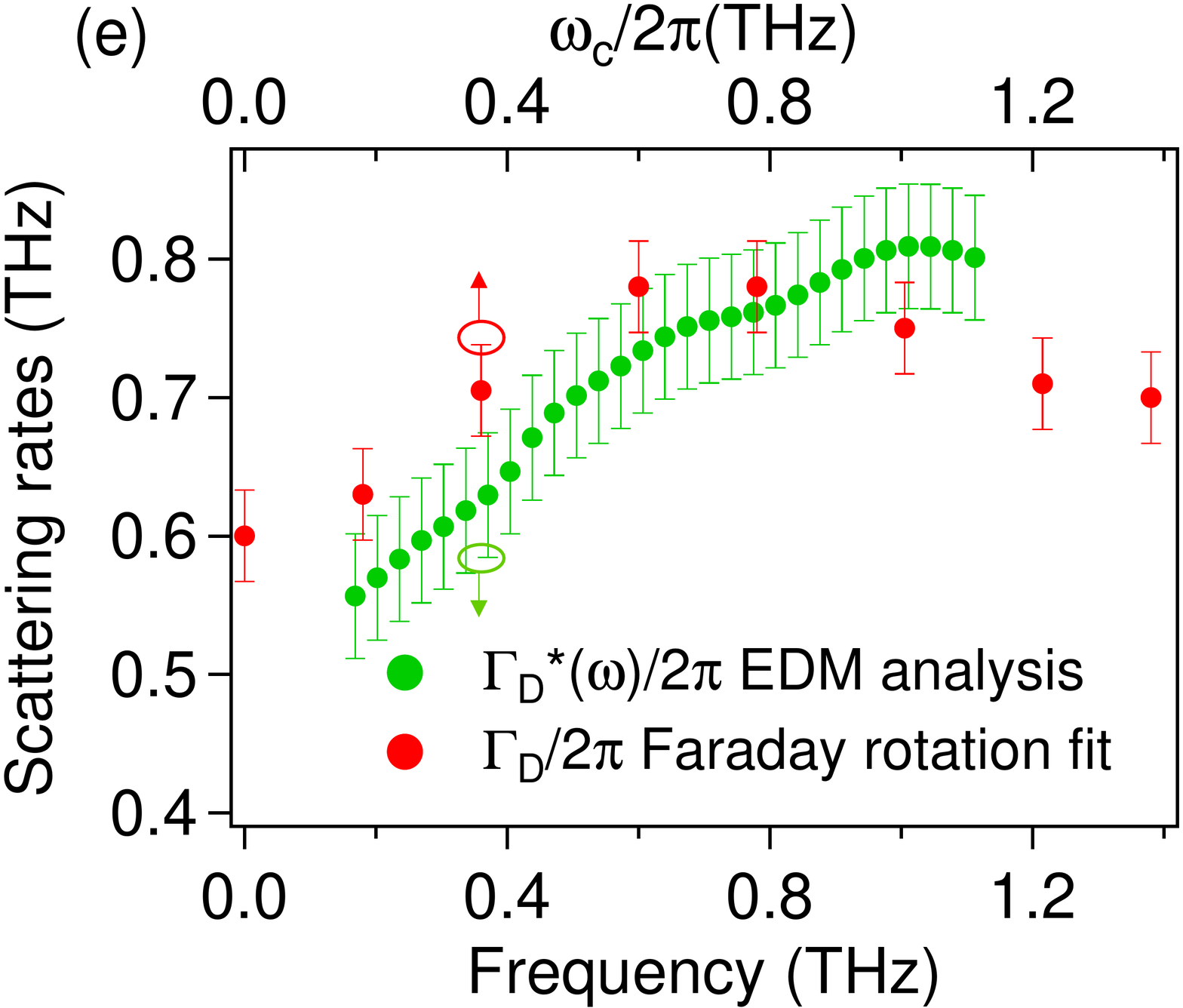}
\includegraphics[trim = 10 5 5 5,width=5.5cm]{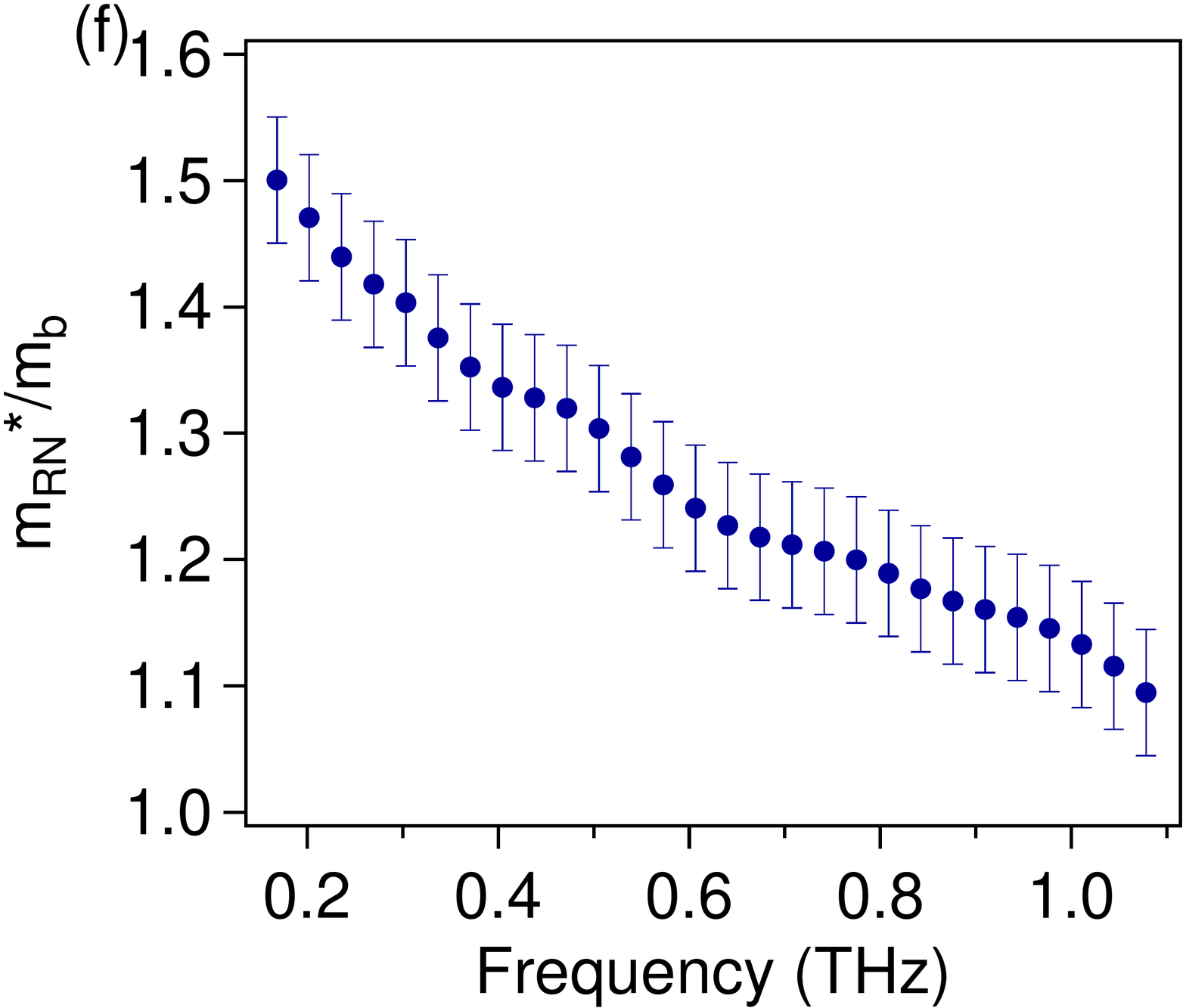}
\caption{(Color online) Data summary on 64 QL Cu$_{0.02}$Bi$_2$Se$_3$  (sample 2) which was exposed in air for 8 months. (a) Complex conductance at 5 K. (b) Real and (c) Imaginary part of complex Faraday rotation data at different fields at 5 K. (d) Cyclotron frequency versus field. The inset is a cartoon indicating $E_F\sim$150 meV, 70 meV below conduction band minimum. (e) Scattering rate as a function of cyclotron frequency (red). Renomalized scattering rate by mass through extended Drude analysis (green) as a function of frequency.  (f) Renormalized mass as a function of frequency.  The error bars express the uncertainty in $\omega_{pD}$ . }
\label{SIFig3}
\end{figure}

In Fig. \ref{SIFig4}, we show the fits for 100 QL Bi$_2$Se$_3$.  This demonstrates that a single channel is the principal contribution to the CR and dominates the Faraday rotation. This channel has spectral weight $(\omega_{p}/2\pi)^{2} d$ = 7.6 $\pm$0.3 $\times$10$^4$ THz$^2$ $\cdot$ nm. We use the spectral weight ( $\omega_{p}^{2} d =\frac{n_{2D}e^{2}}{m^{*}\epsilon_{0}}$) and CR mass 0.20 m$_e$ to extract a total sheet carrier density  $n_{2D}\sim1.9 \pm0.1 \times10^{13}/$cm$^{2}$. If using Eq. \ref{SIEqa2} with the TSS dispersion, from spectral weight we find $k_F\sim$ 0.11 $\AA^{-1}$,  $n_{2D}\sim2.0\times10^{13}/$cm$^{2}$, $m^{*}\sim$0.20 $m_{e}$ and $E_{F}\sim350$ meV, which agrees with ARPES results on the similar samples \cite{CaoNatPhys13}. Agreement between these two analyses self-consistently assign this channel to topological surface states. Considering that the characteristic penetration depth of the TSS is of order 2 nm \cite{WuNatPhys13, ZhangY10}, $\approx$7 $^{\circ}$ rotation per surface is an extremely large rotation per unit length. The second channel gives a small flat background. In the CR fits of 100 QL Bi$_2$Se$_3$, we fixed the scattering rate of the low mobility channel to 4 THz as we obtained from zero-field conductance fits, but the fits are reasonably insensitive to the precise scattering rate of this channel.  Field dependent scattering of the TSS channel of the 100 QL Bi$_2$Se$_3$ is shown in Fig. \ref{SIFig5} and the same basic physics as Cu$_{0.02}$Bi$_2$Se$_3$ was found. Considering twice more free parameters in the fits resulted from multiple channel conduction, we are not able to extract field dependent scattering of the TSS channel of the 100 QL Bi$_2$Se$_3$ as reliable as the Cu$_{0.02}$Bi$_2$Se$_3$ case. Also because of two Drude coexistence,  extended Drude analysis is not reliable to extract coupling strength even though suppression of scattering rate at low frequency is also observed. Another reason is that it is likely that substantial charge impurity scattering comes from Se defects at the back interface and compete with electron-phonon scattering as the former usually give a decreasing scattering rate with magnetic field\cite{YangPRB10}. However, electron-phonon coupling is normally expected not to be sensitive to doping and we believe a similar electron-phonon coupling exists in undoped Bi$_2$Se$_3$. It is just that the development of truly insulating  Cu$_{0.02}$Bi$_2$Se$_3$  system  gives a simpler spectra that allows us to determine the field dependent scattering rate systematically and provide quantitative measures of the electron-phonon coupling.
 

\begin{figure}[htp]
\includegraphics[trim = 10 5 5 5,width=5.5cm]{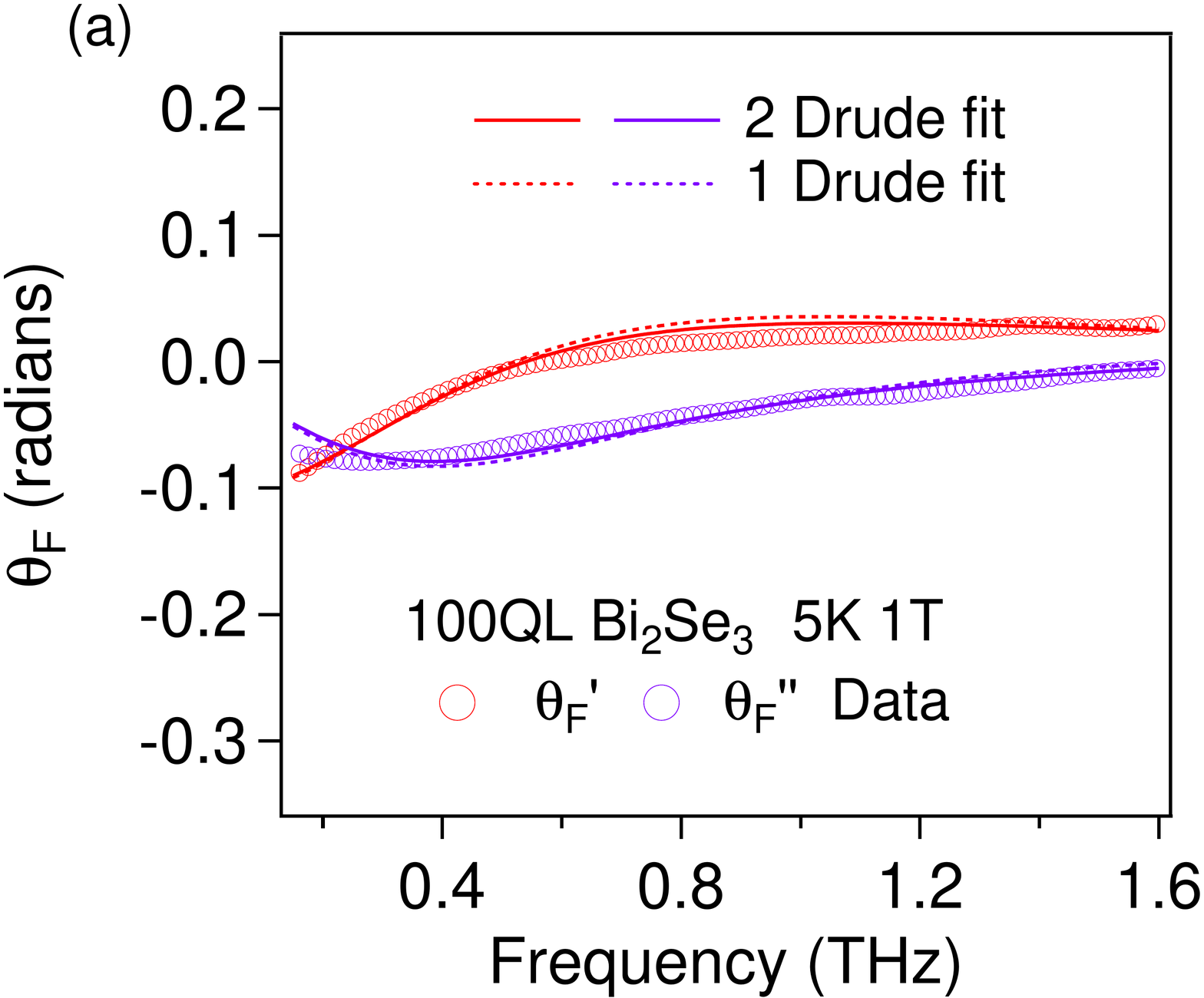}
\includegraphics[trim = 10 5 5 5,width=5.5cm]{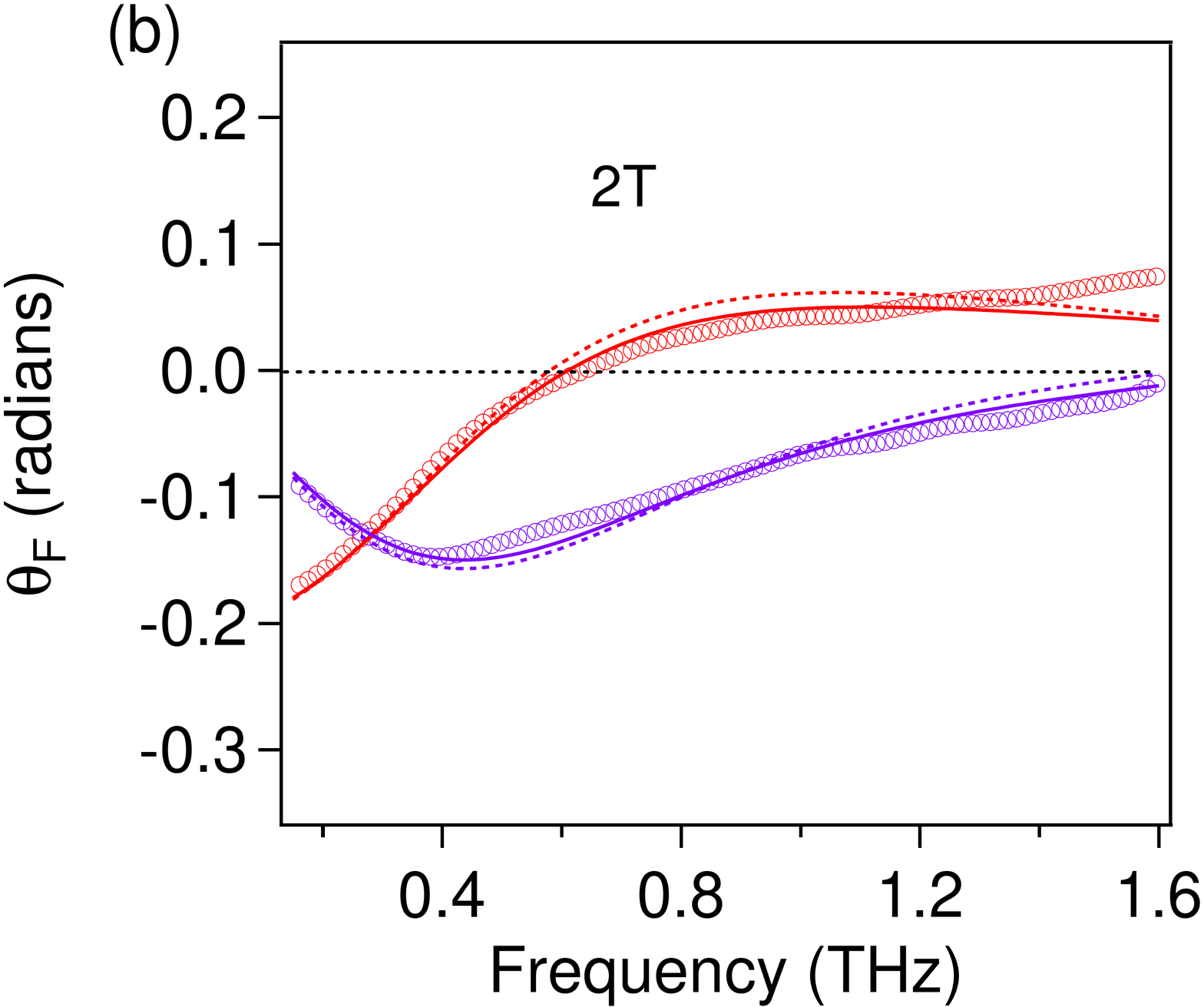}
\includegraphics[trim = 10 5 5 5,width=5.5cm]{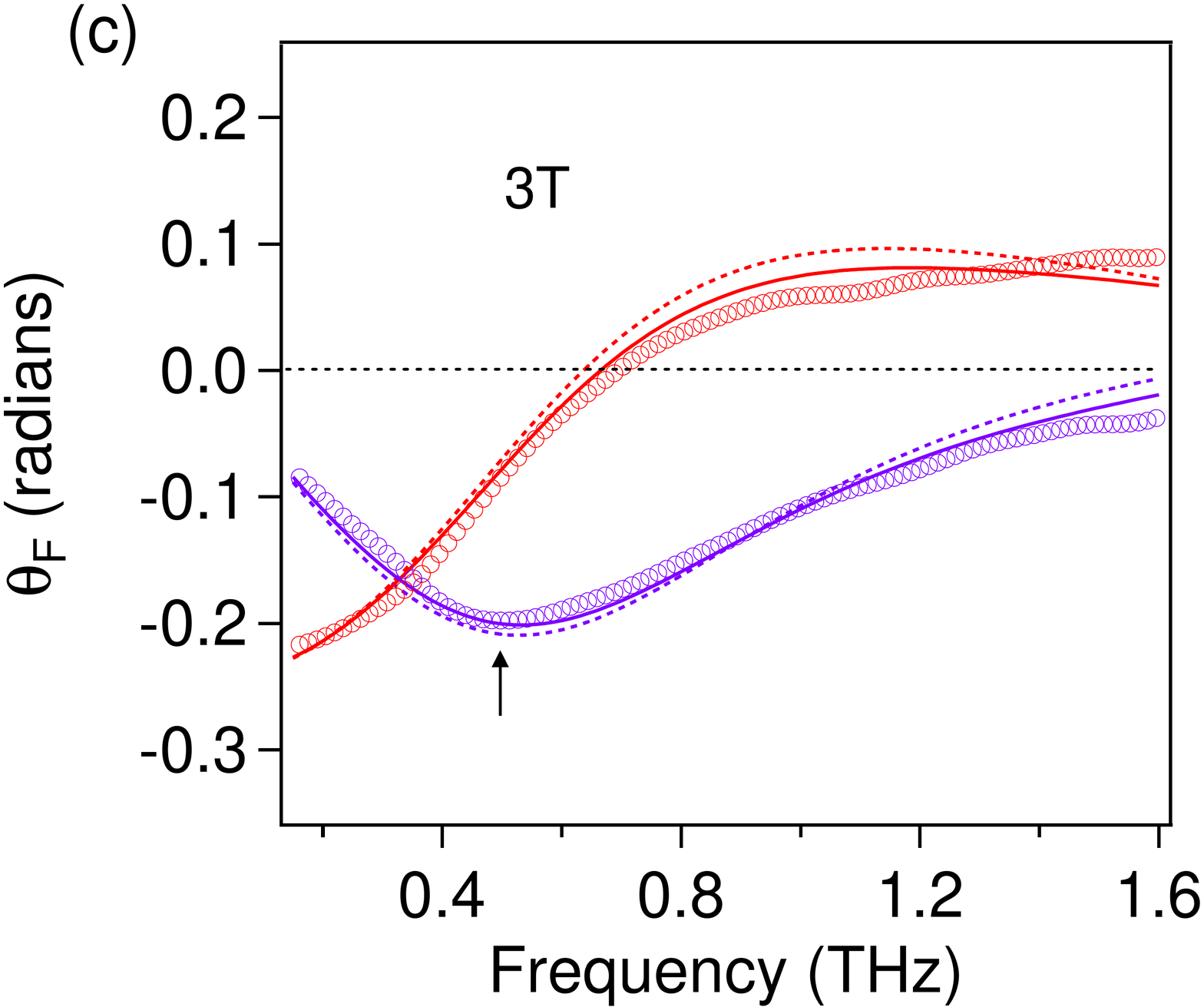}
\includegraphics[trim = 10 5 5 5,width=5.5cm]{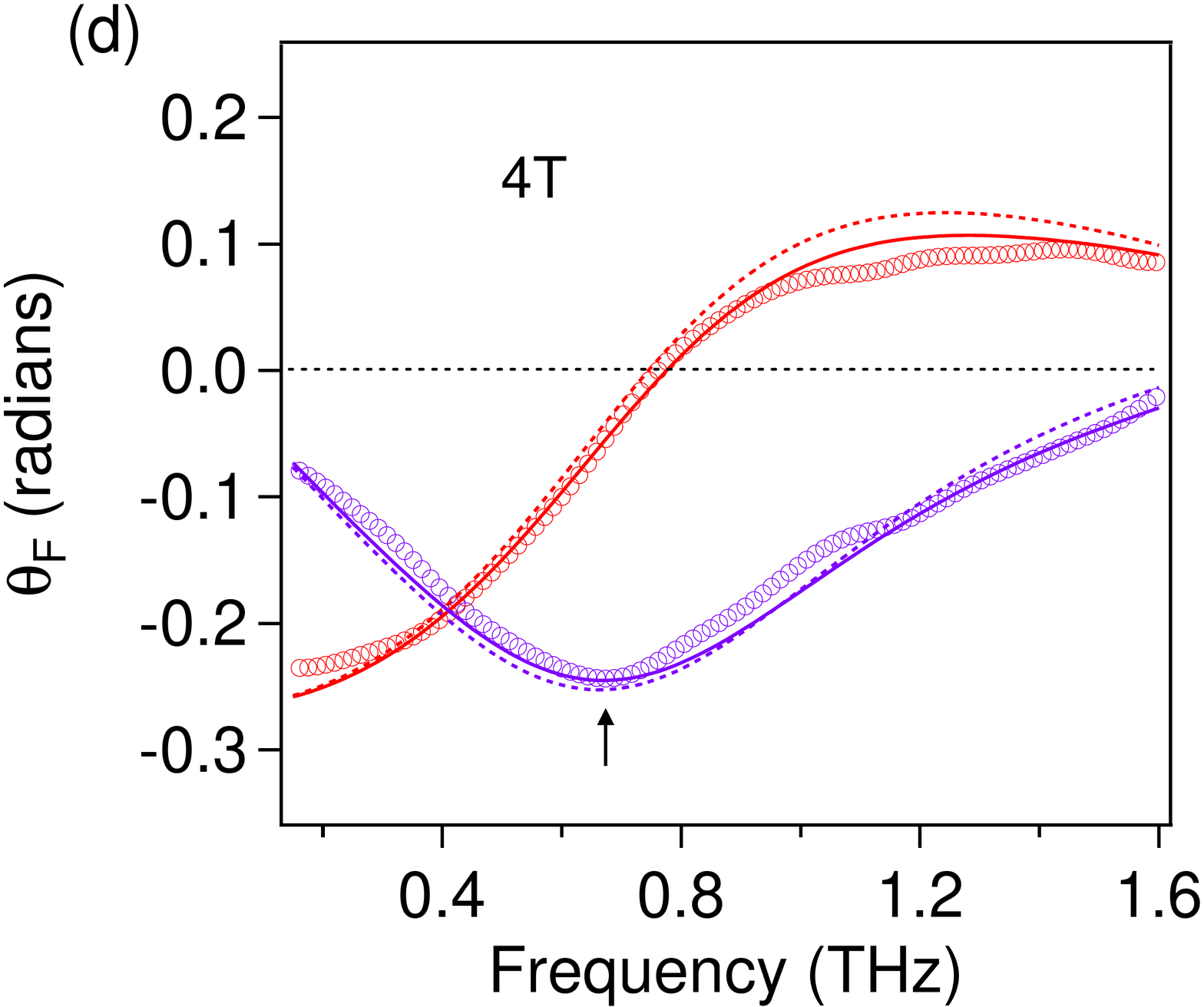}
\includegraphics[trim = 10 5 5 5,width=5.5cm]{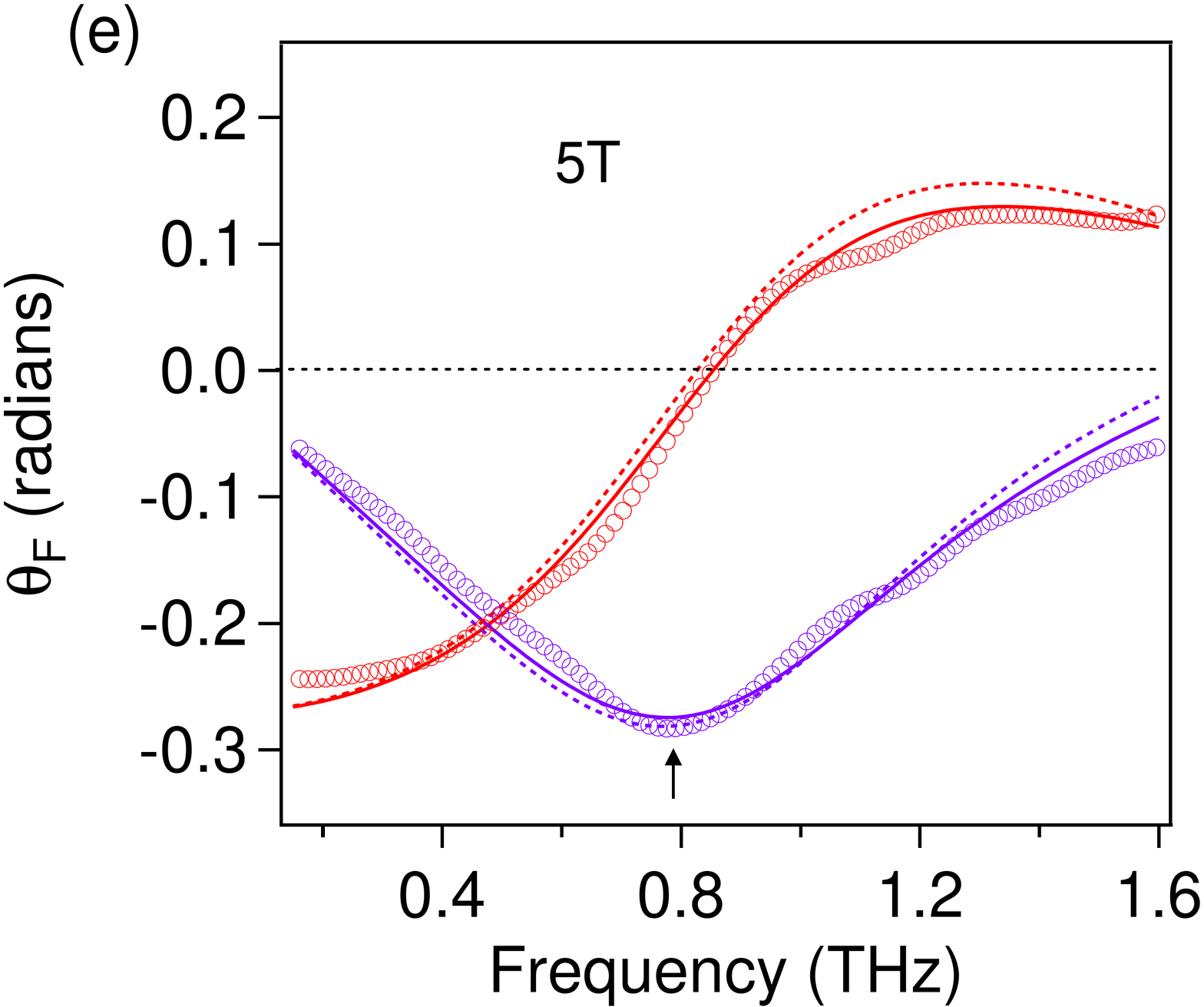}
\includegraphics[trim = 10 5 5 5,width=5.5cm]{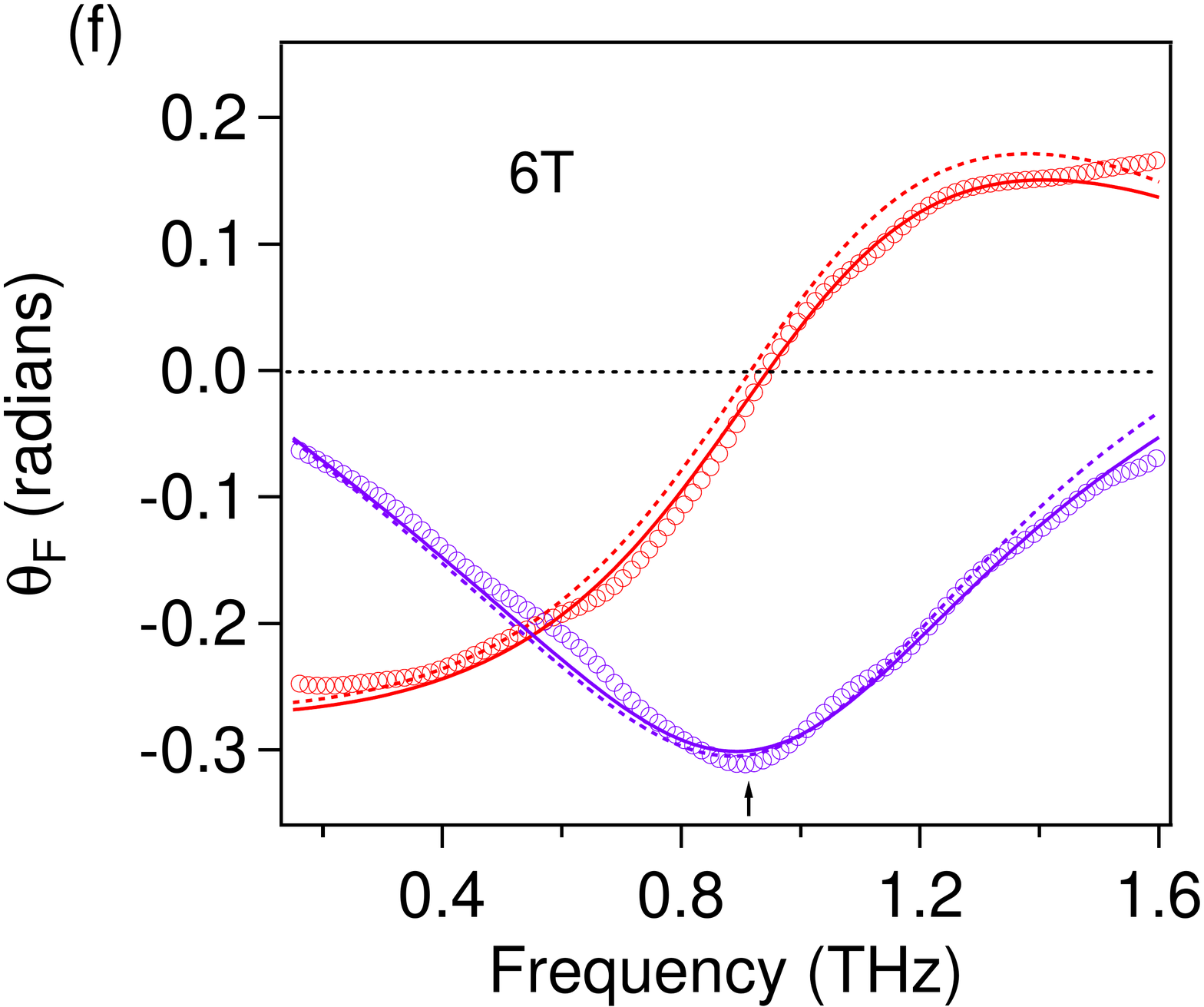}
\caption{(Color online) Fit quality on 100 QL Bi$_2$Se$_3$ at different fields. The solid curves are fits using two Drude terms. Dashed curves are with one Drude term. Arrows are eye guides for  cyclotron frequencies. Accurate numbers were determined by fits.}
\label{SIFig4}
\end{figure} 

\begin{figure}[htp]
\includegraphics[width=0.5\columnwidth,angle=0]{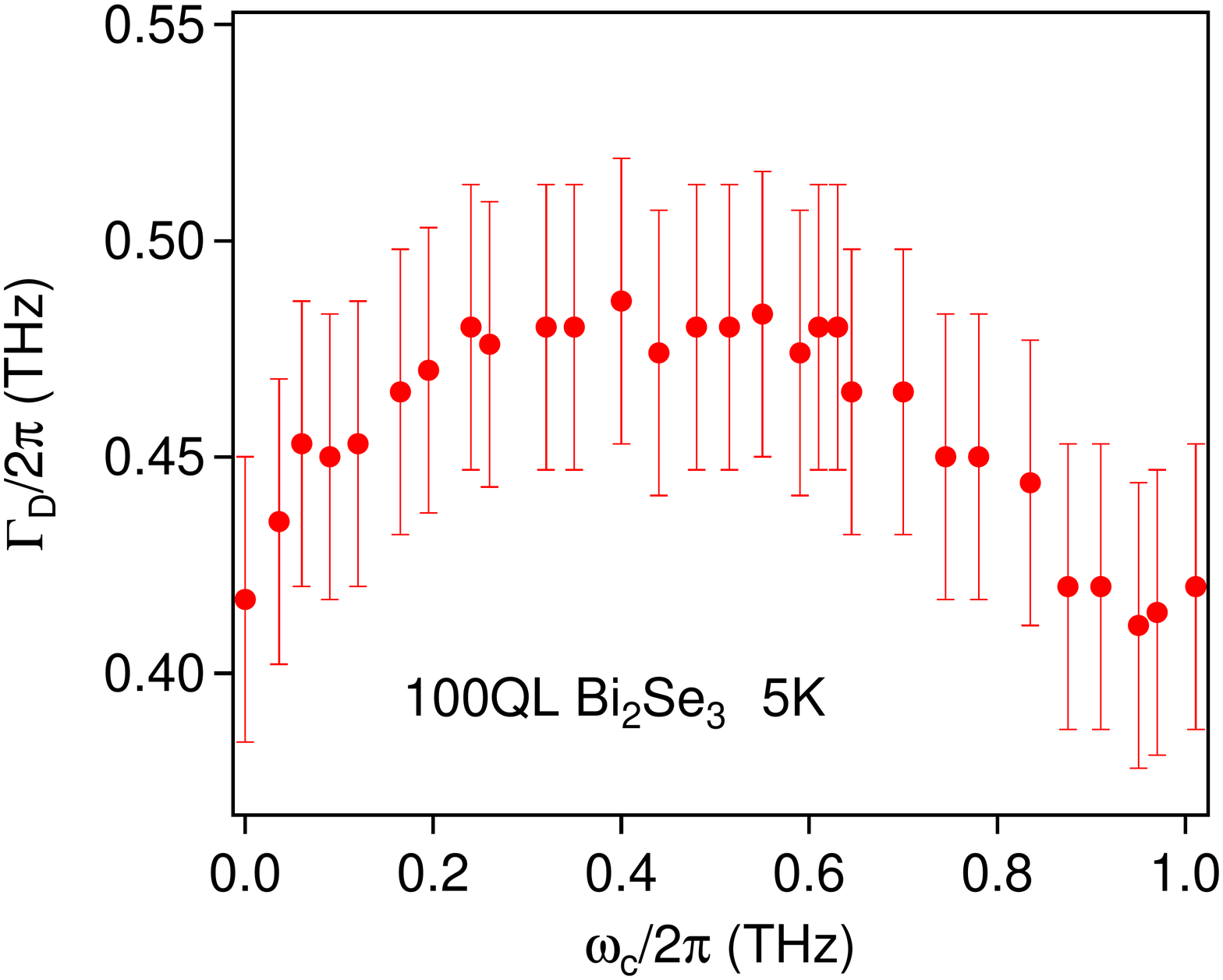}
\caption{(Color online) Drude scattering rate of 100 QL Bi$_2$Se$_3$ as a function of cyclotron frequency at 5 K.}
 \label{SIFig5}
\end{figure} 

\begin{figure}[htp]
\includegraphics[trim = 10 5 5 5,width=5.5cm]{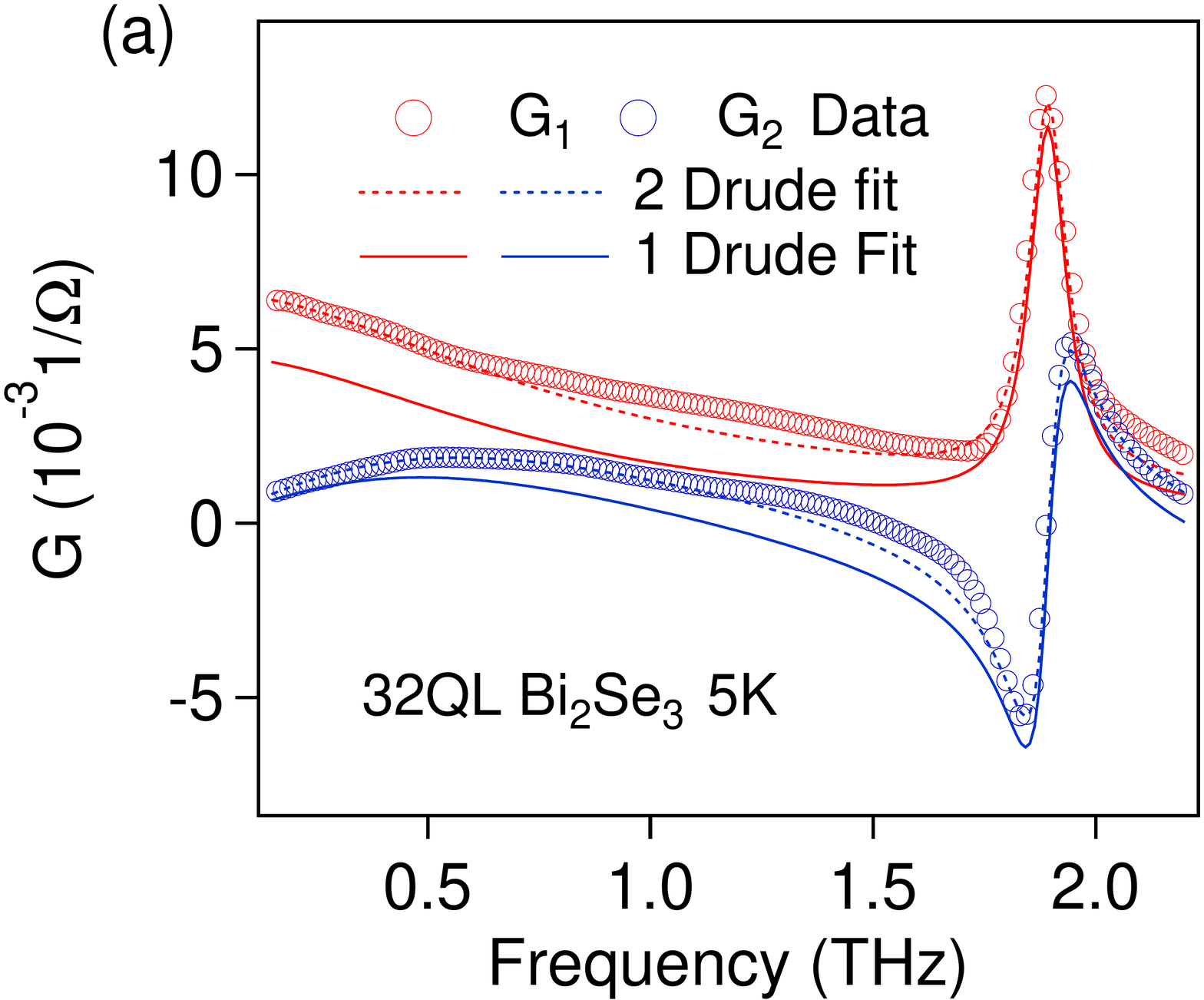}
\includegraphics[trim = 10 5 5 5,width=5.5cm]{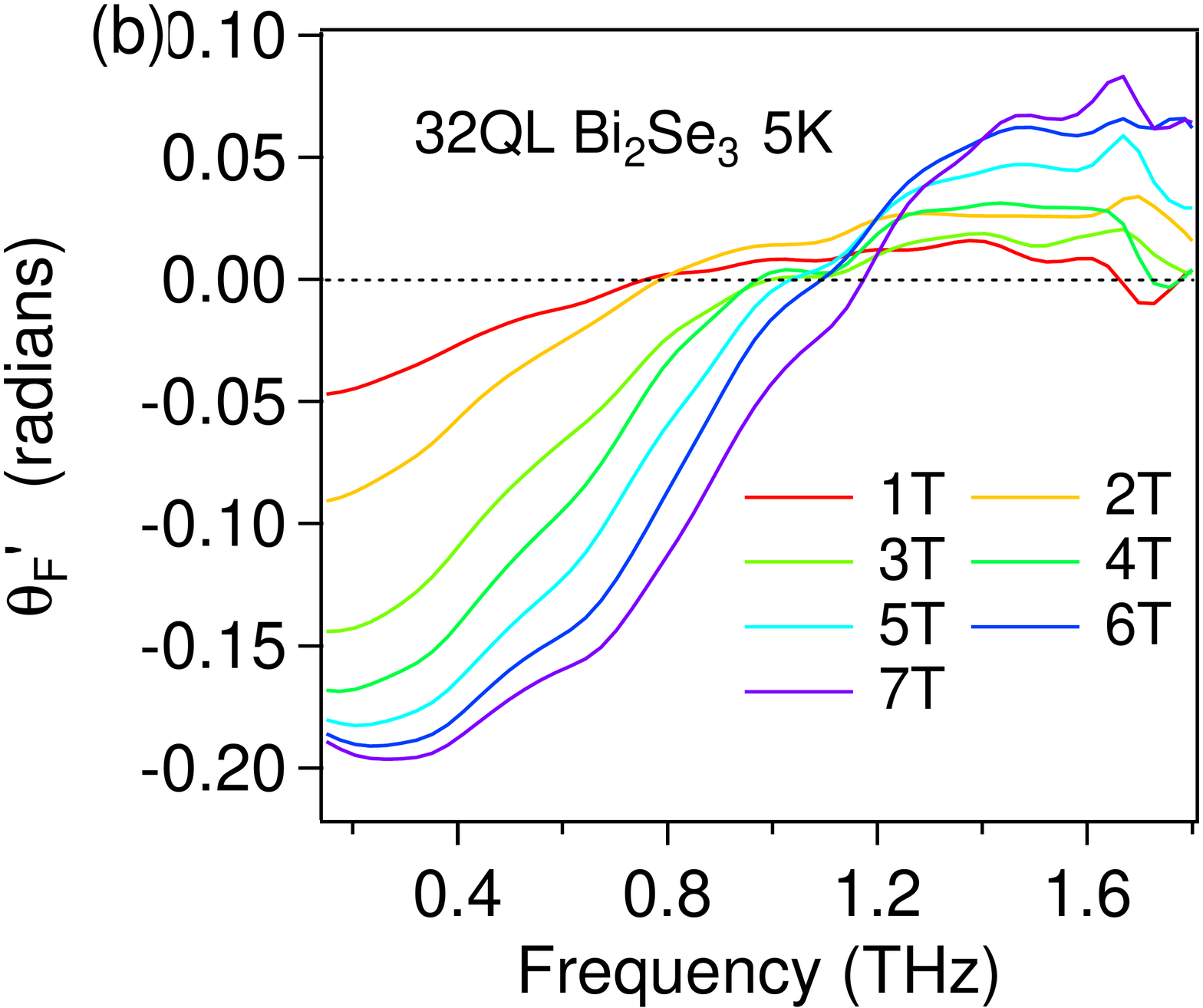}

\includegraphics[trim = 10 5 5 5,width=5.5cm]{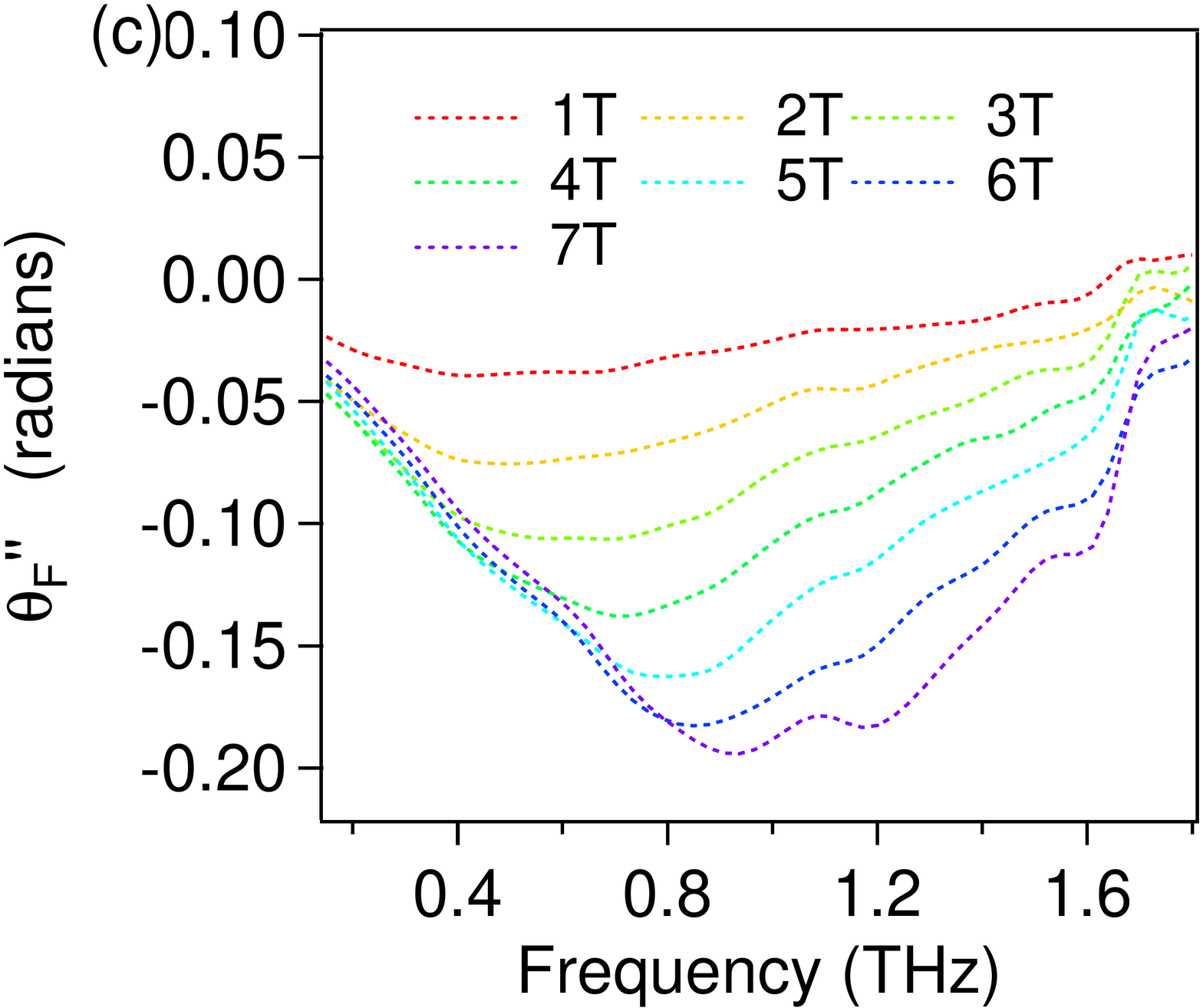}
\includegraphics[trim = 10 5 5 5,width=5.5cm]{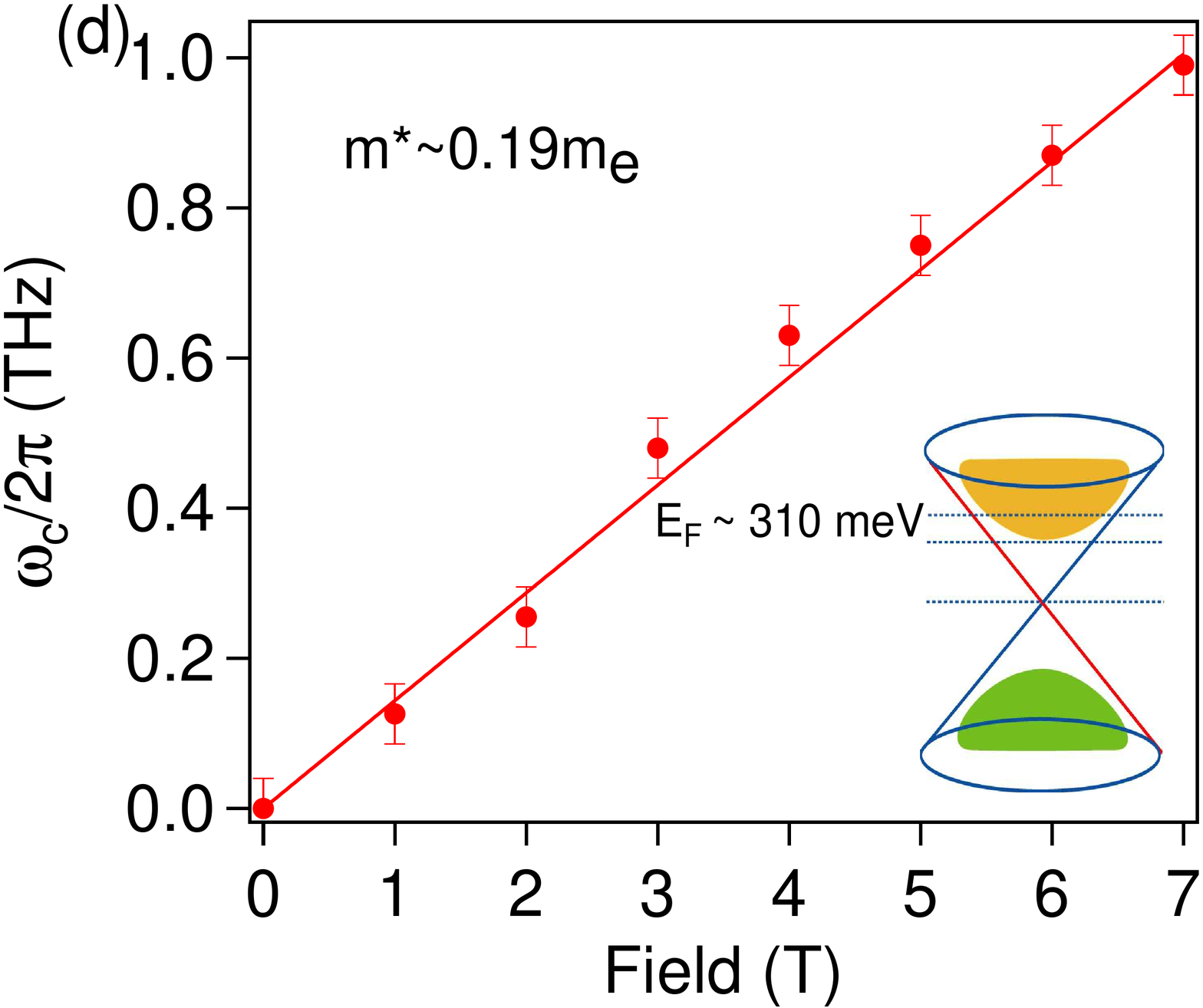}

\caption{(Color online) Data summary on 32 QL Bi$_2$Se$_3$ at 5 K. (a) Complex conductance at 5 K. Solid lines are 1 Drude fits using parameters obtained from fitting the Faradary rotation with 1 Drude term. Dashed lines are 2 Drude fit. (b)Real and (c) Imaginary Faraday rotation at different fields. (d) Cyclotron frequencies as a function of field. Solid line is a linear fit of $\omega_c=eB/m^*$. The inset is a cartoon indicating $E_F\sim$310 meV. }
\label{SIFig6}
\end{figure}

A 32 QL Bi$_2$Se$_3$ was also measured. Data summary is shown as Fig. \ref{SIFig6}. The channel that dominates CR has an effective mass $m^{*}\sim$0.19$m_e$, carrier density $n_{2D}\sim1.90\times10^{13}/$cm$^{2}$ and mobility $\mu\sim$2000 cm$^{2}/$V$\cdot$s, which is consistent with TSSs with $E_F$ $\sim$ 310 meV above the Dirac point. A second channel with scattering rate $\sim$ 4 THz is added to fit zero-field conductance. By using the effective mass of bulk/2DEG from literature,  the second channel has total sheet carrier density $\sim0.8\times10^{13}/$cm$^{2}$ and mobility less than 300 cm$^{2}/$V$\cdot$ s.

Note that the Faraday rotation spectra for Bi$_2$Se$_3$ and Cu$_{0.02}$Bi$_2$Se$_3$ looks similar in shape at first glance, but they are different in two ways. Firstly and obviously, CR frequencies is different, which gives a different mass. Secondly and most importantly, they indicate two samples are in completely different regime in terms of observation of quantized Faraday/Kerr rotation. Ref.\cite{Tse10a} proposed 90 degree Kerr rotation and  Faraday rotation with value of fine structure $\alpha \sim$ 7.3 mrad when chemical potential is in the lowest Landau Level (LL) in free standing films. Kerr rotation and Faraday rotation give the same information because ultimately it traces back to half-integer quantum hall effect in the conductance. In free standing films in the quantum regime, Faraday rotation is $tan(\theta_F)=\alpha(\nu_t+1/2+\nu_b+1/2)$, where $\nu_t$ and $\nu_b$ are LL filling factors for top and bottom surface states. In real experiments, free standing films are difficult to reach. People usually measure films on substrate and therefore renormalization from substrate effects come in. In the quantum hall regime, $tan(\theta_K)=\frac{4n\alpha}{n^2-1}(\nu_t+1/2+\nu_b+1/2)$$\sim$ 10 mrad $\times(\nu_t+1/2+\nu_b+1/2)$, $tan(\theta_F)=\frac{2\alpha}{1+n}(\nu_t+1/2+\nu_b+1/2)$$\sim$3.5 mrad $\times (\nu_t+1/2+\nu_b+1/2)$. In this regards, Ref.\cite{ValdesAguilarPRL12} with 65 degree Kerr rotation has never reached the quantum hall regime because chemical potential is way in the conduction band, which is not a topological insulator at all. From  Bi$_2$Se$_3$ to Cu$_{0.02}$Bi$_2$Se$_3$, we pushed LL filling factor $\nu \sim$ 40 to $\nu \leq$10 with 7 Tesla magnet. The Laudau Level energy spectrum is $E( \pm \nu)=\pm v_F \sqrt{2e \hbar B \mid \nu \mid}$, where $\pm$ stands for electron and holes representatively.  With experimentally achievable magnet field such as pulse magnet facility at Los Alamos National Laboratory, 150 T is enough to push the chemical potential into lowest LL and observe the quantized Faraday rotation. We just need to wait the THz system with pulse field finish its  construction. Recall that scattering rate in Cu$_{0.02}$Bi$_2$Se$_3$  is only less than 2 meV, so LLs will be well separated in energy. The threshold for magnetic field to reach lowest LL can be reduced further by combining with gating.  Also, we reduced resolution from 5 degree in Ref.\cite{ValdesAguilarPRL12} to 0.03 degree (0.5 mrad) in the current work. As discussed in the main text, our high-resolution Faraday rotation spectroscopy on Cu$_{0.02}$Bi$_2$Se$_3$ opens the new era to observed quantized Faraday rotation and topological magneto-electric effect, which are smoking guns for Dirac surface states.

\section{Transport Lifetime of Topological Insulators in an External Magnetic Field}
  
  In this section we investigate various possibilities for the increase in scattering rate at high magnetic fields deriving from effects due to the field itself.

\subsection{Zeeman (Spin) Effect On Transport Lifetime Of Topological Insulators In An External Magnetic Field}

In principle the canting of spins in the surfaces states due to Zeeman coupling under applied magnetic field can cause an increase in the scattering rate because back scattering can occur. In practice this is a very small effect at the current chemical potential levels. One can see this as follows.

In the topological regime where the bulk is insulating, one can
describe the light-matter interaction of a topological insulator by
the Hamiltonian of its surface states:
\begin{equation}
H_{\bm{k}} = v\bm{\sigma}\cdot\left(\bm{k}\times\hat{\bm{z}}\right), \label{Ham}
\end{equation}
where $\bm{k}$ is the electron momentum, $v$ is the Fermi velocity, 
and $\bm{\sigma}$ is the Pauli matrix  corresponding to the spin
degrees of freedom. In an external out-of-plane magnetic field $\bm{B} =
\nabla\times\bm{A}$ (where $\bm{A}$ the corresponding vector
potential), the Hamiltonian Eq.~(\ref{Ham}) couples to the magnetic
field through the orbital degrees of freedom with $\bm{k} \to
\bm{k}-e\bm{A}/c$, and the spin degrees of freedom with $H_{\bm{k}}
\to H_{\bm{k}}+\Delta$ where the Zeeman coupling $\Delta =
-g_s\mu_{\mathrm{B}}B/2$ acts like a Dirac mass term. We examine the
effect due to each coupling on the transport lifetime. 

In low magnetic fields, the effect of magnetic field on the transport 
lifetime is largely due to the spin polarization effect from Zeeman
coupling. Therefore we first consider the regime $\omega_c\tau_{tr},
\vert \Delta\vert\tau_{tr} \ll 1$, where $\omega_c = eB v/\hbar k_F$ is
the cyclotron frequency. It will be convenient to derive the Zeeman field dependence of the
transport lifetime $\tau_{tr}$ using the quantum kinetic theory \cite{QKE_ref}. The kinetic equation for the
momentum-dependent charge-spin density matrix $f_k$ is 
\begin{eqnarray}
\frac{\partial f_k}{\partial
  t}+i[H_k+\Delta\sigma_z,f_k]+\frac{1}{2}\left[e\bm{E}+\frac{e}{c}\bm{v}_k\times\bm{B},
  \frac{\partial f_k}{\partial \bm{k}}\right]_+ = J(f_k\vert \bm{k},
t), \label{QKE}
\end{eqnarray}
where $\bm{E}$ is the external A.C. electric field from the incident
light, $\bm{v}_k = \partial H_k/\partial \bm{k} =
v(\hat{\bm{z}}\times\bm{\sigma})$ is the single-electron velocity operator, and 
$[\,\,\,,\,\,\,]_+$ is the anti-commutator. $J(f_k\vert \bm{k},
t)$ denotes the collision integral containing information about 
scattering rates. Note that Zeeman coupling mixes different spins and
gives rise to a ``spin canting'' effect, opening up a spin-flipping channel through electron-impurity
scattering. A related problem was studied in the literature in the 
context of the anomalous Hall effect of topological insulators due to magnetic impurities 
 \cite{Dimi}. 

We first start by decomposing the density matrix as $f_k = f_k^{(0)}+\delta f_k$, where $f_k^{(0)}$ is the equilibrium 
distribution function and $\delta f_k$ is the non-equilibrium part. We 
can facilitate the solution of Eq.~(\ref{QKE}) by resolving $\delta 
f_k$ into components of projected onto the charge and spin sectors: $\delta
f_k =n_k+\boldsymbol{s}_k\cdot\boldsymbol{\sigma}$, where $n_k$ and
$\boldsymbol{s}_k$ describe the charge and spin density distributions respectively. For the 
semiclassical weak-field regime we are studying in this section, the Drude response is given by the conductivity to 
leading order in $1/E_F\tau_{tr}$, and the 
transport lifetime is determined only by the 
dynamics of $n_k$ and the component of $\boldsymbol{s}_k$ projected along the
electron's spin direction at the momentum $\boldsymbol{k}$. We find
the momentum relaxation rate as
\begin{equation}
\frac{1}{2\tau_k} = \frac{n_i\varepsilon_k}{4 v^2}\int_0^{2\pi}
\frac{\mathrm{d}\phi_{k'k}}{2\pi}\vert u(k,\phi_{k'k})\vert^2
\left(1-\cos\phi_{k'k}\right)\left(1+\cos^2\theta_k+\sin^2\theta_k\cos\phi_{k'k}\right),
 \label{rate1}
\end{equation}
where $\phi_{k'k}$ is the angle between the momenta $\bm{k}$ and
$\bm{k}'$, $\sin\theta_k = vk/\varepsilon_k$ and $\cos\theta_k
  = \Delta/\varepsilon_k$ with $\varepsilon_k=
  \sqrt{(vk)^2+\Delta^2}$, $n_i$ is the impurity concentration, and $\vert u(k,\phi_{k'k})
\vert^2$ is the disorder-averaged impurity potential 
evaluated at momentum $k' = k$.  In the 
absence of a magnetic field such that $\Delta = 0$, the angular form
factors in Eq.~(\ref{rate1})
reduce to the 
standard expression $(1-\cos^2\phi_{k'k})$ which describes the suppression
of back scattering of TSS. 

We can further model the impurity potential as short-range delta-function 
scatters with $\vert u(k,\phi_{k'k}) \vert^2 \equiv u_0^2$ being independent of $k$. The transport lifetime is determined by
Eq.~(\ref{rate1}) at the Fermi level at low temperatures $k_B  T \ll
E_F$. Evaluating the angular integral, we then obtain the transport 
lifetime in the weak magnetic field regime 
\begin{equation}
\frac{1}{2\tau_{tr}} = \frac{n_i u_0^2E_F}{8 v^2}
\left[1+3\left(\frac{g_s\mu_B B}{2E_F}\right)^2\right]. \label{rate2}
\end{equation}
Denoting the zero-field transport scattering rate as $1/2\tau_{tr,0} = n_i
u_0^2E_F/8 v^2$, the weak-field transport
scattering rate $1/2\tau_{tr,B}$ is related to that in zero field as 
\begin{equation}
\frac{1}{\tau_{tr,B}} = \frac{1}{\tau_{tr,0}}
\left[1+3\left(\frac{g_s\mu_B B}{2E_F}\right)^2\right]. \label{rate3}
\end{equation}

If we use $g_s\sim 50$, $E_F \sim 150$meV, the spin effect
only gives a $0.27\%$ increase at $3$ T, which is orders of magnitudes off from experiments. Therefore, we do not believe the Zeeman effect
is the cause for the increase of scattering rate.

\subsection{Orbital Effect On Transport Lifetime Of Topological Insulators In An External Magnetic Field}

In this section, we consider the effect of orbital coupling to the
magnetic field on the transport lifetime. The influence of impurity scattering on the density of states and
magneto-transport properties of the electron system can be captured using the
self-consistent Born approximation (SCBA) \cite{SCBA}. This
approximation captures certain features of Landau level (LL) broadening due
to disorder and is often sufficient to describe magnetotransport
properties when localization effects are not important, as expected in 
optical experiments. 

The SCBA for the case of graphene was discussed extensively in
Ref.~ \cite{Ando1}. For our Cu-doped sample, we estimate from Faraday rotation and SdH oscillations\cite{BrahlekPRL14} that the minimum occupied LL which occurs at the largest field $B = 7$ T is $n \sim10$. Therefore, the Fermi level is far from the $n = 0$ Landau level
(LL), and the effect of inter-LL coupling between opposite
levels $n$ and $-n$ is negligible. The LL spacing at the Fermi level is of the order of meV, which
coincides with the energy scale of the scattering rate we extracted 
from our data [see Fig. 3(a) in main text]. This seems to suggest the sample would
be in the regime where $\omega_c\tau_q \lesssim 1$, where $1/\tau_q$ is the
single-particle lifetime that captures the LL disorder broadening (to
be distinguished from $1/\tau_{tr}$ which is the transport lifetime we
want to calculate. Generally $1/\tau_q > 1/\tau_{tr}$.)


The self-energy matrix is diagonal
in the LL index and given by  
\begin{equation} 
\Sigma = \omega_c^2\rho\sum_{n =
  0}^{N_c}\frac{\varepsilon-\Sigma}{(\varepsilon-\Sigma)^2-\varepsilon_n^2}, \label{SCBA6} 
\end{equation}
where we have defined the dimensionless parameter $\rho = n_i
u_0^2/(4\pi v^2)$ that characterizes the disorder strength. Eq.~(\ref{SCBA6}) defines the
self-consistency condition from which the self-energy can be solved. For high LL filling $\vert N\vert \gg 1$, one can use the
approximation of extending 
the lower limit of the sum to $-N_c$ and setting $N_c \to
\infty$, because the main contribution comes from the Fermi level and
contributions from negative LLs and from $N$ to $\infty$ amount to
negligible errors.

Performing the sum in Eq.~(\ref{SCBA6}) using the Poisson summation
formula,
\begin{equation} 
\Sigma = \pi \rho \left(\varepsilon-\Sigma\right)\mathrm{cot}\left\{\frac{\pi}{\omega_c^2}\left[\left(\varepsilon-\Sigma\right)^2-\Delta^2\right]\right\},
\label{SCBA7}
\end{equation}
and then expanding the $\mathrm{cot}$ function into a Fourier series,
we obtain 
\begin{equation}
\Sigma'+i\Sigma'' = -i
\pi\rho\left(\varepsilon-\Sigma'-i\Sigma''\right)\left\{1+2\sum_{k =
    1}^{\infty}\lambda_D^k \mathrm{exp}\left\{i\frac{2\pi
      k}{\omega_c^2}\left[\left(\varepsilon-\Sigma'\right)^2-\Delta^2-\left(\Sigma''\right)^2\right]\right\}\right\}, \label{SCBA10}
\end{equation}
where we have denoted the real and imaginary parts of the self-energy
as $\Sigma = \Sigma'+i\Sigma''$. $\lambda$ is the so-called Dingle
factor 
\begin{equation}
\lambda_D = \mathrm{exp}\left\{-\frac{4\pi}{\omega_c^2}\vert \Sigma''
  \vert\left(\varepsilon-\Sigma'\right)\right\}.
\label{SCBA11}
\end{equation}
Eq.~(\ref{SCBA10}) is a nonlinear equation that needs to be solved
iteratively. $\lambda_D$ serves as the small parameter with which
Eq.~(\ref{SCBA10}) can be solved for $\Sigma$ up to 
first order in disorder correlation $\sim n_i
u_0^2$ ($\mathcal{O}(\rho^1)$) and to first order 
in $\lambda_D$ ($\mathcal{O}(\lambda_D^1)$). 

First, in the zeroth order $\mathcal{O}(\lambda_D^0)$ we have $\Sigma' = 0$ and
$\Sigma'' = -(n_i u_0^2/4v^2)\varepsilon$. The latter can be
identified as the quasiparticle scattering rate $1/2\tau_{q,0} =  (n_i
u_0^2/4v^2)\varepsilon$ at zero magnetic field. 

To first order $\mathcal{O}(\lambda_D^1)$, we obtain 
\begin{eqnarray}
\Sigma' &=& \frac{1}{\tau_{q,0}}\varepsilon\lambda_D\sin\left[2\pi
    \left(\frac{\varepsilon^2-\Delta^2}{\omega_c^2}\right)\right],
  \nonumber \\
\Sigma'' &=& -\frac{1}{2\tau_{q,0}}\left\{1+2\lambda_D\cos\left[2\pi
    \left(\frac{\varepsilon^2-\Delta^2}{\omega_c^2}\right)\right]\right\}. 
\label{SCBA12}
\end{eqnarray}
The density of states is given by $\nu(\varepsilon) =
2\mathrm{Im}\Sigma/\omega^2\rho$, therefore we obtain from
Eq.~(\ref{SCBA12})  
\begin{equation}
\nu(\varepsilon) = \nu_0(\varepsilon)\left\{1+2\lambda_D
  \cos\left[2\pi\left(\frac{\varepsilon^2-\Delta^2}{\omega_c^2}\right)\right]\right\},
\label{SCBA13}
\end{equation}
with $\nu_0(\varepsilon) = {\varepsilon}/{2\pi v^2}$ being the density 
of states at zero magnetic field. 

Since the ratio of the transport scattering rate at finite field to that at zero
field is equal to that of the density of states \cite{Zudov_RMP},
finally we obtain 
\begin{equation}
\frac{1}{\tau_{tr,B}} = \frac{1}{\tau_{tr,0}}\left\{1+2\lambda_D
  \cos\left[2\pi\left(\frac{E_F^2-(g_s\mu_B B/2)^2}{\omega_c^2}\right)\right]\right\},
\label{SCBA13}
\end{equation}
where $1/\tau_{tr,0} = n_iu_0^2E_F/4v^2$ is the transport 
time at zero field. 

In this scenario, the scattering rate should show 
magneto-oscillations above $3$ T when $\omega_c\tau_{tr,B} \geq 1$. As we saw no such oscillations and only an increase in the scattering rate with field, we also rule out this possibility.



\section{Effect of S\lowercase{e} capping}

\begin{figure}[htp]
\includegraphics[trim = 10 5 5 5,width=8cm]{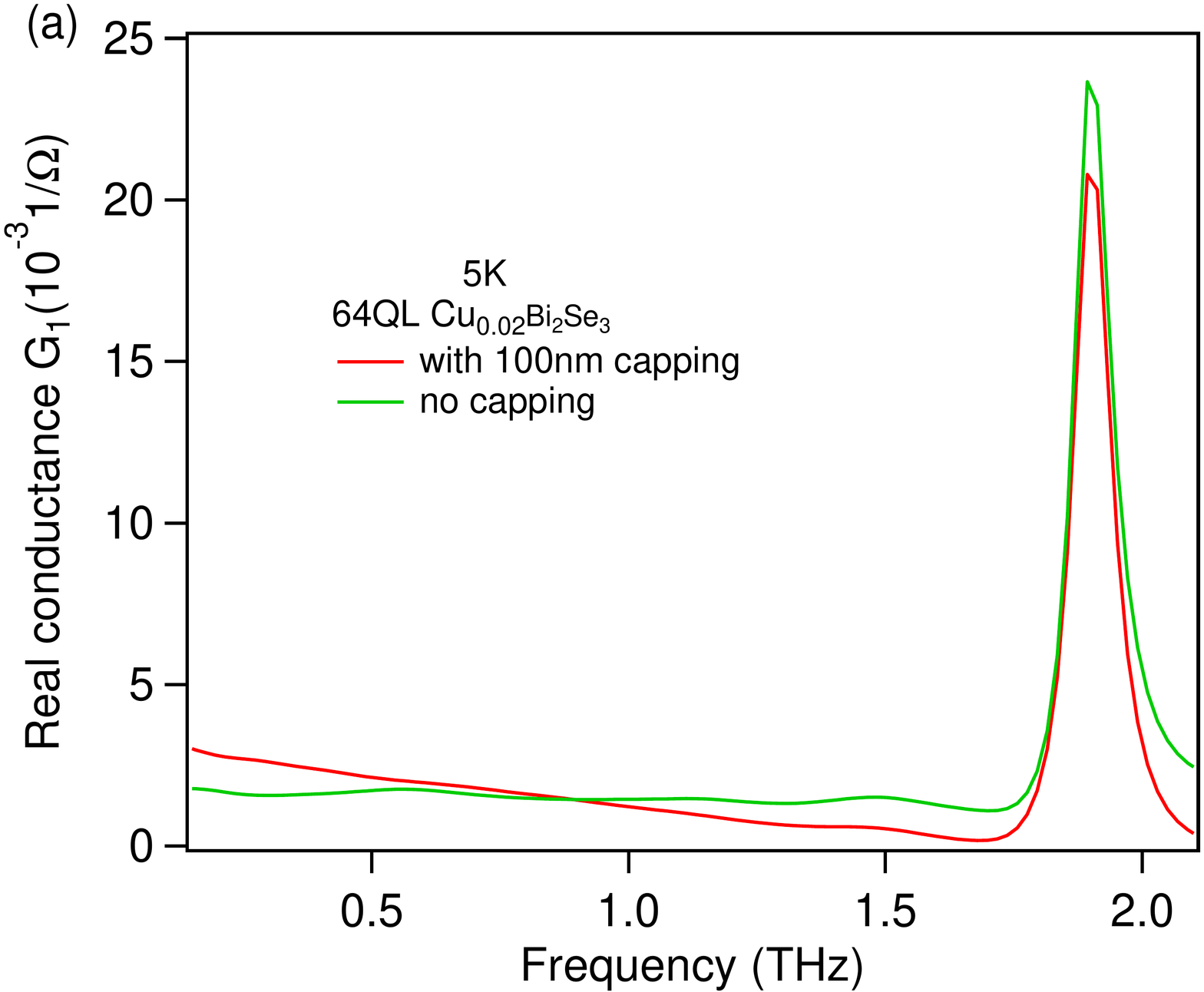}
\includegraphics[trim = 10 5 5 5,width=8cm]{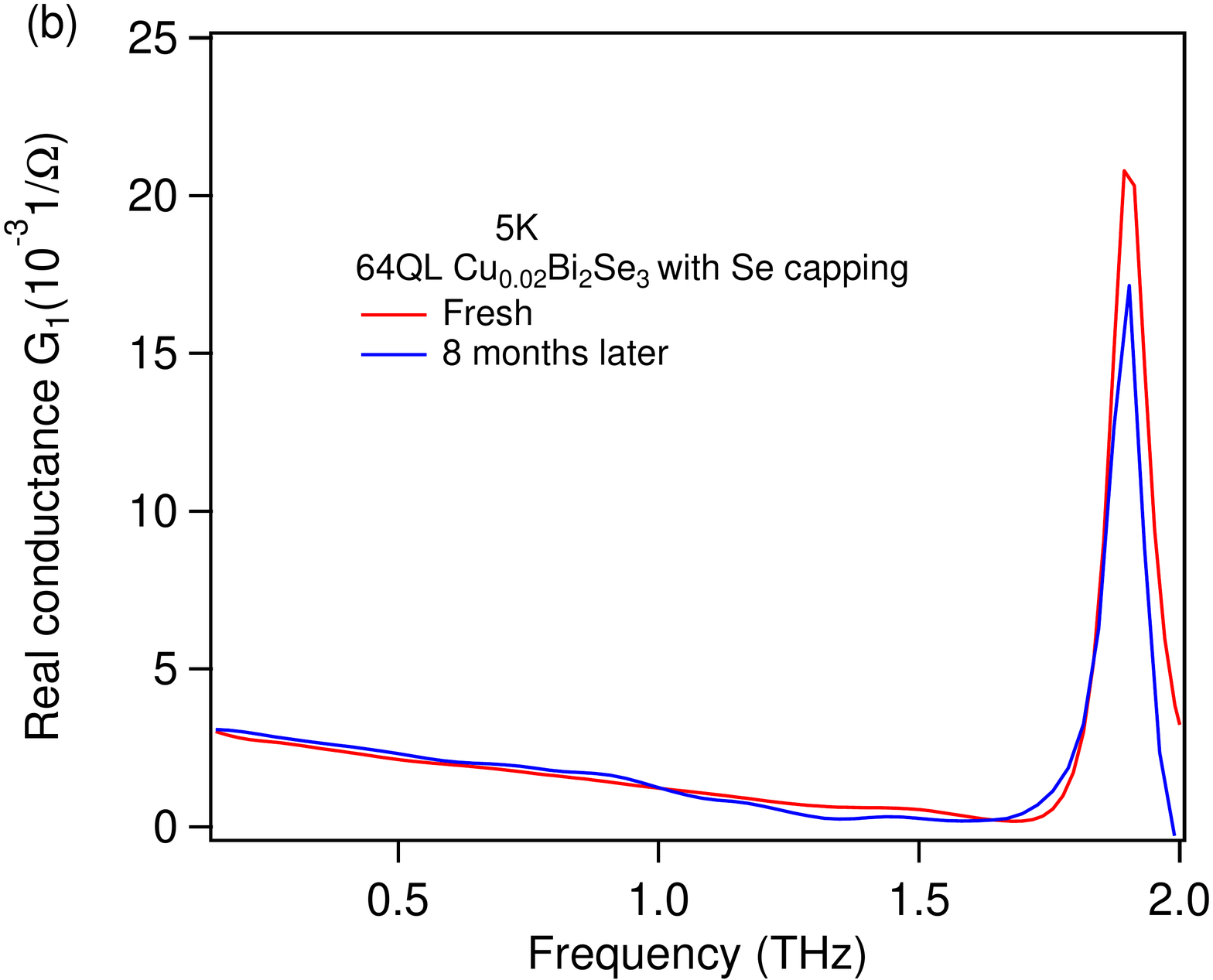}
\caption{(Color online) (a) Real conductance of 64 QL Cu$_{0.02}$Bi$_2$Se$_3$ (sample 2) with and without 100nm Se capping at 5 K . (b) Real conductance of 64 QL Cu$_{0.02}$Bi$_2$Se$_3$ (sample 2) with 100nm Se capping measured right after receiving the sample and 8 months later.}
 \label{SIFig7}
\end{figure} 

Our previous work on pure Bi$_2$Se$_3$ showed that Se capping had a negligible effect on the optical properties of Bi$_2$Se$_3$, but does decrease the scattering rate somewhat\cite{WuNatPhys13}. However, it is of great importance in the present case of Cu doped Bi$_2$Se$_3$ as these samples are very air sensitive.   Transport measurements in Ref. \cite{BrahlekPRL14} were performed right after samples were taken out of ultra-high vacuum MBE chamber. Samples for THz measurements were sealed in vacuum bags and shipped overnight to JHU. The total exposure to atmosphere is around 30 mins before loading into the cryostat and plus the overnight in the raw-vacuum bag. As one can see in Fig. \ref{SIFig7}(a), Se capping protects the mobility of the sample well as opposed to non-capped samples with a flat Drude component. 100nm Se capping still protects the sample well even after 8 months explosure in air as shown in Fig.\ref{SIFig7}(b) on sample 2.

\section{Limitations of using one Drude component fit in previous work \cite{ValdesAguilarPRL12, WuNatPhys13}}

For the 100 QL Bi$_2$Se$_3$ sample, if we fit the zero field conductance data by only a single Drude term, a phonon term and a $\epsilon_{\infty}$ term,  we find the Drude spectral weight $(\omega_{pD}/2\pi)^{2} d$ is $1.25  \times 10^5$ THz$^{2}$ $\cdot$ nm.  Using Eq. \ref{SIEqa2}, one gets $k_F\sim$0.14 $\AA^{-1}$, $m^{*}\sim$0.22 $m_{e}$ and $E_{F}\sim$480 meV.  We believe these values are overestimated and the low frequency spectral weight is actually comprised of two terms.   As detailed above, a fit to the Faraday rotation data by one Drude term alone shows that the spectral weight  $(\omega_{pD}/2\pi)^{2} d$ contributing to the TSSs is only $7.6  \times 10^4$ THz$^{2}$ $\cdot$nm.     Using Eq. \ref{SIEqa2},  $k_F\sim$0.11 $\AA^{-1}$, $m^{*}\sim$0.20 $m_{e}$ and $E_{F}\sim$ 350 meV are obtained, as  already mentioned above.  One can see $E_F$ and $k_F$ are overestimated by $\sim$ 30$\%$, while $m^{*}$ is overestimated by $\sim$ 10$\%$ if one associates all the Drude spectral weight in the zero-field conductance to the TSSs.  These quantities are overestimated in the single Drude component model as one assigns all the bulk/2DEG spectral weight as the TSSs spectral weight. The second Drude channel (bulk/2DEG) is flat in our measurement frequency regime and only contributes a smooth background. However, considering the fact that the TSSs contribute more than 90$\%$ to the total conductance at low frequencies, a single Drude model analysis is still a good approximation and a thickness independent Drude peak as observed previously in Bi$_2$Se$_3$ \cite{ValdesAguilarPRL12, WuNatPhys13} is not surprising.  The presence of this very flat subdominant contribution does not call into question any of the conclusions of our previous works \cite{ValdesAguilarPRL12, WuNatPhys13}.  Here we just clarify this point and provide a method to separate contributions from TSSs and bulk/2DEG in magneto-THz measurements.   

\section{Effective mass inconsistency with previous work\cite{ValdesAguilarPRL12}}

\begin{figure}[htp]
\includegraphics[width=0.5\columnwidth,angle=0]{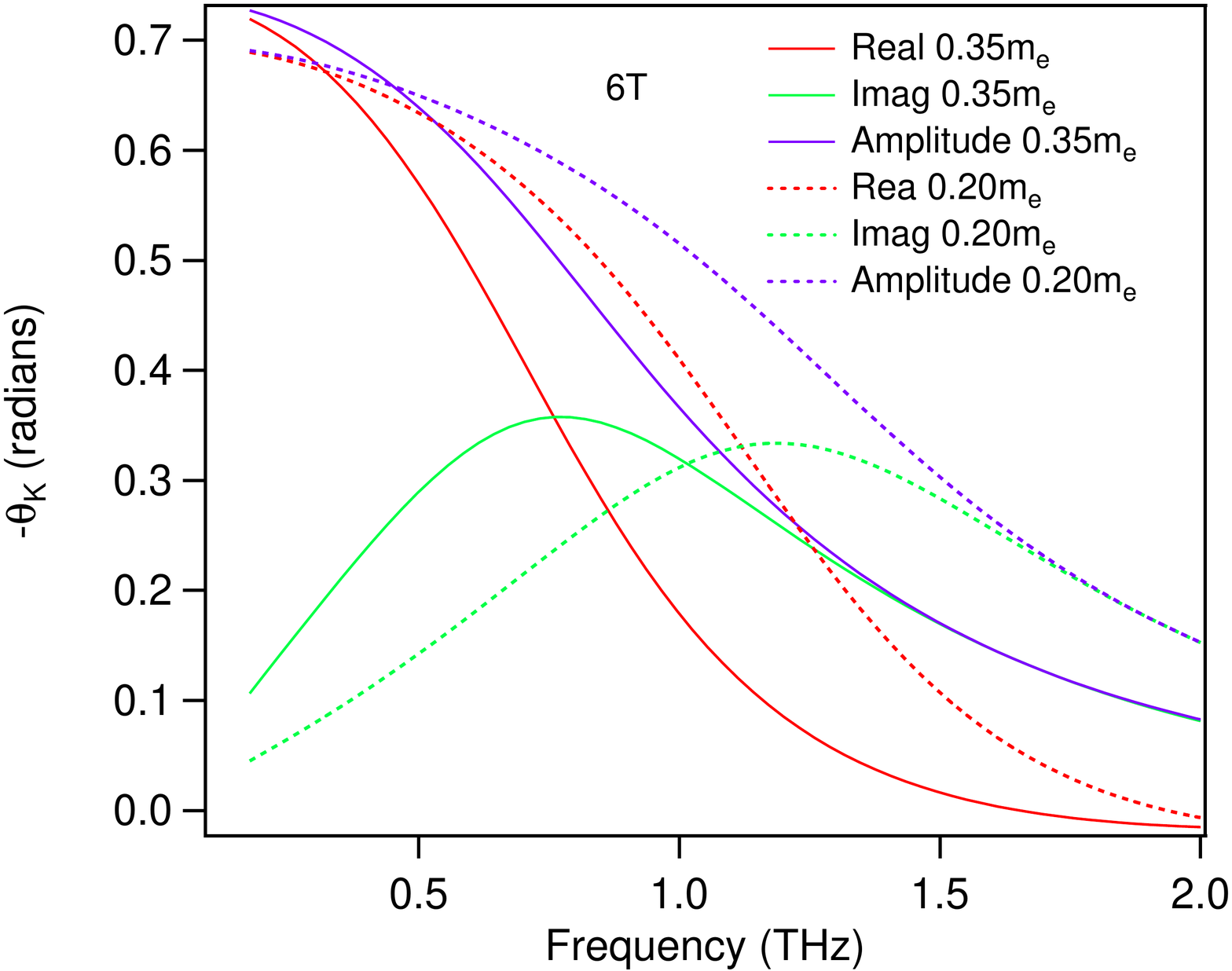}
\caption{(Color online) Calculation of various measures of the complex Kerr rotation with effective mass 0.35 $m_e$ and mass 0.20 $m_e$ with carrier density $n_{2D}=2\times10^{13}/$cm$^{2}$ and scattering rate $\Gamma_D$=1 THz at 6 T.}
 \label{SIFig8}
\end{figure} 

In our earlier work \cite{ValdesAguilarPRL12}, based on a fit with just a single Drude term, a heavier 0.35 $m_{e}$ cyclotron mass for TSSs was reported.  This was judged to be a reasonable number based on a linear dispersion of the TSSs as it gave a reasonable $E_F\sim$0.5 eV using $m^{*}=E_F/v_F^2$ and $v_{F}= 5\times10^{5}$ m/s.  However,  there are a number of inconsistencies that this mass presents.  If one considers quadratic corrections to the TSS dispersion, one finds that this mass does not give a consistent $k_F$.  Moreover, if one uses a linear dispersion $E_F=\hbar v_{F}k_{F} $ and uses the spectral weight of a single Drude component to estimate the chemical potential, the relation is:

\begin{equation}
 \omega_{pD}^{2} d  =  \frac{\hbar v_{F}k_{F}e^2 }{ 2\pi \epsilon_0 \hbar^2}=\frac{E_F e^2}{2\pi \epsilon_0 \hbar^2}
\label{Eqa16}
\end{equation}

\noindent Here we assume two almost identical surface states. An analysis of the spectral weight gives  $E_{F}\sim$0.75 eV, which is unreasonably high. Therefore, when  $E_{F}$ is in the conduction band, quadratic corrections must be significant.  One may note that in order that the wave functions of the TSSs vanish at the TI/vacuum interface, a quadratic correction is required \cite{LiuPRB10,RefaelPriviteComm}.  In contrast, when we use Eq. \ref{Eqa16} based on linear dispersion to estimate the chemical potential of Cu$_{0.02}$Bi$_2$Se$_3$, $E_{F}$ is found to be 170 meV $\pm$ 10 meV. This better agreement is expected because surface state dispersions deviate from linearity less when $E_F$ is closer to Dirac point.  In summary, we believe that the CR mass 0.35 m$_e$ given in previous work Ref. \cite{ValdesAguilarPRL12} was incorrect for pure Bi$_2$Se$_3$ films and the revised value of 0.19-0.20 $m_e$ given in this work is correct.

The reasons for this correction is multifold.   1.) Signal to noise has been improved by orders of magnitude in the current generation of experiments with polarization modulation technique with 0.5 mead resolution\cite{MorrisOE12} allowing more reliable fits to the spectra.   2.) The current generation of Bi$_2$Se$_3$ films have  mobilities of the order 2000 - 3200 cm$^{2}/$V$\cdot$ s \citep{BansalPRL12}. In order to see a well-defined cyclotron resonance in the dissipative response, the relation $\omega_{c}\tau=\mu B \geqslant1$ needs to be satisfied. $\omega_{c}\tau=\mu B \sim$ 1.5-2 is still too small to see very sharp peaks in the amplitude of Kerr/Faraday rotation in our field range.   3.)  In the previous work, due to (now overcome) uncertainties in the measured phase in the polarization modulation experiment, only the amplitude of the Faraday/Kerr rotation was considered reliable.   Unfortunately this is a somewhat insensitive method of probing the cyclotron frequency.  As one can see from Fig. \ref{SIFig8}, the rotation spectrum does not differ much using 0.35$m_{e}$  or 0.20$m_{e}$ in the amplitude plot.  To fit the data accurately when $\omega_{c}\tau \sim 1$, one needs to fit the complex Faraday rotation; the peak in imaginary part ($-\theta_K^{''}$) approximately gives the direction evidence for cyclotron resonance. Ref.\cite{ValdesAguilarPRL12} did not observe obvious signature for CRs because of measuring amplitude of Kerr rotation with poor resolution. In this regard, phase sensitive time-domain THz spectroscopy is a powerful tool to study cyclotron resonances in topological insulators of current generation. Despite the differences in these numbers, the Bi$_2$Se$_3$ samples we measured in this paper are similar to those in Ref \cite{ValdesAguilarPRL12} and the qualitative conclusion that TSSs dominates made in Ref \cite{ValdesAguilarPRL12} remains.  

\section{Assumption of nearly equal contribution of the top and bottom surface states}

\begin{figure}[htp]
\includegraphics[width=0.5\columnwidth,angle=0]{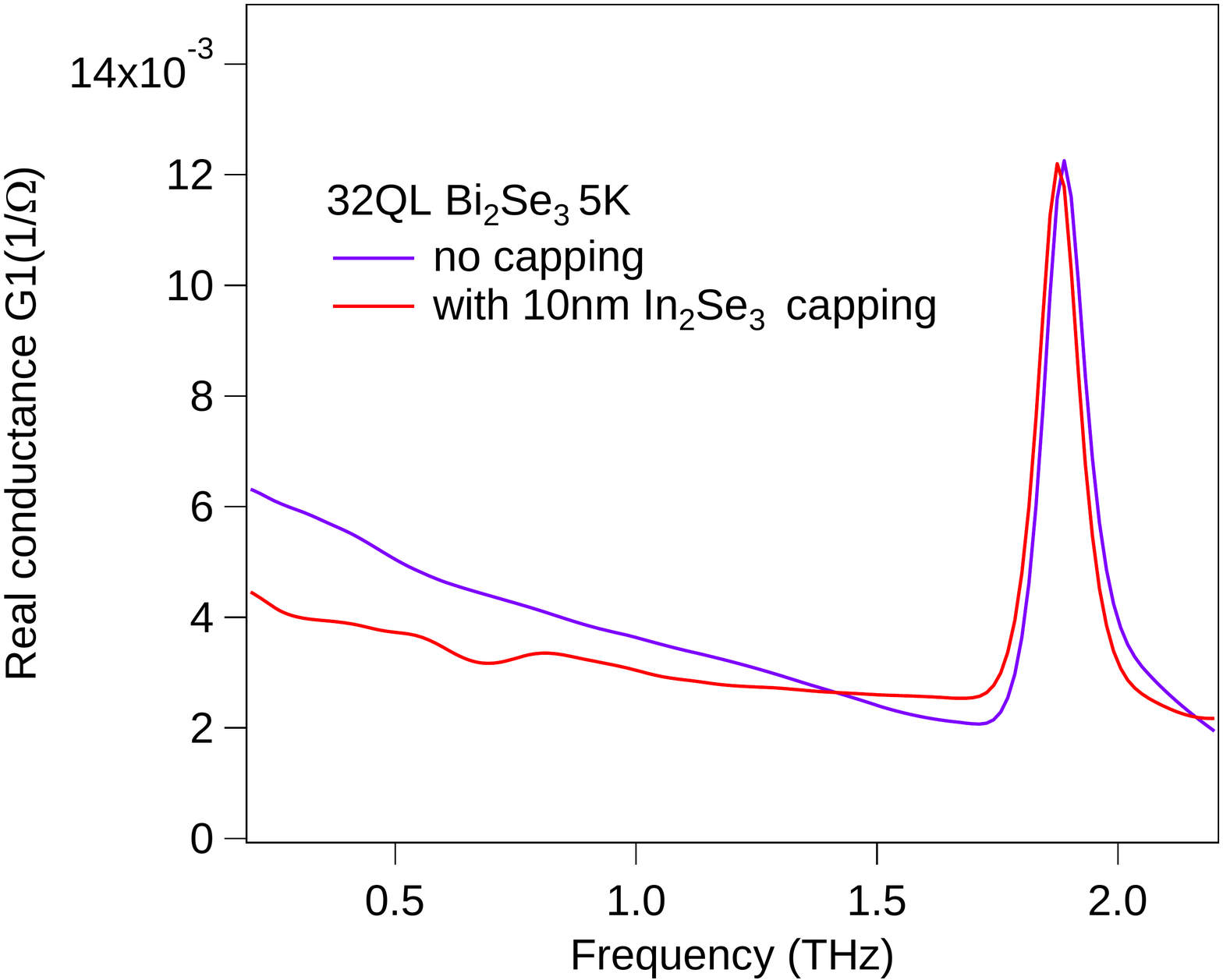}
\caption{(Color online) Real conductance of 32 QL Bi$_2$Se$_3$ with and without 10nm In$_2$Se$_3$ capping at 5 K.}
 \label{SIFig9}
\end{figure} 

In the main text, we showed that the carrier density estimated by using the spectral weight and mass from the Faraday rotation fits is close to the value obtained using spectral weight from Faraday rotation fit and surface state dispersion. This shows the assumption of nearly equal contributions from the top and bottom surface states is a good approximation.  Similarly, DC transport have both shown the chemical potential of the two surface states differed by around 50 meV \cite{XuNatPhys14}. 

A method was discussed in Ref. \cite{JenkinsPRB13} to distinguish the contribution of the two surfaces from each other. Unfortunately, it appears that the interface between In$_2$Se$_3$ and Bi$_2$Se$_3$ which present in the In$_2$Se$_3$ capped samples used in that work is not as simple as assumed in Ref. \cite{JenkinsPRB13}.   It was been found that there is 20-30$\%$ Indium diffusion into the Bi$_2$Se$_3$ layer in a recent study \cite{LeeThinFilms14}.  The topological phase transition occurs near x$\sim$ 6 $\%$ in (Bi$_{1-x}$In$_x$)$_2$Se$_3$ \cite{WuNatPhys13, BrahlekPRL12}. Therefore, the interface of In$_2$Se$_3$ and Bi$_2$Se$_3$ is not the boundary of normal band insulator and topological insulator. Indium probably diffuses into Bi$_2$Se$_3$ layer with a gradient. The true interface TSS must be buried deeply and exist in a background of high disorder. Due to Indium diffusion, the interface which hosts SSs could be  (Bi$_{0.06}$In$_{0.94}$)$_2$Se$_3$ /  (Bi$_{0.045}$In$_{0.955}$)$_2$Se$_3$. This hypothesis may explain that the conduction band minimum is positioned only $\sim$ 80 meV above the Dirac point, because the bulk gap decreases when approaching the topological phase transition point at $x$ $\sim$ 6 $\%$. Also, more Indium substitution reduces the total carrier density \cite{BrahlekPRL12}, which could be the reason that the top surface has a lower carrier density as discussed in Ref. \cite{JenkinsPRB13}.  We also measured a 32 QL Bi$_2$Se$_3$ film capped by 10 nm In$_2$Se$_3$. This sample has lower spectral weight  and a larger scattering rate, as shown in Fig. \ref{SIFig9}.

\section{Role of copper}
The majority of Cu was found to be Cu$^{0+}$ (neutral) by X-ray photo-emission spectroscopy in this batch of samples \cite{BrahlekPRL14}. Our data supports this scenario. If 2$\%$ Cu substitutes for Bi, then we would observe the shift of phonon frequency as we did in the (Bi$_{1-x}$In$_x$)$_2$Se$_3$ case. In (Bi$_{1-x}$In$_x$)$_2$Se$_3$, we can observe a shift of the phonon at x$\sim$ 1$\%$. Also note that Cu has around half of the atomic number of In, which means 2$\%$ Cu substitution would  shift the phonon frequency more than In substituted case. Normal  Bi$_2$Se$_3$ is known to have a  conducting bulk with chemical potential 350$\sim$450 meV due to Se vacancies. In keeping with Ref. \cite{BrahlekPRL14}, we believe Cu incorporation passivates Se vacancies, so the carrier density is reduced. We encourage more work to be done to resolve the role of copper by other methods such as STM measurements and first-principle DFT calculations. 

\section{ Band bending}
Topological Insulators can be understood as special narrow-gap semiconductors with surface states. Therefore, band bending effects can be important in such systems. Ref. \cite{BrahlekPRL14} has a nice summary about band bending in TIs. Here we re-emphasize its importance.  The bulk chemical potential for normal Bi$_2$Se$_3$ is pinned near the conduction band minimum (therefore $\sim$ 220 meV above Dirac point) \cite{AnalytisPRB2010, BrahlekPRL14}, while the surface state chemical potential is $\sim$ 350 meV in our samples. Downward band bending results in accumulation layers. From magneto-THz measurements, we concluded that any accumulation layer carriers have a large scattering rate, are essentially featureless in the Faraday rotation and count for less than 10 $\%$ of total conductance at low frequencies. For Cu$_{0.02}$Bi$_2$Se$_3$, the chemical potential at the surface is $\sim$ 150 meV above the Dirac point,  and therefore upwards band bending must occur.  ARPES can play an essential role in determining whether a material is a topological insulator by counting if the number of surface state branches is odd or even, but it is not the best tool to conclude whether a TI is bulk-insulating or not due to its extreme surface sensitivity and band bending. For example, upwards band bending was reported in bulk-conducting TIs \cite{AnalytisPRB2010} where ARPES did not observe bulk states but SdH saw a dominating bulk contribution. The most effective way to probe bulk-insulating TIs is through transport measurements. In DC transport, if the carrier density contributing to SdH oscillations is equal to the total carrier density measured by the Hall effect, then the bulk is insulating \cite{BrahlekPRL14}. In AC optics, if the carrier density contributing to cyclotron resonance  is equal to the total carrier density in the Drude term of the zero field conductance, one can conclude that the bulk is insulating.

\end{widetext}

\end{document}